\documentstyle [12pt]{article}
\newcommand{\agt}{{\lower 2pt\hbox{$>$} \atop \raise 2pt\hbox{$\sim$}}}
\newcommand{\alt}{{\lower 2pt\hbox{$<$} \atop \raise 2pt\hbox{$\sim$}}}

\begin{document}

\rightline{CMU-HEP94-01}
\rightline{DOE-ER/40682-54}
\rightline{hep-ph/9404241, 1994}
\vspace{5cm}
\begin{center}
{\LARGE\bf A  MODEL OF CP VIOLATION}
\end{center}
\bigskip
\begin{center}
{\large\bf YUE-LIANG WU} \\
 Department of Physics, \\ Carnegie-Mellon University, \\
Pittsburgh, Pennsylvania 15213, U.S.A.
\end{center}
\newpage
\rightline{CMU-HEP94-01}
\rightline{DOE-ER/40682-54}
\rightline{hep-ph/9404241, 1994}
\bigskip
\begin{center}
{\Large\bf A  Model of CP Violation}
\end{center}
\bigskip
\begin{center}
{\bf Yue-Liang Wu \\
\small Department of Physics, Carnegie-Mellon University, \\
\small Pittsburgh, Pennsylvania 15213, U.S.A.}
\end{center}
\bigskip
\date{March, 1994}
\bigskip
\begin{abstract}

    It is shown that a two-Higgs doublet model  with Vacuum CP Violation
and Approximate Global $U(1)$ Family Symmetries (AGUFS) may provide one of
the simplest and attractive models in understanding  origin and mechanisms of
CP violation at the weak scale. The whole new interactions of the model are
explicitly presented here.  It is seen that CP violation can occur, after
spontaneous symmetry breaking, everywhere it can from a single CP-phase in
the vacuum. It is also shown  that "the mechanism of spontaneous symmetry
breaking provides not only a mechanism for giving mass to the bosons and
the fermions, but also a mechanism for generating CP-phase of the bosons and
the fermions".  Four types of CP-violating mechanism are classified. A new
type of CP-violating mechanism is emphasized and  can provide a consistent
application to both the established and the reported CP-violating phenomena.
The smallness of the CKM mixing angles and the induced KM-type CP-violating
effects as well as the suppression of flavor-changing neutral scalar
interactions can be attributed to the AGUFS and Partial Conservation of
Neutral Flavor (PCNF).  This suggests that if one Higgs doublet is necessary
for generation of mass, two Higgs doublets are then needed for origin and
phenomenology of CP violation. Various interesting phenomenological features
arising from this model are analyzed. Their experimental implications and
importance are discussed and emphasized. Directly searching for the exotic
Higgs bosons introduced in this model is worthwhile at both $e^{+}e^{-}$ and
hadron colliders.
\end{abstract}

\newpage

{\bf Contents:}

\begin{enumerate}
\item Introduction
\item The Model \\
      2.1. \  The Lagrangian, AGUFS and PCNF \\
      2.2. \  Vacuum CP Violation (VCPV) \\
      2.3. \  New interactions
\item General Features of the Model \\
      3.1. \  Origin of CP violation \\
      3.2. \  Classification of CP-violating mechanisms \\
      3.3. \  General Features of the Physical Parameters
\item Neutral Meson Mixings in the Model\\
      4.1. \  $K^{0}- \bar{K}^{0}$ mixing \\
      4.1. \  $B^{0}- \bar{B}^{0}$ mixing \\
      4.1. \  $B_{s}^{0}- \bar{B}_{s}^{0}$ mixing and its implication \\
      4.1. \  $D^{0}- \bar{D}^{0}$ mixing and its implication
\item CP-violating Phenomenology of the Model \\
      5.1. \  Indirect CP violation in kaon decay ($\epsilon$) \\
      5.2. \  Direct CP violation in kaon decay ($\epsilon'/\epsilon$) \\
      5.3. \  Direct CP violation in B-system \\
      5.4. \  CP violation in hyperon decays
\item Neutron and Lepton EDMs $d_{n}$ and $d_{l}$ in the Model \\
      6.1. \  $d_{n}$ from one-loop contributions \\
      6.2. \  $d_{n}$ from Weinberg gluonic operator and quark CEDM \\
      6.3. \  $d_{n}$ from FCNSI \\
      6.4. \  One-loop contribution to $d_{l}$ \\
      6.5. \  $d_{l}$ from two-loop Barr-Zee mechanism \\
      6.6. \  $d_{l}$ from FCNSI
\item CP Violation in Other Processes \\
      7.1. \  Muon polarization in $K_{L}\rightarrow \mu^{+}\mu^{-}$ decay \\
      7.2. \  T-odd and CP-odd triple momentum correlation in $B\rightarrow
              D^{\ast} (D\pi) \tau \nu$ decay \\
      7.3. \  Triple spin-momentum correlation in the inclusive decay
               $b\rightarrow c\tau \nu$
\item Are Higgs Bosons Heavy or Light ?  \\
      8.1. \  Implications from experiments at LEP \\
      8.2. \  Implications from $b\rightarrow s \gamma$ decay \\
      8.3. \  Where are the Higgs bosons ?
\item Conclusions and Remarks
\end{enumerate}

 {\bf Appendix}

 {\bf References}

\newpage

\section{Introduction}.

    The 30th anniversary of the discovery of CP violation renews the activity
on
the field of CP violation. Two B-meson factories have been approved to be built
at SLAC and KEK. Two improved experiments NA48 at CERN and E832 at Fermilab
will commence their measurements this year and the goals are a
determination of the direct CP violation in kaon decay with a precision in the
range of $(1-2)\times 10^{-4}$. The experiments: CPLEAR at CERN and the
$\Phi$ Factory will also provide a measurement on the direct CP violation
in kaon decay. Before sheding light on this important field,  which is going
to be discussed in detail in this paper, it should be worthwhile
to briefly review several outstanding landmarks made during this period.

 Originally, it may go back to the discovery of parity nonconservation in
weak interactions by T.D. Lee and C.N. Yang \cite{LY} in 1956 and its
experimental test by C.S. Wu {\it et al} \cite{CSWU} in 1957, which opened
a new field in the particle physics \cite{LOY}. Later on (in 1964),
the experiment by Christenson {\it et al} \cite{CP} further established
CP violation in kaon decay. For history of this development,  it is refered
to the article in \cite{TDL0}.

 In the same year as the one of the discovery of CP violation, another
important observation was known the  so-called
Higgs mechanism  \cite{HIGGS}, i.e.,
a spontaneous symmetry breaking mechanism by which the gauge bosons can get
mass.  It is then clear that the Yang-Mills gauge fields \cite{YM} can become
massive through the mechanism of spontaneous symmetry breaking.
Three years late (in 1967), this mechanism was extended by Weinberg and
Salam \cite{WS} to a realistic $SU(2)\times U(1)$ gauge theory to generate the
mass of the gauge bosons and the fermions. Thus, the electromagnetic and weak
interactions have been well formulated since Glashow's work \cite{GLASH}.
This model, which is usually called Standard Model (SM), is known a very
successful theory in which presently no experimental result contradicts its
predictions. In the SM, a single Higgs doublet of $SU(2)$ was introduced,
which is  sufficient to break the $SU(2)_{L}\times U(1)_{Y}$ symmetry to
$U(1)_{em}$ and give mass to the gauge bosons and the fermions. Nevertheless,
the Higgs sector of the SM has not yet been experimentally tested although
enormous efforts  have been made \cite{LEP,PDG,KANE}.

    On origin and mechanisms of CP violation,  many attempts have been
made  by both theoretists and experimentalists since the discovery of CP
violation. For the history and developments, one can consult many excellent
review articles\cite{CPRV} and references therein. Here,
we only briefly outline the most interesting observations which are relevant
to our discussions in this paper. It is known that a superweak interaction
was suggested, soon  after the experiment, by Wolfenstein\cite{WOLF1} to
account for the observed indirect CP violation (parameterized
by $\epsilon$ with $\epsilon = 2.27\times 10^{-3}$).
In this theory, the direct CP violation ( parameterized by $\epsilon'/
\epsilon$) and the neutron Electric Dipole Moment (EDM) ($d_{n}$) were
predicted to  be unobservable small ($\epsilon'/\epsilon < 10^{-8}$).
In 1973, Kobayashi-Maskawa (KM) \cite{KM} pointed out that CP
violation  can occur in the SM if there exist more than two
families which is nowadays known true \cite{LEP,PDG}. With three families,
there exists, in the extended Cabibbo \cite{CAB} quark mixing matrix
( so-called CKM matrix), one physical KM-phase which comes originally from the
complex Yukawa couplings. With this mechanism, the indirect CP-violating
parameter $\epsilon$ can be fitted and the direct CP violation has been
predicted to be of order $\epsilon'/\epsilon \sim 10^{-3}$ \cite{YLWU1}
which is consistent with the current experimental data
\begin{equation}
Re(\epsilon'/\epsilon) = (2.3\pm 0.7)\times 10^{-3}
\end{equation}
by NA31 Group at CERN\cite{NA31} and
\begin{equation}
Re(\epsilon'/\epsilon) = (0.6\pm 0.69)\times 10^{-3}
\end{equation}
by E731 Group at Fermilab \cite{E731}.

  Whereas with the KM-phase, the neutron EDM was
predicted to be smaller by of order five than the current
experimental limit \cite{SMITH}
\begin{equation}
d_{n} < 1.2 \times 10^{-25}\ \mbox{e cm}
\end{equation}

   CP symmetry in the above two schemes was considered to be explicitly
violated in the interactions, i.e. so-called {\it hard CP violation}.
Therefore we are still uncertain about its origin.  This question has been
, in the meantime, challenged by T.D. Lee
\cite{TDL}. He pointed out that in the gauge theories of spontaneous symmetry
breaking (SSB), CP symmetry could be violated spontaneously, i.e. so-called
Spontaneous CP Violation (SCPV). Thus, the Higgs bosons are responsible for
a CP violation, which provides a new scheme alternative to KM-model to
understand the mechanism of CP violation.

  Later on, Weinberg \cite{SW1} proposed a CP-violating model by extended the
SM
(with a single Higgs doublet) to a three Higgs-doublet model (3HDM).  In this
model the Neutural Flavor Conservation (NFC) which was suggested by Glashow and
Weinberg \cite{GW}, and Paschos\cite{EAP}, was ensured by imposing additional
discrete symmetries  based on the Glashow-Weinberg criterion\cite{GW} for the
Higgs sector. CP violation in this model was assumed to occur only in the
Higgs potential, which provided a new mechanism of CP violation  alternative
to KM-model, namely CP violation occurs through the mixings among the scalar
bosons. In the charged scalar sector, such a CP violation requries at least
three scalars to  have a nontrival physical phase ( We may refer it as
Weinberg-phase which is similar to the KM-phase in the quark mixing matrix).
The charged-scalar mixing matrix in the Weinberg 3HDM has explicitly been
presented in \cite{AST}. CP violation in the Weinberg 3HDM can also occur
through the neutral scalar mixings which was first pointed out by Deshpande
and Ma\cite{DM}.  It was of interest that the Weinberg 3HDM in which CP can
be violated spontaneously may accommodate both indirect- and direct-CP
violation in kaon decay\cite{DPD,DH,HYC1}. Nevertheless, this model seems
to run into difficulties as the predicted neutron EDM appears to be
somewhat too large \cite{SW2,HYC2} when $\epsilon$ fits to the experimental
data, or other way round,  the $\epsilon$ and $\epsilon'/\epsilon$ become
too small once the neutron EDM is accommodated. The most stringent constraint
for its inconsistent may arise from the inclusive decay $b\rightarrow s
\gamma$.  This is because all of these observables receive contributions
from a single Weinberg-phase and are strongly correlated each other.

  The difficulties encountered in the Weinberg 3HDM likely result from
the strong conditions imposed by the Glashow-Weinberg criterion.
Recently, this criterion was challenged in an article
\cite{AHR} in which approximate global flavor symmetries have been
introduced to replace  discrete symmetries.  The parameters
associated with these symmetry breakings have been estimated by Hall and
Weinberg \cite{HW} from observed masses and mixing angles. Based on this,
they found that within their scheme and neglecting possible theoretical
uncertainties from the hadronic matrix elements, the well-established
$K^{0}-\bar{K}^{0}$ and $B^{0}-\bar{B}^{0}$ mixings will restrict the
neutral scalar mass to be of order of
magnitude $m_{H^{0}} > (600-1000)$ GeV. Furthermore, the smallness of the
established indirect CP violation ($\epsilon$) implies that in their model
CP symmetry should be a good approximate symmetry if the scalar masses are
not much larger than those values constrained from $K^{0}-\bar{K}^{0}$ and
$B^{0}-\bar{B}^{0}$ mixings. With these considerations, the direct CP
violation in kaon decay was predicted to be unobservable small
($\epsilon'/\epsilon \sim 10^{-6}$) in their model.

 One of the most interesting models is the simplest extension of the
SM with the introduction of two Higgs doublets, i.e. a general 2HDM which
has been extensively investigated for a long time \cite{2HDM}.
In particular, the SCPV can be realized in such a simple model. Nevertheless,
If SCPV occurs at the electroweak scale, there exists so-called domain-wall
problem\cite{KOZ}. In general, domain-wall problem occurs whenever the
discrete symmetries are broken spontaneously at the electroweak scale.
This problem is therefore encountered by the most of the SCPV models
and the models with spontaneous discrete symmetry breaking at the scale
which is much below the grand unification scale. Nevertheless, the basic
principle that CP violation solely originates from the vacuum remains
so fascinating for origin of CP violation that one may keep it.

  It has recently been shown \cite{YLWU3} that one can relax the
SCPV with a simple {\it demanded} condition, so that the domain-wall
problem does not arise explicitly, but CP violation remains solely
originating  from the vacuum, namely if vacuum has no CP violation,
then the theory becomes CP invariant. Such a CP violation may be simply
refered as a Vacuum CP Violation (VCPV) for convenience of mention
(for a detailed analysis see below).
With VCPV, an alternative 2HDM by considering Approximate Global $U(1)$
Family Symmetries (AGUFS) has been briefly discussed in \cite{YLWU3}.  It
has been pointed out that in general four types of CP-violating
mechanism can be induced from a single CP-phase in the vacuum after
Spontaneous Symmetry Breaking (SSB). In particular, a new type of
CP-violating mechanism has been observed in the model, by which both the
indirect- and direct-CP violation (i.e. $\epsilon$ and $\epsilon'/\epsilon$)
in kaon decay and the neutron EDM can be consistently accommodated.
The smallness of the quark mixing angles in the CKM matrix and the suppression
of the Flavor-Changing Neutral Scalar Interactions (FCNSI) have been
attributed to the AGUFS and can be regarded as being naturally in the sense
of the 't Hooft's criterion \cite{THOOFT}.  In general, without imposing any
additional conditions, the AGUFS will
naturally lead to a Partial Conservation of Neutral Flavor (PCNF).

 These observations suggest that such an
2HDM with VCPV and AGUFS may provide one of the simplest
and attractive models in understanding origin and mechanisms
of CP violation at the weak scale. Therefore, before having any conclusive
experimental test on origin and mechanisms of CP violation, we should
consider all possibilities. These points of view come to our main purpose to
present a detailed description on this model and make a systematic and general
analysis for possible interesting physical phenomena arising from this model.
For Higgs physics, we only briefly discuss the features appearing in
this model. The general perspective on Higgs physics is refered to
the excellent articles in \cite{KANE} and references therein.

\section{The Model}

\subsection{The Lagrangian, AGUFS and PCNF}

  Let us start with the general 2HDM in the  $SU(2)\times
U(1)$ gauge theory, its lagrangian reads

\begin{equation}
L = L_{GF}^{SM} + L_{H} + L_{Y} + V(\phi_a)
\end{equation}
where $L_{GF}^{SM}$ contains the pure Yang-Mills gauge interactions
and the gauge interactions of the fermions and has the same form as the
one in the SM.  $L_{H}$ and  $L_{Y}$ are the gauge
interactions of the scalars and the scalar interactions of the fermions
respectively
\begin{eqnarray}
L_{H} & = & \sum_{a=1}^{n_{H}} (D_{\mu}\phi_{a})^{\dagger} (D^{\mu}\phi_{a}) \\
L_{Y} & = & \bar{q}_{L}\Gamma^{a}_{D} D_{R}\phi_{a} + \bar{q}_{L}\Gamma^{a}_{U}
U_{R}\bar{\phi}_{a} + \bar{l}_{L}\Gamma^{a}_{E} E_{R}\phi_{a} + H.C.
\end{eqnarray}
where $q^{i}_{L}$, $l^{i}_{L}$ and $\phi_{a}$ are $SU(2)_{L}$ doublet quarks,
leptons and Higgs bosons, while $U^{i}_{R}$, $D^{i}_{R}$ and $E^{i}_{R}$
are $SU(2)_{L}$ singlets. $i = 1,\cdots , n_{F}$ is a family label and
$a = 1, \cdots , n_{H}$ is a Higgs doublet label. $\Gamma^{a}_{F}$
($F= U, D, E$) are the arbitrary real Yukawa coupling matrices.

   We now assume that the lagrangian possesses approximate global $U(1)$
family symmetries (AGUFS) which act only on the fermions. This assumption
is based on the known fact that in the limit that CKM matrix is unity, any
model with NFC at tree level generates global $U(1)$ family symmetries, i.e.,
under the global U(1) transformations for each family of the fermions

\[ (U,\ D )_{i}\rightarrow e^{i\alpha_{i}} (U, \ D)_{i} \]
the lagrangian is invariant. Where $\alpha_{i}$ are the constants and depend
only on the family index, namely the up-type and down-type quarks have the
same $U(1)$ charges for each family. Note that these global $U(1)$
transformations are vector-like and act on both the left- and
right-handed fermions.

In the lepton sector, there are exact
global $U(1)$ family symmetries in the SM, i.e. under the transformation

\[ (N,\ E )_{i}\rightarrow e^{i\beta_{i}} (N, \ E)_{i} \]
the lagrangian is invariant. This is because the neutrinos are massless.

In the realistic case, it is known that CKM
matrix deviates only slightly from unity. This implies that at the
electroweak scale any successful models can only possess approximate
global $U(1)$ family symmetries.  Without imposing any additional conditions,
the AGUFS should naturally lead to a Partial Conservation of Neutral Flavor
(PCNF).

 With this assumption, we are motivated to parameterize the Yukawa coupling
matrices in such a convenient way that the global
$U(1)$ family symmetry violations in the charged currents and the neutral
currents can be easily distinguished and the magnitudes of their violations
are characterized by the different sets of parameters.
This consideration  is found to be implemented explicitly by
parameterizing the matrices $\Gamma^{a}_{F}$ in terms of
the following general structure
\begin{equation}
\Gamma^{a}_{F}  =  O_{L}^{F} \sum_{i,j=1}^{n_{F}}\{ \omega_{i} (g_{a}^{F_{i}}
\delta_{ij}  + \zeta_{F} \sqrt{g^{F_{i}}} S_{a}^{F}\sqrt{g^{F_{j}}} )
\omega_{j} \} (O_{R}^{F})^{T}
\end{equation}
with $g^{F_{i}} = | \sum_{a} g^{F_{i}}_{a} \hat{v}_{a} |
/ (\sum_{a} |\hat{v}_{a}|^{2})^{\frac{1}{2}}$ and $\{\omega_{i}, i=1,
\cdots,n_{F}\}$ the set of diagonalized projection matrices
$(\omega_{i})_{jj'} = \delta_{ji}\delta_{j'i}$. $\hat{v}_{a}\equiv
<\phi^{0}_{a}>$ ($a=1, \cdots , n_{H}$) are Vacuum Expectation Values (VEV's)
which will develop from the Higgs bosons after SSB. $g^{F_{i}}_{a}$ are the
arbitrary real Yukawa coupling constants. By convention, we choose
$S_{a}^{F}=0$ for $a=n_{H}$ to eliminate the non-independent parameters.
$S_{a}^{F}$ ($a\neq n_{H}$) are the arbitrary off-diagonal real matrices.
$g^{F_{i}}$ are  introduced so that a comparison between the diagonal and
off-diagonal matrix elements becomes  available. $\zeta_{F}$ is a conventional
parameter introduced to scale the off-diagonal matrix elements with
the normalization $(S_{1}^{F})_{12}\equiv 1$ and $(S_{a}^{F})_{ij}$ being
expected to be of order unity (some elements of $S_{a}^{F}$ may be off
by a factor of 2 or more).  $O_{L,R}^{F}$ are the arbitrary orthogonal
matrices. Note that the above parameterization is general and
applicable to any real matrices,  in the meantime it is  also found to be
very useful and powerful for our purposes in analysing various
interesting physical phenomena (see below).

 In general, one can always choose, by a redifinition of the  fermions,
a basis so that $O_{L}^{F}=O_{R}^{F}\equiv O^{F}$ and $O^{U} = 1$ or
$O^{D} = 1$ as well as $O^{E}=1$ as the neutrinos are considered to be
massless in our present consideration. In this basis, the AGUFS and
PCNF then imply that
\begin{equation}
(O^{F})_{ij}^{2} \ll 1 \  , \qquad i\neq j \ ; \qquad  \zeta_{F}^{2}  \ll 1 \
{}.
\end{equation}
where $O^{F}$ describe the AGUFS in the charged currents and $\zeta_{F}$
mainly characterizes the PCNF. Obviously, if taking $\zeta_{F} =0$ , it turns
to the case  of NFC at tree level\cite{YLWU2}. Furthermore, when
$\zeta_{F} =0$ and $O^{D} = O^{U} = 1$, the theory possesses global
$U(1)$ family symmetries, namely the above lagrangian becomes invaraint
under the global $U(1)$ family
transformation $(U, D)_{i}\rightarrow e^{i\alpha_{i}} (U, D)_{i}$.
We then conclude that the smallness of the CKM mixing angles and the
suppression of the FCNSI can be attributed to the AGUFS and PCNF.
It is actually manifest since  exact global U(1) family symmteries demand
that all the off-diagonal interactions should disappear from the model.

 $V(\phi_{a})$ is the most general Higgs potential subject to the gauge
invariance. As it plays an important role on  origin of CP violation,
we shall seperately discuss it in detail below.

\subsection{Vacuum CP Violation}

 To understand origin of CP violation and prevent the
domain-wall problem from arising explicitly, we observe the following fact that

   {\it In the gauge theories of spontaneous symmetry breaking (SSB),
CP violation can be required solely originating from the vacuum after SSB,
even if CP symmetry is not good prior to the symmetry breaking.} The
{\it demanded condition} for such a statement is:
{\it CP nonconservation occurs only at one place of the interactions in the
Higgs potential}.  In fact, it is this requirement which results that
the vacuum must violate CP symmetry. In particular, this condition may be
simply realized by an {\it universal rule}. That is in a renormalizable
lagrangian all the  interactions with dimension-four  conserve CP and
only  interactions with dimension-two possess CP nonconservation.
It may also be naturally implemented through imposing some symmetries.

   This theorem can be simply demonstrated based on the following observations.
As one knows that the stable conditions
of the Higgs potential after SSB will result in certain relations among the
coefficients of the potential and the VEV's. Suppose that there is only one
place in the potential violating CP, the minimal conditions which contain the
single complex coefficient then result that the imaginary part of this single
complex coefficient has to be zero when all the VEV's are real or have the same
phase. This is because in this case there is no any other complex parameters to
match the single complex coefficient in the equation. With this in mind,
it is not difficult to deduce, in order to have CP violation take place,
the following conseqences:   i)  at least two places in the potential should
violate CP when all the VEV's are real or have the same phase,
ii)  if only one place in the potential
is required to violate CP, then the relative phase of the VEV's must not
be zero, namely vacuum must violate CP. This implies that such an explicit
CP violation is eventually attributed to the one in the vacuum after SSB,
since if vacuum conserves CP, such an explicit CP violation disappears
automatically, iii) if no place violates CP, i.e. CP is good prior to
the symmetry breaking, then CP can only be violated spontaneously, iv)
CP violation occurs everywhere it can in the potential, i.e. both from the
coefficients of the potential and from the VEV's.  Case ii) is the one that
we are interested in, this is because in this case the domain-wall problem
does not arise explicitly but CP violation remains solely originating
from the vacuum.

For convenience of mention, we have refered such a CP violation  as a
Vacuum CP Violation (VCPV). Based on this, it is
of interest in noticing that Weinberg 3HDM can be regarded as a remarkable
example of the VCPV, this is because one can always choose a basis so that
there is only one place in the potential allowing to violate CP  because of
the additional discrete symmetries.

 For a 2HDM which we want to consider in this paper, the most interesting
case is the one with the universal rule stated above. Thus, the
Higgs potential can be simply written in the following general form

\begin{eqnarray}
V(\phi) & = & \lambda_{1}(\phi_{1}^{\dagger}\phi_{1}- \frac{1}{2} v_{1}^{2})^2
+ \lambda_{2}(\phi_{2}^{\dagger}\phi_{2} - \frac{1}{2} v_{1}^{2})^2 \nonumber
\\
& & + \lambda_{3}(\phi_{1}^{\dagger}\phi_{1} - \frac{1}{2} v_{1}^{2})
(\phi_{2}^{\dagger}\phi_{2} - \frac{1}{2} v_{2}^{2})
+ \lambda_{4}[(\phi_{1}^{\dagger}\phi_{1})(\phi_{2}^{\dagger}\phi_{2})
-(\phi_{1}^{\dagger}\phi_{2})(\phi_{2}^{\dagger}\phi_{1})]  \nonumber  \\
& &  + \frac{1}{2}\lambda_{5}(\phi_{1}^{\dagger}\phi_{2} +
\phi_{2}^{\dagger}\phi_{1} - v_{1}v_2 \cos\delta )^2
+ \lambda_{6}(\phi_{1}^{\dagger}\phi_{2}- \phi_{2}^{\dagger}\phi_{1} -
v_{1}v_{2} \sin\delta )^{2}  \\
& & + [\lambda_{7}(\phi_{1}^{\dagger}\phi_{1} - \frac{1}{2} v_{1}^{2})
+ \lambda_{8} (\phi_{2}^{\dagger}\phi_{2} - \frac{1}{2} v_{2}^{2})]
[\phi_{1}^{\dagger}\phi_{2} + \phi_{2}^{\dagger}\phi_{1} -
v_{1}v_2\cos\delta ]  \nonumber
\end{eqnarray}
where the $\lambda_i$ ($i=1, \cdots, 8$) are all real parameters.
If all the $\lambda_i$ are non-negative,  the minimum
of the potential then occurs at $<\phi_{1}^{0} > = v_1 e^{i\delta}/\sqrt{2} $
and $<\phi_{2}^{0} > = v_2/\sqrt{2}$. It is clear that in the above potential
CP nonconservation can only occur through the vacuum, namely $\delta \neq 0$.
Obviously, such a CP violation appears as an explicit one in the potential
when $\lambda_{6} \neq 0$, so that the domain-wall problem does not explicitly
arise. Note that in general one can also demand one of other terms, such as
$\lambda_{5}$ or $\lambda_{7}$ or $\lambda_{8}$ to be complex in a general
potential, whereas such cases seem less interesting comparing to the above case
which obeys the universal rule.

\subsection{New Interactions}

     The physical interactions  are usually given in the mass basis of
the particles. For the simplest 2HDM, the physical basis  after SSB is defined
through
\begin{equation}
f_L = (O_{L}^{F}V_{L}^{f})^{\dagger}F_L \ , \qquad f_R = (O_{R}^{F}P^f
V_{R}^{f})^{\dagger}F_R
\end{equation}
with $V_{L,R}^{f}$ being unitary matrices and introduced to diagonalize
the mass matrices
\begin{equation}
(V_{L}^{f})^{\dagger}(\sum_{i}m_{f_{i}}^{o}\omega_{i} + \zeta_{F} c_{\beta}
\sum_{i,j} \sqrt{m_{f_{i}}^{o}} \omega_{i} S_{1}^{F} \omega_{j}
\sqrt{m_{f_{j}}^{o}} e^{i\sigma_{f}(\delta - \delta_{f_{j}})}) V_{R}^{f} =
\sum_{i} m_{f_{i}}\omega_{i}
\end{equation}
with $m_{f_{i}}$ the masses of the physical states $f_{i}= u_i, d_i, e_i$.
Where $m_{f_{i}}^{o}$ and $\delta_{f_{i}}$ are defined via
\begin{equation}
(c_{\beta} g_{1}^{F_{i}}e^{i\sigma_{f}\delta} +
s_{\beta} g_{2}^{F_{i}})v \equiv \sqrt{2}m_{f_{i}}^{o} e^{i\sigma_{f}
\delta_{f_{i}}}
\end{equation}
 with $v^2 = v_{1}^{2} + v_{2}^{2}= (\sqrt{2}G_{F})^{-1}$,
$ c_{\beta}\equiv \cos\beta = v_1/v$ and $s_{\beta}\equiv \sin\beta
= v_2/v$. $P^{f}_{ij} = e^{i\sigma_{f} \delta_{f_{i}}}\delta_{ij}$,
with $\sigma_{f} =+$, for $f= d, e$, and  $\sigma_{f} = - $, for $f = u$.
Note that by convention we have taken $S_{2}^{F}=0$ to eliminate the
non-independent parameters.

  For convenience of discussions, it is simple to make the
phase convention by writting
\begin{equation}
 V_{L,R}^{f} \equiv  1 + \zeta_{F} T_{L,R}^{f}
\end{equation}

 In a good approximation, to the first order in $\zeta_{F}$ and the lowest
order in $m_{f_{i}}/m_{f_{j}}$ with $i < j$, we find
that $m_{f_{i}}^{2} \simeq (m_{f_{i}}^{o})^{2} + O(\zeta^{2}_{F})$ and
for $i < j$
\begin{eqnarray}
(T_{L}^{f})_{ij} & \simeq & - (T_{L}^{f})_{ji}^{\ast} \simeq c_{\beta}
\sqrt{\frac{m_{f_{i}}}{m_{f_{j}}}} (S_{1}^{F})_{ij} e^{-i\sigma_{f} (\delta
- \delta_{f_{j}})} + O((\frac{m_{f_{i}}}{m_{f_{j}}})^{3/2}, \zeta_{F}) \\
(T_{R}^{f})_{ij} & \simeq & - (T_{R}^{f})_{ji}^{\ast} \simeq c_{\beta}
\sqrt{\frac{m_{f_{i}}}{m_{f_{j}}}} (S_{1}^{F})_{ji} e^{-i\sigma_{f} (\delta
- \delta_{f_{j}})} + O((\frac{m_{f_{i}}}{m_{f_{j}}})^{3/2}, \zeta_{F})
\end{eqnarray}

  In the physical basis, the CP violation is known to occur in the
 {\it charged gauge boson interactions of the quarks } through complex
quark mixing matrix
\begin{equation}
L_{W^{\pm}}^{SM} = \frac{g}{\sqrt{2}}\bar{u}_{L}^{i}V_{ij} \gamma^{\mu}
d_{L}^{j} W_{\mu}^{+} + H.C.
\end{equation}
with $V$ the CKM matrix which has the following form in our model when the
phase convention is given by the eqs.(10) and (13)
\begin{eqnarray}
V  & = & (V_{L}^{U})^{\dagger} (O_{L}^{U})^{T} O_{L}^{D} V_{L}^{D} \equiv
V^{o} + V'  \\
V^{o} & \equiv & (O_{L}^{U})^{T} O_{L}^{D}\ , \qquad
V' \equiv \zeta_{D}[V^{o} T_{L}^{d}] + \zeta_{U}[V^{o T}T_{L}^{u}]^{\dagger}
+ \zeta_{U} \zeta_{D}[ (T_{L}^{u})^{\dagger}V^{o}T_{L}^{d}]
\end{eqnarray}
where $V^{o}$ is a real matrix and $V'$ is a complex matrix.

{\it  The scalar interactions of the fermions} read in the physical basis
\begin{eqnarray}
L_{Y}^{I} & = & (2\sqrt{2}G_{F})^{1/2}\sum_{i,j,j'}^{3}\{
\bar{u}_{L}^{i} V_{ij'} (m_{d_{j'}}\xi_{d_{j'}}\delta_{j'j} + \zeta_{D}
\sqrt{m_{d_{j'}}m_{d_{j}}} S_{j'j}^{d}) d^{j}_{R}H^+
- \bar{d}_{L}^{i} V_{ij'}^{\dagger} (m_{u_{j'}}\xi_{u_{j'}}\delta_{j'j}
\nonumber \\
& & + \zeta_{U}\sqrt{m_{u_{j'}}m_{u_{j}}} S_{j'j}^{u}) u^{j}_{R}H^-
+ \bar{\nu}_{L}^{i} (m_{e_{i}}\xi_{e_{i}}\delta_{ij} + \zeta_{E}
\sqrt{m_{e_{i}}m_{e_{j}}} S_{ij}^{e}) e^{j}_{R}H^+  + H.C. \} \nonumber \\
& & + (\sqrt{2}G_{F})^{1/2}\sum_{i,j}^{3} \sum_{k}^{3}\{ \bar{u}^{i}_{L}
(m_{u_{i}} \eta_{u_{i}}^{(k)} \delta_{ij} + \zeta_{U}\sqrt{m_{u_{i}}m_{u_{j}}}
S_{k,ij}^{u}) u^{j}_{R} + \bar{d}^{i}_{L}
(m_{d_{i}} \eta_{d_{i}}^{(k)} \delta_{ij}  \\
 & & +\zeta_{D}\sqrt{m_{d_{i}}m_{d_{j}}} S_{k,ij}^{d}) d^{j}_{R}
+ \bar{e}^{i}_{L}(m_{e_{i}} \eta_{e_{i}}^{(k)} \delta_{ij} +
\zeta_{E}\sqrt{m_{e_{i}}m_{e_{j}}} S_{k,ij}^{e}) e^{j}_{R}
+ H.C. \} H_{k}^{0} \nonumber
\end{eqnarray}
with
\begin{eqnarray}
 (2\sqrt{2}G)^{1/2} \xi_{f_{i}} m_{f_{i}} & = & g_{1}^{F_{i}}s_{\beta}^{-1}
e^{i \sigma_{f}(\delta - \delta_{f_{i}})} - v^{-1}\sqrt{2}m_{f_{i}} \cot\beta
 \\
\sqrt{m_{f_{i}}m_{f_{j}}} S_{ij}^{f} & = & \sum_{i'j'}s_{\beta}^{-1}\{
\sqrt{m^{o}_{f_{i'}}m^{o}_{f_{j'}}} e^{i \sigma_{f}(\delta - \delta_{f_{j'}})}
[(V_{L}^{F})^{\dagger}\omega_{i'}S_{1}^{F}\omega_{j'}V_{R}^{F}]_{ij}
\nonumber \\
& & + (\zeta_{F}\sqrt{2})^{-1} v g_{1}^{F_{i'}} e^{i \sigma_{f}(\delta -
\delta_{f_{i'}})} [(V_{L}^{F})^{\dagger}\omega_{i'}V_{R}^{F} -
\omega_{i'}]_{ij} \} \ ; \\
\eta_{f_{i}}^{(k)}  = O_{2k}^{H} & + & (O_{1k}^{H} + i \sigma_{f} O_{3k}^{H} )
\xi_{f_{i}} \ ; \qquad S_{k,ij}^{f} = (O_{1k}^{H} + i \sigma_{f}
O_{3k}^{H} )S^{f}_{ij}\ .
\end{eqnarray}
where $O_{ij}^{H}$ is the orthogonal matrix introduced to redefine the
three neutral scalars $ \hat{H}_{k}^{0} \equiv (R, \hat{H}^{0}, I)$
into their mass eigenstates $H_{k}^{0} \equiv (h, H, A)$, i.e. $\hat{H}_{k}^{0}
 =  O^{H}_{kl} H_{l}^{0}$ with
\begin{equation}
(R+iI)/\sqrt{2} = s_{\beta}\phi^{0}_{1}e^{-i\delta} - c_{\beta} \phi^{0}_{2}
\  , \qquad   (v+\hat{H}^{0} +iG^{0})/\sqrt{2} = c_{\beta}\phi^{0}_{1}
e^{-i\delta} + s_{\beta} \phi^{0}_{2}
\end{equation}
here $H_{2}^{0}\equiv H^{0}$ plays the role of the Higgs boson in the
standard model. $ H^{\pm}$ are the charged scalar pair with $ H^{\pm} =
s_{\beta}\phi^{\pm}_{1}e^{-i\delta} - c_{\beta} \phi^{\pm}_{2}$.

In the approximations considered in eqs.(14) and (15), the equations (20) and
(21) are simplified (with convention $i < j$) to
\begin{eqnarray}
& & \xi_{f_{i}} \simeq \frac{\sin\delta_{f_{i}}}
{s_{\beta}c_{\beta}\sin\delta}e^{i \sigma_{f}(\delta - \delta_{f_{i}})}
 - \cot\beta \  ,  \\
& & S^{f}_{ij} \simeq s_{\beta}^{-1} (e^{i\sigma_{f}(\delta - \delta_{f_{j}})}
- \frac{\sin\delta_{f_{j}}}{\sin\delta} ) (S_{1}^{F})_{ij} \  , \qquad
S^{f}_{ji} \simeq  s_{\beta}^{-1} (e^{i\sigma_{f}(\delta - \delta_{f_{i}})}
- \frac{\sin\delta_{f_{j}}}{\sin\delta} ) (S_{1}^{F})_{ji} \  .
\end{eqnarray}

 {\it The gauge interactions of the scalars} read in the unitary gauge
$ L_{H}^{I} = L_{H}^{+} + L_{H}^{-}$ with $L_{H}^{+}$ the CP-even part
\begin{eqnarray}
L_{H}^{+} & = & -\frac{1}{2} \sqrt{g^{2} + g'^{2}} (H^-i\partial^{\mu} H^+ -
H^+i\partial^{\mu} H^-) (\cos 2\theta_{W} Z_{\mu} - \sin 2\theta_{W} A_{\mu})
\nonumber  \\
& & + \frac{1}{4} (g^2 + g'^{2}) \{ \sum_{k=1}^{3} (H_{k}^{0}H_{k}^{0})
[ \frac{1}{2} Z_{\mu}Z^{\mu} + \cos^{2} \theta_{W} W_{\mu}^{+} W^{\mu -} ]  \\
 & & + H^+ H^- [ W_{\mu}^{+} W^{\mu -} + \frac{1}{2} (\cos 2\theta_{W}
Z_{\mu} - \sin 2\theta_{W} A_{\mu})^{2} ] \}  \nonumber
\end{eqnarray}
and $L_{H}^{-}$ the CP-odd part
\begin{eqnarray}
L_{H}^{-} & = & \frac{1}{2} (g^2 + g'^{2}) (\sqrt{2G_{F}})^{-\frac{1}{2}}
\sum_{k=1}^{3} (O_{2k}^{H}H_{k}^{0}) [ \frac{1}{2} Z_{\mu}Z^{\mu} +
\cos^{2} \theta_{W} W_{\mu}^{+} W^{\mu -} ] \nonumber  \\
& & + \frac{1}{2} \sqrt{g^{2} + g'^{2}} \{ -i \sum_{k,l=1}^{3} O_{1k}^{H}
O_{3l}^{H} (H^{0}_{k}i\partial^{\mu} H^{0}_{l} -
H^{0}_{l}i\partial^{\mu} H^{0}_{k}) Z_{\mu} \nonumber  \\
 & & + \cos \theta_{W} [\sum_{k=1}^{3}
(O_{1k}^{H}+ i O_{3k}^{H} ) (H^{0}_{k}i\partial^{\mu} H^{-} -
H^{-}i\partial^{\mu} H^{0}_{k}) W_{\mu}^{+} + H.C. ] \}  \\
& & + \frac{1}{4} (g^2 + g'^{2}) \cos \theta_{W} [\sum_{k=1}^{3}
(O_{1k}^{H}+ i O_{3k}^{H} ) H_{k}^{0}H^{-}W_{\mu}^{+}Z^{\mu} + H.C. ] \nonumber
\end{eqnarray}

   {\it The pure scalar interactions } in the mass eigenstates have the
following general form
\begin{eqnarray}
V_{I}(H) & = & \sum_{k,l,m} \mu_{klm} H_{k}^{0} H_{l}^{0}
H_{m}^{0} + \sum_{k} \mu_{k} H_{k}^{0} H^{+}H^{-} + \lambda_{+}
(H^{+}H^{-})^{2}
\nonumber \\
& & + \sum_{k,l} \lambda_{kl} H_{k}^{0}
H_{l}^{0}H^{+}H^{-} + \sum_{k,l,m} \lambda_{klm} H_{k}^{0} H_{k}^{0}
H_{l}^{0}H_{m}^{0}
\end{eqnarray}
where all the couplings are real and given as  functions of the parameters
$\lambda_{i}$ and VEV's as well as the phase $\delta$, they should not be
presented here as they are too complicated.

\section{General Features of the Model}.

\subsection{Origin of CP Violation}

  From the interactions  given above, it is easily noticed
that from a single CP-phase ($\delta$) in the vacuum, CP violation can occur,
after SSB, everywhere it can in the physical basis.   Such rich vacuum-induced
CP-violating sources will result various significant phenomenological
effects (see below).

   Before proceeding, it may be useful to remark, after having given all
the interactions in the previous section, why origin of CP violation may be
 simply understood by VCPV with only adding one Higgs doublet to the SM,
and how the vacuum-induced CP-violating sources become so rich after SSB.
Firstly, the idea that CP violation solely originates from the vacuum has
actually been motivated from the SCPV,  one has already known that two Higgs
doublets are the minimal number required for SCPV, and so does for VCPV to
take place in the $SU(2)\times U(1)$ gauge theory.  It is of interest in
noticing that if CP is assumed to be good prior to symmetry breakdown,
two cases can occur, i.e. the theory can be either having SCPV or still
conserving CP symmetry after SSB. Whereas the VCPV motivated from solving
the domain-wall problem  requires that the vacuum must violate CP symmetry.
Secondly, the reason for having rich vacuum-induced CP-violating sources in
our 2HDM is also simple, this is because all the quarks and leptons couple to
two Higgs doublets, after SSB, they receive contributions to their mass
from two VEV's and the corresponding two Yukawa couplings,  this can be
explicitly seen from the eqs.(11) and (12). We observe that if the relative
phase between the two VEV's is nonzero,  each fermion ($f_{i}$) is then
characterized not only by its physical mass ($m_{f_{i}}$) but also by
a physical phase

\[ \delta_{f_{i}} \equiv  arg [(c_{\beta} g_{1}^{F_{i}}e^{i\sigma_{f}
\delta} + s_{\beta} g_{2}^{F_{i}})v ]  \sigma_{f} \]

 In general, all the fermions will have different phases and masses since
their two diagonal Yukawa couplings ($g_{1}^{F_{i}}$ and $g_{2}^{F_{i}}$)
are in principle all free parameters.  Thirdly, the AGUFS which we consider
in our model only act on the fermions,  the Higgs potential then has the most
general form with only subject to the gauge invariance and VCPV, therefore the
mass matrix among the three neutral Higgs bosons  is  arbitrary, and so does
the mixing angles. These mixing angles then generate CP-phases

\[ \delta_{H_{k}^{0}} \equiv arg(O_{1k}^{H} + i \sigma_{f} O_{3k}^{H}) \]
which associate to the corresponding neutral scalars $H_{k}^{0}$
(see eqs.(19)-(27)).

 With these observations,  we may conclude that the Higgs mechanism provides
not only a mechanism for generating mass of the bosons and the fermions,
but also a mechanism for creating CP-phase of the bosons and the fermions.

 \subsection{Classification of CP-violating Mechanisms}.

 All these vacuum-induced CP violations can be classified  into four
types of mechanism according to their origins and/or interactions.
To be more clear, we emphasize as follows

 {\bf Type-I.} \  The new type of CP-violating mechanism \cite{YLWU3} which
arises from the induced complex diagonal Yukawa couplings $\xi_{f_{i}}$.
Such a CP violation can occur through both charged- and neutral-scalar
exchanges.

{\bf Type-II.}\  Flavor-Changing SuperWeak (FCSW)-type  mechanism.
This type of mechanism also occurs through both charged- and neutral-scalar
exchanges and is described by the coupling matrices $S_{ij}^{f}$ in our model.

 {\bf Type-III.}\  The induced KM-type CP-violating mechanism  which
is characterized in our model by the complex parameters $\zeta_{F} T_{L}^{f}$
and occurs in the charged gauge boson and  charged scalar interactions of the
quarks.

 {\bf Type-IV.}\  The Scalar-Pseudoscalar Mixing (SPM) mechanism
which is described by the mixing matrix $O_{kl}^{H}$ and the phases
$arg(O_{1k}^{H}+i\sigma_{f}O_{3k}^{H})$. This type of CP
violation appears in the purely bosonic interactions and also in the
neutral-scalar-fermion interactions in our model (in general it can also
occur in the charged-scalar-fermion interactions when there are more than
two charged scalars, for example, the Weinberg 3HDM).

    It should be pointed out that these four types of CP-violating mechanism
are in principle correlated in our model since they originate from one
single CP-phase in the vacuum. Nevertheless, for a given vacuum CP-phase
$\delta$ their induced effective CP-violating phases in the physical
processes are quite independent for various mechanisms. This is because all
the coupling constants of the interactions are in general all different.

\subsection{General Features of the Physical Parameters}

  1) Without making any  additional assumptions, $m_{f_{i}}$, $V_{ij}$,
$\delta_{f_{i}}$ (or $\xi_{f_{i}}$), $\delta$, $\tan\beta$, $\zeta_{F}$,
$(S_{1}^{F})_{ij}$ (or $S_{ij}^{f}$), $m_{H_{k}^{0}}$, $m_{H^{+}}$ and
$O_{kl}^{H}$ are in principle all the
free parameters and will be determined only by the experiments. From the AGUFS
and PCNF,  we can only draw the general features that $V_{ij}^{2} \ll 1$ for
$i \neq j$ and $\zeta_{F}^{2} \ll 1$. The $m_{f_{i}}$ and $V_{ij}$
already appear in the SM and have been extensively investigated. For the
other parameters, it is expected that  $(S_{1}^{F})_{ij}$ are of order unity (
by convention $(S_{1}^{F})_{12}\equiv 1$ ). Moreover, in order to have
the FCNSI be suppressed manifestly, it is in favor of having  $\tan\beta > 1$
and  $|\sin\delta_{f_{j}}/\sin\delta | \alt 1 $ (see eq.(25)).
This means that $v_{1}< v_{2}$ and $|g_{1}^{F_{i}} v_{1} | \alt
\sqrt{2}m_{f_{i}}$. Note that this depends on the convention in the
Yukawa coupling matrices parameterized by the eq.(7), we have already chosen
$S_{2}^{F} = 0$ and $S_{1}^{F}\neq 0$. The diagonal scalar-fermion Yukawa
couplings $\eta_{f_{i}}^{(k)}$ and/or $\xi_{f_{i}}$ can be, for the light
fermions,  much larger than those in the SM  and may, of course, also be
smaller than those in the SM (the latter case appears to happen for heavy
top quark see below). Nevertheless, the former case is more
attractive since it will result in significant interesting phenomenological
effects (see below).  There are actually
several choices for large $\xi_{f_{i}}$ (see eq.(24)): a)  $\tan\beta \gg 1$
with $|\sin\delta_{f_{j}}/\sin\delta | \alt 1 $, b) $\cot\beta \gg 1$ with
$|\sin\delta_{f_{j}}/\sin\delta | \alt 1 $, c) $|\sin\delta_{f_{j}}/
\sin\delta | \gg 1 $ with $\tan\beta \simeq 1$ (i.e $v_{1}\simeq v_{2}$).
We see that case a) coincides with the suppression of the FCNSI.  Case b)
does not favor the suppression of the FCNSI. Case c) requires relative
small $|\sin\delta|$  and becomes possible only if a big cancellation occurs
between two terms in the combination factor
($v_{1}g_{1}^{F_{i}}\cos\delta + v_{2}g_{2}^{F_{i}}$), so as to
generate the known physical mass of the fermions. It appears  that the case a)
is a more favorable choice for large $\xi_{f_{i}}$.  Note that for
a given large value of $\tan\beta$, the values of $\xi_{f_{i}}$ are still
governed by the ratio $\sin\delta_{f_{i}}/\sin\delta$.

  2), one may notice that to the lowest order of $m_{f_{i}}/m_{f_{j}}$
with ($i < j$), the CKM matrix only depends on the neutral flavor-changing
couplings $(S_{1}^{F})_{ij}$ with $i < j$. The other three coupling parameters
$(S_{1}^{F})_{ji}$ with $i < j$ entering into the CKM matrix will be suppressed
by the higher order terms of $m_{f_{i}}/m_{f_{j}}$ with ($i < j$).
Therefore, $(S_{1}^{F})_{ji}$ with $i < j$ are in general less restricted from
the CKM mixings. Nevertheless, it is expected that $(S_{1}^{F})_{ji}$ with
$i < j$  are of the same order of magnitude as the $(S_{1}^{F})_{ij}$ with
$i < j$, we will make further remarks on this point from
discussions of the neutral meson mixings and the neutron and lepton EDMs.

  3), when  $(S_{1}^{F})_{ij}$ are  of order unity,
 the features of the model are then mainly
characterized  by  the important parameters $\tan\beta$ and  $\zeta_{F}$ and
the phases $\delta_{f_{i}}$ (or $\sin\delta_{f_{i}}/\sin\delta $),
and described by the known hierarchic properties of the
fermion masses (i.e.  $m_{f_{i}} \ll m_{f_{j}}$ for $i<j$)  and the
CKM matrix (i.e.  $|V_{12}|\sim |V_{21}| \sim \lambda = 0.22$,
$|V_{23}|\sim |V_{32}| \sim \lambda^{2}$ and $|V_{13}|\sim |V_{31}|
\sim \lambda^{3}/2$).

 4), an interesting feature which has been investigated in \cite{YLWU3}
results from the ansatz that
$\zeta_{D} \ll |V_{13}| \ll |V_{23}| \ll |V_{12}|$ and $\tan\beta \gg 1$.
With this ansatz, the CP-violating effects from the induced KM-type mechanism
and FCSW-type mechanism are negligible, whereas the indirect- and direct-CP
violation (i.e. $\epsilon$ and $\epsilon'/\epsilon$ ) in kaon decay and
the neutron EDM  can be consistently accommodated by the new type of
CP-violating mechanism (i.e. type-I) (a more detailed analysis will
be given below). We will further show  that in this case the
$K^{0}-\bar{K}^{0}$, $B_{d}^{0}-\bar{B}_{d}^{0}$ and
$B_{s}^{0}-\bar{B}_{s}^{0}$  mixings can also receive large contributions
from  short-distance box diagrams through the charged-scalar exchange. In
addition, for such a hierarchy, the masses of the exotic scalars are
unconstrained from the $K^{0}-\bar{K}^{0}$ and $B_{d}^{0}-\bar{B}_{d}^{0}$
mixings. It should be remarked that this simplest extended model,  like the SM,
gives no explanation for such a hierarchic property and also for the
hierarchy of the fermion masses. In particular, we do not yet
understand why $m_{e}/m_{t} < 5\times 10^{-5}$ in the SM.

5), the phase convention in this model becomes nontrivial due to the existence
of the FCNSI, which is unlike the case with neutral flavor conservation.
Consequently, all the six independent phases in the unitary matrix $V$ are
now all physical observables, which distinguishes from the standard KM-model
in the SM.  In the phase convention given by the eqs. (10) and (13), all
of the CKM matrix elements $V_{ij}$ in the eq.(17) are  complex. One can, of
course,  rephase the quark fields to make $V$   have the same phase convention
as the one in the standard KM-model, i.e.  with only one physical CP-phase,
while the phases of $S^{f}_{ij}$ in the flavor-changing scalar interactions
will be changed correspondingly.  This requires that one should first specify
the phase convention before discussing the contributions to physical
observables
from various mechanisms.

 6) The naturalness of the small CP-violating effects in the induced
KM-type mechanism can now be simply understood because it is directly
related to the FCNSI, whose suppression has been attributed to the PCNF
resulting naturally from the AGUFS. It becomes clear based on the fact that
when the FCNSI at tree level disappear, namely the theory has NFC at tree
level, the CKM matrix is known to be real \cite{BRANKO}.

  7),  when $(S_{1}^{F})_{ij} \sim O(1)$ ($(S_{1}^{F})_{12} \equiv 1$),
the complex part $V'$ of the CKM matrix $V$ possesses a similar
hierarchic structure as the $V$ due to the hierarchic properties of the
quark masses (see eqs.(18) and (14)), so does the real part $V^{o}$.
When $\zeta_{F} c_{\beta}$ is very small
( $\zeta_{F} c_{\beta} < 10^{-2}$), the amplitudes of the measured CKM matrix
elements are mainly attributed to the real part $V^{o}$. However, as
$\zeta_{F} c_{\beta}$ increase to the values that
$\zeta_{D} c_{\beta}\sim 0.1$ and/or $\zeta_{U} c_{\beta}\sim 0.3$ for
$m_{t} \sim 150$ GeV,  another interesting feature appears in our model.
That is,  for those orders of magnitude  of the $\zeta_{F} c_{\beta}$ the
amplitudes of the complex elements $|V'_{23}|$, $|V'_{32}|$, $|V'_{13}|$
and $|V'_{31}|$ approach to the values of the precent experimental
measurements (note that $|V'_{12}|$ is still much smaller than the Cabibbo
angle so that  $|V_{12}|$ is still dominated by $V^{o}_{12}$). This implies
that when the parameters $\zeta_{F} c_{\beta}$ become relative large,
the CP-violating effects from the induced KM-type mechanism could be
comparable with those from the standard KM-model. This is because the
corresponding vacuum-induced KM-phase  in this case could be as large as
possible. It should be noted that before extract the KM-phase one should
change the previous phase convention to a  CKM matrix with a standard phase
convention. Actually,  the above naive phase convention is closed to the
standard phase convention in the KM-model, this can be easily seen by
expressing $V_{ij} =|V_{ij}| e^{i\theta_{ij}}$, since $\theta_{ij}$ have
opposite hierarchic properties to the $|V_{ij}|$, namely $|\theta_{ii}| \ll
|\theta_{ij}|$ for $i\neq j$ and $|\theta_{12}| < |\theta_{23}| \alt
|\theta_{13}|$ if all the elements $(S_{1}^{F})_{ij}$ have values which are of
the same order of magnitude.

8),  one may notice that: i) when $(S_{1}^{F})_{ij}\sim O(1)$
and $|\sin\delta_{f_{i}}/\sin\delta | \alt 1 $, the FCNSI mainly  rely on
the ratio $\zeta_{F}/s_{\beta}$  and its suppression is in favor of the
small values of $\zeta_{F}/s_{\beta}$, ii) the new type of CP-violating
mechanism also favors to have large $s_{\beta}$ (i.e. $\tan\beta \gg 1$) which
coincides with the i), iii) the induced KM-type mechanism favors
 large values of $\zeta_{F} c_{\beta} \equiv (\zeta_{F}
/s_{\beta}) c_{\beta} s_{\beta}$. When the values of $\zeta_{F}/s_{\beta}$ are
fixed from the requirement of the suppression of the FCNSI,
the maximum of the quantity $\zeta_{F} c_{\beta}$ occurs when $c_{\beta} =
s_{\beta} = 1/\sqrt{2}$, i.e. $v_{1} = v_{2}$. From these observations together
with the feature 6), we see that the induced
KM-type mechanism is going to become important when
$\zeta_{D}/s_{\beta}\agt 0.2$ and $\zeta_{U}/s_{\beta}\agt 0.6$ for
$c_{\beta}\sim s_{\beta}$. To fit the $K^{0}-\bar{K}^{0}$ and
$B^{0}-\bar{B}^{0}$ mixings, it will be seen that the constraint on
$\zeta_{D}/s_{\beta}$ is propotional to the neutral scalar mass. With this
in mind, we can conclude quanlitatively that for relatively small mass of
the scalars, namely the ratio $\zeta_{F}/s_{\beta}$ must also be small,
the induced KM-type mechanism becomes unimportant and the new type of CP
violating mechanism should play an important role, while for large mass of
the  scalars  either the new type of the mechanism or the induced KM-type
mechanism can be significant.

9) It is seen that in this model CP violation also occurs in the purely
bosonic and leptonic interactions.

10)   we would like to comment that if further imposing  additional
approximate discrete symmetries  on the fermions and the Higgs
bosons, it then results in more stringent constraints on the diagonal Yukawa
coupling constants. Specifically, one type of the diagonal Yukawa couplings
must be small.  For example, considering the following approximate
discrete symmtery: $\phi_{1} \rightarrow - \phi_{1}$, $\phi_{2} \rightarrow
\phi_{2}$ , $D_{R} \rightarrow - D_{R}$, $U_{R} \rightarrow U_{R}$, it then
implies that $g_{1}^{U_{i}} \ll g_{2}^{U_{i}}$ and
$g_{2}^{D_{i}} \ll g_{1}^{D_{i}}$. In addition, the coefficients
$\mu_{12}^{2}$, $\lambda_{7}$ and $\lambda_{8}$ in the potential must also
be small. To consider a more general case, we should not impose any additional
approximate discrete symmetries on the lagrangian. Only the AGUFS
will be considered in our model.

\section{Neutral Meson Mixings in the Model}

 In the standard model, it is known that the neutral meson mixings arise from
the familar box diagram through two-W-boson exchange. In our model,
more significant contributions to the neutral meson mixings can arise from
the new scalar interactions of the fermions.  One such a contribution arises
from the box diagrams with charged-scalar exchange. Another one
comes from the neutral scalar
exchange at tree level because of the existance of the FCNSI in the model, its
contributions to  the mixing mass matrix between the neutral meson $P^{0}$
and $\bar{P}^{0}$ can be easily obtained.  Let $P^{0}$  the bound state of
two quarks with quantum number $P^0 \equiv (\bar{f}_{i}\gamma_{5}f_{j})$,
we then find

\begin{eqnarray}
M_{12}^{H^{0}} & = & <P^0 | H_{eff}^{H^{0}} | \bar{P}^{0} > \nonumber \\
& = & \frac{G^{2}}{12\pi^{2}} f_{P^{0}}^{2} \tilde{B}_{P^{0}}m_{P^{0}}
(\sqrt{\frac{m_{f_{i}}}{m_{f_{j}}}})^{2}
(1+\frac{m_{f_{i}}}{m_{f_{j}}})^{-1} m_{f'_{j}}^{2}\sum_{k}
(\frac{2\sqrt{3} \pi v m_{P^{0}}} {m_{H_{k}^{0}}m_{f'_{j}}}
\frac{\zeta_{F}}{s_{\beta}})^{2} (Y_{k,ij}^{f})^{2}
\end{eqnarray}
with

\[ (Y_{k,ij}^{f})^{2}  =  (Z_{k,ij}^{f})^{2} + \frac{1}{2}r_{P^{0}}
S_{k,ij}^{f} S_{k,ji}^{f \ast}, \qquad Z_{k,ij}^{f} =
-\frac{i}{2} (S_{k,ij}^{f} - S_{k,ji}^{f \ast}) \]
where the formula is expressed in a form which is convenient in comparison
with the one obtained from the box diagram in the standard model, here
$\sqrt{m_{f_{i}}/m_{f_{j}}}$ with convention $i<j$ plays the role of
the CKM matrix element $V_{ij}$, and $m_{f'_{j}}$ is introduced to
correspond to the loop-quark mass of box diagram. Namely $f'_{j}$ and
$f_{j}$ are the two quarks in the same weak isospin doublet.  Note that the
result is actually independent of $m_{f'_{j}}$. $m_{f_{i}}$ are understood the
current quark masses. In our follwoing nemerical estimations we will use
$m_{u}= 5.5$ MeV, $m_{d}= 9$ MeV, $m_{s}= 180$ MeV, $m_{c}= 1.4$ GeV  and
$m_{b}= 6$ GeV with being defined at a renormalization
scale of $1$ GeV. $f_{P^{0}}$ and $m_{P^{0}}$ are the leptonic decay
constant (with normalization $f_{\pi}=133$MeV) and the mass of the meson
$P^{0}$ respectively. $\tilde{B}_{P^{0}}$ and $\tilde{r}_{P^{0}}$ are bag
parameters defined by
\begin{eqnarray}
& & <P^0 |(\bar{f}_{i}(1\pm \gamma_{5})f_{j})^{2} | \bar{P}^{0} > =
- \frac{f_{P^{0}}m_{P^{0}}^{3}}{(m_{f_{i}}+m_{f_{j}})^{2}} \tilde{B}_{P^{0}} \\
& & 1 + \tilde{r}_{P^{0}} = - \frac{<P^0 |\bar{f}_{i}(1\pm \gamma_{5})f_{j}
\bar{f}_{i}(1\mp \gamma_{5})f_{j} | \bar{P}^{0} >}{<P^0 |\bar{f}_{i}(1\pm
\gamma_{5})f_{j} \bar{f}_{i}(1\pm \gamma_{5})f_{j} | \bar{P}^{0} >}
\end{eqnarray}
In the vacuum saturation and factorization approximation with the limit of
a large number of colors, we have  $\tilde{B}_{P^{0}}\rightarrow 1$ and
$\tilde{r}_{P^{0}}\rightarrow 0$, thus  $Y_{k,ij}^{f} = Z_{k,ij}^{f}$.

 Let us now discuss in detail for all the neutral meson mixings.

\subsection{$K^0-\bar{K}^{0}$ Mixing}

 The mass and width differences have been well established in the
$K^0-\bar{K}^{0}$ system \cite{PDG}. As $\Delta m_{K} \simeq
\Delta \Gamma_{K}/2$, the mass difference is then given by
\begin{equation}
\Delta m_{K} \simeq 2Re M_{12} \equiv 2Re (M_{12}^{WW} + M_{12}^{HH} +
M_{12}^{HW} + M_{12}^{H^{0}} + M'_{12} )
\end{equation}
where $M_{12}^{WW}$,  $M_{12}^{HH}$ and $M_{12}^{HW}$ are the contributions
from box diagrams through two $W$- boson, two charged-scalar $H^{+}$  and
one $W$- boson and one charged-scalar exchanges, respectively.
$M_{12}^{H^{0}}$ is the one from the FCNSI
through one neutral scalar exchange at tree level and is easily read off from
the previous general result. $M'_{12}$ presents other possible contributions,
such as two-coupled penguin diagrams and nonperturbative effects.  Their
result is expressed in terms of an effective Hamiltonian
\begin{eqnarray}
H_{eff}^{WW} & = & -\frac{G^{2}}{16\pi^{2}} m_{W}^{2} \sum_{i,j}^{c,t}
\eta_{ij}\lambda_{i} \lambda_{j} \sqrt{x_{i}x_{j}} B^{WW}(x_{i}, x_{j})
\bar{d}\gamma_{\mu}(1-\gamma_{5})s \bar{d}\gamma^{\mu}(1-\gamma_{5})s \\
H_{eff}^{HH} & = & -\frac{G^{2}}{16\pi^{2}} m_{W}^{2} \sum_{i,j}^{u,c,t}
\eta_{ij}^{HH} \lambda_{i} \lambda_{j} \frac{1}{4}\{ B_{V}^{HH}(y_{i}, y_{j})
[ \sqrt{x_{i}x_{j}}\sqrt{y_{i}y_{j}} |\xi_{i}|^{2}|\xi_{j}|^{2} \nonumber \\
& & \cdot \bar{d}\gamma_{\mu}(1-\gamma_{5})s \bar{d}\gamma^{\mu}
(1-\gamma_{5})s  + \sqrt{x_{s}x_{d}}\sqrt{y_{s}y_{d}}
\xi_{s}^{2}\xi_{d}^{\ast 2} \bar{d}\gamma_{\mu}(1+\gamma_{5})s
\bar{d}\gamma^{\mu}(1+\gamma_{5})s  \nonumber \\
& & + 2 \sqrt{x_{i}x_{j}} \sqrt{y_{s}y_{d}}
\xi_{s} \xi_{d}^{\ast} \xi_{i} \xi_{j}^{\ast} \bar{d}\gamma_{\mu}
(1+\gamma_{5})s \bar{d}\gamma^{\mu}(1-\gamma_{5})s ] \\
& & + B_{S}^{HH}(y_{i}, y_{j}) \sqrt{x_{i}y_{j}}[ x_{d} \xi_{d}^{\ast 2}
\xi_{i}^{\ast} \xi_{j}^{\ast} \bar{d}(1-\gamma_{5})s \bar{d}(1-\gamma_{5})s
\nonumber \\
 & & + x_{s}\xi_{s}^{2}\xi_{i}\xi_{j} \bar{d}(1+\gamma_{5})s
\bar{d}(1+\gamma_{5})s + 2 \sqrt{x_{s}x_{d}}\xi_{s} \xi_{d}^{\ast}
\xi_{i} \xi_{j}^{\ast} \bar{d}(1+\gamma_{5})s \bar{d}(1-\gamma_{5})s ] \}
\nonumber \\
H_{eff}^{HW} & = & -\frac{G}{16\pi^{2}} m_{W}^{2} \sum_{i,j}^{u,c,t}
\eta_{ij}^{HW}\lambda_{i} \lambda_{j} \{ 2 \sqrt{x_{i}x_{j}}\sqrt{y_{i}y_{j}}
\xi_{i}\xi_{j}^{\ast} B_{V}^{HW}(y_{i}, y_{j}, y_{w}) \nonumber \\
& & \cdot \bar{d}\gamma_{\mu}
(1-\gamma_{5})s \bar{d}\gamma^{\mu}(1-\gamma_{5})s  + (y_{i}+y_{j})
\sqrt{x_{d}x_{s}} \xi_{s}\xi_{d}^{\ast} [B_{T}^{HW}(y_{i}, y_{j},y_{w}) \\
& & \cdot \bar{d}\sigma_{\mu\nu}(1-\gamma_{5})s \bar{d}\sigma^{\mu\nu}
(1+\gamma_{5})s  + B_{S}^{HW}(y_{i},y_{j},y_{w}) \bar{d}(1-\gamma_{5})s
\bar{d}(1+\gamma_{5})s ] \} \nonumber
\end{eqnarray}
where the $B^{WW}$, $B_{V}^{HH}$, $B_{S}^{HH}$, $B_{V}^{HW}$, $B_{S}^{HW}$
and $B_{T}^{HW}$ arise from the loop integrals and are the functions of
$x_{i}=m_{i}^{2}/m_{W}^{2}$ and $y_{i}=m_{i}^{2}/m_{H}^{2}$ with $i=u,c,t,W$,
their explicit expressions are presented in the Appendix. $\eta_{ij}$,
$\eta_{ij}^{HH}$ and $\eta_{ij}^{HW}$  are the possible QCD corrections
and $\lambda_{i} = V_{is} V_{id}^{\ast}$.
Note that in obtaining above results
the external momentum of the d- and s-quark has been neglected. Except this
approximation which is reliable as their current mass is small, we keep all the
terms. This is because all the couplings $\lambda_{i}$ and $\xi_{i}$ are
complex in our model,  even if some terms are small, they can still play an
important role on CP violation since the observed CP-violating effect in
kaon decay is of order $10^{-3}$.

  It is known that $H_{eff}^{WW}$ contribution to $\Delta m_{K}$ is dominated
by the c-quark exchange and its value is still uncertain due to the large
uncertainties of the hadronic matrix element
\begin{equation}
 <K^0 |(\bar{d}\gamma_{\mu}(1-\gamma_{5})s)^{2} | \bar{K}^{0} > =
- \frac{8}{3} f_{K}^{2}m_{K}^{2} B_{K}
\end{equation}
where $B_{K}$ ranges from $1/3$ \cite{DGH1} (by the PCAC and $SU(3)$ symmetry),
$3/4$ \cite{BAG} (in the limit of a large number of colors) to $1$ \cite{GL}
(by the vacuum insertion approximation). The results from QCD sum rule and
Lattice calculations lie in this range. For small $B_{K}$, the
short-distance $H_{eff}^{WW}$ contribution to $\Delta m_{K}$ fails badly to
account for the measured mass difference.

  The $H_{eff}^{HH}$ and $H_{eff}^{HW}$ contributions to $\Delta m_{K}$ depend
on the couplings $m_{f_{i}}\xi_{f_{i}}$ and the mass of the charged scalar
($m_{H^{+}}$). The result can be of the same order of magnitude as the one
from the standard box diagram contribution or even bigger.  To have a numerical
intuition, let us assume that  $|\xi_{d}|\sim |\xi_{s}|\sim |\xi_{c}|$,
so that the terms proportional to the masses $m_{s}$ and $m_{d}$ are
negligible relative to those proportional to $m_{c}$. Since
$|\lambda_{t}|\ll |\lambda_{c}|$, it is easily seen that the charm-quark
contribution to  $\Delta m_{K}$ is then dominant.  For convenience, we
introduce a useful positive quantity via
\begin{equation}
\xi_{o}^{2}(m_{i}, m_{H^{+}}) \equiv -\frac{4B_{V}^{HW}(y_{i}, y_{w})}
{B_{V}^{HH}(y_{i})} + \frac{1}{y_{i}B_{V}^{HH}} \{[4y_{i}B_{V}^{HW}(y_{i},
y_{w})]^{2} + 4 y_{i}B_{V}^{HH}(y_{i})B^{WW}(x_{i}) \}^{1/2}
\end{equation}
It is not difficult to show that when
\begin{equation}
|\xi_{c}|^{2} \geq ( \leq )\  \xi_{o}^{2}(m_{c}, m_{H^{+}}) \simeq -
4B_{V}^{HW}
(y_{c}, y_{w}) + \{[4B_{V}^{HW}(y_{c}, y_{w})]^{2} +
4m_{H^{+}}^{2}/m_{c}^{2}\}^{1/2}
\end{equation}
it then correspondingly has
\begin{equation}
\Delta m_{K}^{(HH+HW)} \geq ( \leq )\  \Delta m_{K}^{WW}
\end{equation}
where a good approximation that $B_{V}^{HH}(y_{c})
\sim 1$ for $y_{c}\ll 1$ and $B^{WW}(x_{c})\sim 1$ for $x_{c}\ll 1$ have been
used. We have also neglected the diffrences of the QCD corrections among the
three types of the box diagrams.  Numerically, we find that
\begin{eqnarray}
|\xi_{c}| & \geq ( \leq ) & \xi_{o}(m_{c}, m_{H^{+}}) \simeq 7\ , \qquad
for \qquad m_{H^{+}} = 50 GeV\ , \nonumber \\
|\xi_{c}| & \geq (\leq ) & \xi_{o}(m_{c}, m_{H^{+}}) \simeq 12\ , \qquad
for \qquad m_{H^{+}} = 100 GeV\ .
\end{eqnarray}
this requirement can be easily implemented by simply taking $\tan\beta \gg 1$.
when $|\xi_{c}| > \xi_{o}(m_{c}, m_{H^{+}})$, the contribution to
$\Delta m_{K}$ from box diagram with charged-scalar exchange will dominate over
the one with W-boson exchange in the standard model.

 We then believe that by including the new contributions to $K^{0}-\bar{K}^{0}$
mixing from the box diagram through the charged-scalar exchange, the
experimental data for the mass difference can be reproduced by a purely
short-distance analysis.  For instance, taking $\eta_{cc} = 0.66$ and
$B_{K}=0.7$, we then find that when

\[|\xi_{c}| = \xi_{o}(m_{c}, m_{H^{+}})
\simeq 9 (13), \qquad  for \qquad m_{H^{+}} = 50 (100) GeV \]
the experimental data is then fitted via the contributions

\[ \Delta m_{K}^{WW} \simeq
1/3 \Delta m_{K}^{exp.}, \qquad  \Delta m_{K}^{HH+HW} \simeq
2/3 \Delta m_{K}^{exp.} \]
it is seen that the contribution to $\Delta m_{K}$ is dominated in this case
by the box diagrams with charged scalar exchange.

  So far we do not yet consider the $M_{12}^{H^{0}}$ contribution to
$\Delta m_{K}$. Depending on the parameter $\zeta_{D}$ and the mass of the
neutral scalars, its magnitude can be in general large because it arises
from a tree level diagram
\begin{equation}
M_{12}^{H^{0}} = \frac{G^{2}}{12\pi^{2}} f_{K}^{2} \tilde{B}_{K}m_{K}
(\sqrt{\frac{m_{d}}{m_{s}}})^{2}
(1+\frac{m_{d}}{m_{s}})^{-1} m_{c}^{2}\sum_{k}
(\frac{2\sqrt{3}\pi  v m_{K}} {m_{H_{k}^{0}}m_{c}}
\frac{\zeta_{D}}{s_{\beta}})^{2} (Z_{k,12}^{d})^{2}
\end{equation}
For a numerical estimation,
let us assume that $\Delta m_{K}^{H^{0}} = 2Re M_{12}^{H^{0}}\equiv
D_{K}^{H^{0}} \Delta m_{K}^{exp.}$ with $\Delta m_{K}^{exp.} =
3.5\times 10^{-6}$ eV. Without accidental cancellations, it is expected that
$Re (Y_{k,12}^{d})^{2} \sim O(1)$ (note that $(S_{1}^{D})_{12}\equiv 1$).
This yields
\begin{equation}
\Delta m_{K}^{H^{0}} = D_{K}^{H^{0}} \Delta m_{K}^{exp.} = 3.5 \times 10^{-6}
\tilde{B}_{K} (\frac{10^{3} GeV}{m_{H_{k}^{0}}}
\frac{\zeta_{D}}{s_{\beta}})^{2} \frac{Re (Y_{k,12}^{d})^{2}}{1} eV
\end{equation}
Assuming that this contribution is a dominant one, i.e. $D^{H^{0}}_{K}\sim 1$,
which may occur  when $B_{K}$ takes its small value and  $|\xi_{c}| <
\xi_{o}(m_{c}, m_{H^{+}})$. Taking $\tilde{B}_{K}=1$ ( i.e. the value in the
vacuum insertion approximation), we then obtain a constraint
\begin{equation}
\zeta_{D}/s_{\beta} < 10^{-3} m_{H_{k}^{0}}/GeV
\end{equation}
 Whereas, when the mass
difference is accommodated by the box diagrams through the W-boson and
charged-scalar exchanges, either $\zeta_{D}$ should be  much smaller
or the neutral scalars must be very heavy.  From our considerations, the former
is more natural because it can be attributed to the AGUFS and PCNF. In fact,
we do not expect that within the framework of the electroweak theory the
scalar mass would be larger than the symmetry breaking scale.

  In general, we obtain
\begin{eqnarray}
\Delta m_{K} & = & \frac{G^{2}}{6\pi^{2}} f_{K}^{2} B_{K}m_{K} m_{c}^{2}
\sin^{2} \theta \{ \eta_{cc} B^{WW}(x_{c}) + \frac{1}{4} \eta_{cc}^{HH} y_{c}
|\xi_{c}|^{4}  B_{V}^{HH}(y_{c}) \nonumber \\
& & + 2 \eta_{cc}^{HW} y_{c} |\xi_{c}|^{2}  B_{V}^{HW}(y_{c}, y_{w})
+\frac{\tilde{B}_{K}}{B_{K}} \sum_{k}
(\frac{2\sqrt{3} \pi v m_{K}} {m_{H_{k}^{0}}m_{c}}
\frac{\zeta_{D}}{s_{\beta}})^{2} Re (Y_{k,12}^{d})^{2} \}
\end{eqnarray}
which is subject to the experimental constraint
\begin{equation}
\Delta m_{K} = 3.5 \times 10^{-6} eV \simeq \frac{G^{2}}{6\pi^{2}} f_{K}^{2}
 m_{K} m_{c}^{2} \sin^{2} \theta \  \sqrt{2}
\end{equation}

 \subsection{$B^{0}_{d}-\bar{B}^{0}_{d}$ Mixing}

The effective Hamiltonian for  $B^{0}_{d}-\bar{B}^{0}_{d}$ Mixing is
calculated  with the aid of the box diagrams in full analogy to the
treatment of the $K^{0}-\bar{K}^{0}$ system. Its explicit expression can be
simply read off from the one for $K^{0}-\bar{K}^{0}$ system by a corresponding
replacement: $s\leftrightarrow b$. The "standard approximation" made there,
namely neglecting the external momenta of the quarks,  is also reliable
since dominant contributions come from the intermediate top quark.
With this analogy, the considerations and discussions on $K^{0}-\bar{K}^{0}$
mixing  can be applied to the $B^{0}_{d}-\bar{B}^{0}_{d}$
mixing for the contributions from box diagrams. As it is expected that
$|\Gamma_{12}|/2 \ll |M_{12}|$ in the B-system (which is different from
K-system), the mass difference for $B^{0}_{d}-\bar{B}^{0}_{d}$ system is
given by $\Delta m_{B} \simeq 2 |M_{12}|$.  Similar to the K-system, we have
that when
\begin{equation}
|\xi_{t}|^{2} \sim \xi_{o}^{2}(m_{t}, m_{H^{+}})
\end{equation}
the $ |(M_{12}^{HH} + M_{12}^{HW})|$ becomes comparable with
$|M_{12}^{WW}|$ in $B^{0}_{d}-\bar{B}^{0}_{d}$ mixing.
Numerically, we find that for top quark mass $m_{t} = 150$ GeV
\begin{eqnarray}
|\xi_{t}| & \sim & \xi_{o}(m_{t}, m_{H^{+}}) \simeq 0.84\ , \qquad for \qquad
m_{H^{+}} = 50 GeV\ ,
\nonumber \\
|\xi_{t}| & \sim & \xi_{o}(m_{t}, m_{H^{+}}) \simeq 0.90\ , \qquad for \qquad
m_{H^{+}} = 100 GeV\ .
\end{eqnarray}
it indicates that when top quark mass around $m_{t} \sim 150$GeV,
$\xi_{o}(m_{t}, m_{H^{+}})$ is not so sensitive to the
charged scalar mass as it varies from  $m_{H^{+}}\sim 50$ GeV to
$m_{H^{+}}\sim 150$ GeV.

  We then conclude that for  $m_{t} \sim 150$GeV and $|\xi_{t}|\simeq 1 $
(if the top quark mass is relatively small one can
take relatively large $|\xi_{t}|$ ),
the contribution to the mass  difference $\Delta m_{B_{d}}$ from box digrams
through the charged-scalar exchange is of the same order of magnitude as the
one through two-W-boson exchange when the charged-scalar mass lies in
a reasonable range $m_{H^{+}}\sim (50-150)$GeV. This shows that
once top quark is discovered,  such a conclusion implies
that either the CKM matrix element $|V_{td}|$ or the hadronic
matrix element $\sqrt{B_{B}\eta_{QCD}} f_{B}$ is not necessary to be as large
as the one required from fitting the observed
$B^{0}_{d}-\bar{B}^{0}_{d}$ mixing in the standard model, or in other way
round,
if $|V_{td}|$ and $\sqrt{B_{B}\eta_{QCD}} f_{B}$ are known, thus
$B^{0}_{d}-\bar{B}^{0}_{d}$ mixing will provide a constraint on either the
mass of the charged scalar or the coupling constant $\xi_{t}$.

   Let us now discuss the effect from the FCNSI. Assuming that $M_{12}^{H^{0}}$
is dominant, which could happen only when the CKM matrix element
$|V_{td}|$ and/or the hadronic matrix element $\sqrt{B_{B}\eta_{QCD}} f_{B}$
and/or top quark mass take(s) their (its) minimal values as well as
$|\xi_{t}|\ll \xi_{o}(m_{t}, m_{H^{+}})$. Anyway, with this assumption and
taking $f_{B}/f_{K}\sim 1$, its mass difference is found to be
\begin{equation}
\Delta m_{B_{d}}^{H^{0}} \simeq 37 |Y_{k,13}^{d}/Y_{k,12}^{d}|^{2}
\Delta m_{K}^{H^{0}}
\end{equation}
 when $\Delta m_{K}^{H^{0}} \simeq \Delta m_{K}^{exp.}$ and $|Y_{k,13}^{d}|
\sim 1.6 |Y_{k,12}^{d}| $  it then reproduces the experimental
result $\Delta m_{B_{d}}^{exp.} \simeq (100\pm 20) \Delta m_{K}^{exp.}$.
It is clear that  when $|Y_{k,13}^{d}|$ is of the same order as
$|Y_{k,12}^{d}| $, $B^{0}_{d}-\bar{B}^{0}_{d}$ mixing  then
gives a weaker constraint on the scalar mass.  Based on this result,  we
may conclude that if the contribution $\Delta m_{K}^{H^{0}}$ to the
mass difference in the  $K^{0}-\bar{K}^{0}$ system  is not dominant,  so does
$\Delta m_{B_{d}}^{H^{0}}$ in the $B^{0}_{d}-\bar{B}^{0}_{d}$ system except
the ratio $|Y_{k,13}^{d}/Y_{k,12}^{d}|^{2}$ becomes unexpected large and
the contributions from the box diagrams through W-boson and charged-scalar
exchanges are unexpected small. Note that one may also keep in mind
that a possible cancellation may occur between the contributions from the box
diagrams and the FCNSI when they become comparable each other and have the
opposite sign.

 The general form for the mass difference in the $B^{0}_{d}-\bar{B}^{0}_{d}$
system  can be written
\begin{eqnarray}
\Delta m_{B} & \simeq & \frac{G^{2}}{6\pi^{2}} (f_{B}\sqrt{B_{B}\eta_{tt}})^{2}
m_{B} m_{t}^{2} |V_{td}|^{2} \frac{1}{\eta_{tt}} |\{ \eta_{tt} B^{WW}(x_{t})
+ \frac{1}{4} \eta_{tt}^{HH} y_{t} |\xi_{t}|^{4}  B_{V}^{HH}(y_{t}) \\
& & + 2 \eta_{tt}^{HW} y_{t} |\xi_{t}|^{2} B_{V}^{HW}(y_{t}, y_{w}) ]
+ \frac{\tilde{B}_{B}}{B_{B}} \sum_{k}
(\frac{2\sqrt{3} \pi v m_{B}} {m_{H_{k}^{0}}m_{t}}
\frac{\zeta_{D}}{s_{\beta}})^{2} \frac{m_{d}}{m_{b}}\frac{1}{V_{td}^{2}}
 (Y_{k,13}^{d})^{2} \}| \nonumber
\end{eqnarray}
which is subject to the experimental constraint
\begin{equation}
\Delta m_{B} = (3.6 \pm 0.7)\times 10^{-4} eV \simeq \frac{G^{2}}{6\pi^{2}}
(120MeV)  ^{2}  m_{B} (140GeV)^{2} (\sin\theta =0.22)^{6}
\end{equation}

\subsection{$B^{0}_{s}-\bar{B}^{0}_{s}$ Mixing and its implications}

All the considerations and discussions on
$B^{0}_{d}-\bar{B}^{0}_{d}$ system can be applied to
$B^{0}_{s}-\bar{B}^{0}_{s}$ system with simply replacing $d$-quark  by
$s$-quark. By a parallel analysis, we may directly conclude that

i) if  contributions to $\Delta m_{B_{s}}$ are dominated by the box
diagrams through W-boson and charged-scalar exchanges, it is known that
a maximal mixing  will occur
\begin{equation}
\Delta m_{B_{s}} \simeq \Delta m_{B_{s}}^{box} \simeq
\frac{f_{B_{s}}^{2}B_{B_{s}}m_{B_{s}}}
{f_{B_{d}}^{2}B_{B_{d}}m_{B_{d}}} \frac{|V_{ts}|^{2}}{|V_{td}|^{2}}
\Delta m_{B_{d}} \sim 7.4 \times 10^{-3} eV
\end{equation}

ii) if  contributions to $\Delta m_{B_{s}}$ are dominated by the
FCNSI through neutral-scalar exchange at tree level, we find that
\begin{equation}
\Delta m_{B_{s}} \simeq \Delta m_{B_{s}}^{H^{0}} \simeq \frac{f^{2}_{B_{s}}
\tilde{B}_{B_{s}}m_{B_{s}}}{f^{2}_{B_{d}}\tilde{B}_{B_{d}}m_{B_{d}}}
\frac{m_{s}}{m_{d}}(\frac{|Y_{k,23}^{d}|}{|Y_{k,13}^{d}|}
)^{2}\Delta m_{B_{d}}^{H} \sim 7.4\times 10^{-3} eV
\end{equation}
as an order-of-magnitude prediction, the numerical values for the above
two cases are estimated by taking $f_{B_{s}}^{2}B_{B_{s}}m_{B_{s}}
/f_{B_{d}}^{2}B_{B_{d}}m_{B_{d}}\sim 1$ (i.e. in the approximation of
$SU(3)$ symmetry) and assuming $|Y_{k,23}^{d}|\sim
|Y_{k,13}^{d}|$ as well as using a phenomenological feature
$|V_{ts}|/|V_{td}|\simeq \sqrt{m_{s}/m_{d}}$. Therefore it should not be
surprised that the predictions from the two limit cases provide almost the
same result. It is seen that in these two limit cases
$B^{0}_{s}-\bar{B}^{0}_{s}$ mixing always approaches to its maximal value.

iii)  With the above observations, it is not difficult to expect that the
following conseqences  may occur: a) $B^{0}_{s}-\bar{B}^{0}_{s}$ mixing
is larger than the one expected from the standard model when
$(M_{12}^{WW} + M_{12}^{HH} + M_{12}^{HW})$ and $M_{12}^{H^{0}}$ have
the same phase, b)   $B^{0}_{s}-\bar{B}^{0}_{s}$ mixing becomes
much smaller than its maximal value. This happens when there is a big
cancellation between $(M_{12}^{WW} + M_{12}^{HH} + M_{12}^{HW})$ and
$M_{12}^{H^{0}}$, c) $B^{0}_{s}-\bar{B}^{0}_{s}$ mixing
can be any values.  Anyway, any significant deviations from its expected
maximal value in the standard model should provide a clear signature on new
physics which may be explained by this model.

Before proceeding, it may be interesting in noticing that when assuming
all the $|Y_{k,ij}^{d}|$ have values which are of the same oder of magnitude,
i.e., $|Y_{k,23}^{d}|\sim |Y_{k,13}^{d}|\sim |Y_{k,12}^{d}|\sim 1$, the
predicted relative values for the three mass differences $\Delta m_{K}$,
$\Delta m_{B}$ and $\Delta m_{B_{s}}$ from the FCNSI are closed to those
from the standard model. We also observe that if the mass differences of the
neutral mesons are dominated by the FCNSI, thus, with the well parameterized
Yukawa couplings  in our model and also the above
assumption,   the mass difference of the neutral mesons only relies on the
masses of those quarks of which the meson consists. Consequently, the
relative values of the mass differences for various neutral mesons are then
mainly characterized by the hierarchic properties of the quark masses of
which the neutral mesons consist.  This case is quite different from the one
in the standard model in which the neutral meson mixings depend
on the CKM matrix elements and the loop-quark masses of the box diagram.

\subsection{$D^{0}-\bar{D}^{0}$ Mixing and its implications}

    It is known that in the standard model the short-distance  contribution
to $\Delta m_{D}$ from the box diagram with W-boson exchange is of order
of magnitude  $\Delta m_{D}^{Box} \sim O(10^{-9})$ eV, here the external
momentum effects have to be considered and were found to suppress the
contribution by two orders of magnitude\cite{DK}. This is because of the
low mass of the intermediate state. It is not difficult to see that the
additional box diagram with charged-scalar gives even smaller contribution
except $|\xi_{s}|$ is as large as $|\xi_{s}|\sim 2 m_{H^{+}}/m_{s}$ which
is unreliable large for the present bound $m_{H^{+}}> 41$ GeV.  It has been
shown that dominant contribution to $\Delta m_{D}$ may come from the
long-distance effect since the intermediate states in the box diagram are
$d$- and $s$-quraks. The original estimations were found that
$\Delta m_{D}\sim 3 \times 10^{-5}$ eV \cite{WOLF2} and  $\Delta m_{D}\sim 1
\times 10^{-6}$ eV \cite{DGH2}.  A recent study \cite{HG} using the heavy quark
effective theory showed that large cancellations among the intermediate states
may occur so that the long-distance standard model contribution to
$\Delta m_{D}$ is only larger by about one order of magnitude than the short-
distance contribution, which was also supported in a subsequent calculation
\cite{ORS}.

    With this in mind, we now consider the contribution to $\Delta m_{D}$
from the FCNSI in our model. It is easy to read off  from eq.(29)
\begin{eqnarray}
\Delta m_{D}^{H} & = & 2 |M_{12}^{H}|
=\frac{G^{2}}{6\pi^{2}} f_{D}^{2} \tilde{B}_{D}m_{D}
(\sqrt{\frac{m_{u}}{m_{c}}})^{2} m_{s}^{2}\sum_{k}
(\frac{2\sqrt{3} \pi v m_{D}} {m_{H_{k}^{0}}m_{s}}
\frac{\zeta_{U}}{s_{\beta}})^{2} |Y_{k,12}^{u}|^{2} \nonumber \\
& = & 0.64 \times 10^{-4} (\frac{f_{D}\sqrt{\tilde{B}_{D}}}{210 MeV})^{2}
\sum_{k=1}^{3}(\frac{\zeta_{U}}{0.1 s_{\beta}}\frac{50 GeV}
{m_{H^{0}_{k}}})^{2} \frac{|Y_{k,12}^{u}|^{2}}{1}
\end{eqnarray}
With the above expected values in the second line for various parameters,
the predicted value for $\Delta m_{D}$ can be closed to the
current experimental limit $|\Delta m_{D}| < 1.3 \times 10^{-4}$ eV,
this implies that a big $D^{0}-\bar{D}^{0}$ mixing which is larger than the
standard model prediction does not excluded. With this analysis, we come to the
conclusion that a positive signal of neutral
$D$ meson mixing from the future experiments at Fermilab, CESR at Cornell and
at a $\tau$-charm factory would be in favor of our model especially when the
exotic neutral scalars are not so heavy (or even when $m_{H^{0}_{k}} <
m_{Z}/2 $).

 Alternatively, we may consider that the current experimental limit in fact
provides a constraint on the parameter $\zeta_{U}/s_{\beta}$ when other
parameters are assumed to take the above appropriate values, i.e.
$f_{D}\sqrt{\tilde{B}_{D}} = 210$ MeV and $|Y_{k,12}^{u}| \sim O(1)$. The
constraint is
\begin{equation}
 \zeta_{U}/s_{\beta} < 3\times 10^{-3} m_{H_{k}^{0}}/GeV
\end{equation}
which is closed to the constraint on $ \zeta_{D}/s_{\beta}$ obtained from
the $K^{0}-\bar{K}^{0}$.

  It should be emphasized that  the measurement of $\Delta m_{D}$ may
provide  a good candidate channel  for probing the FCNSI in our model.

\section{CP-violating Phenomenology of the Model}

\subsection{Indirect CP Violation in Kaon Decay ($\epsilon$)}

  CP-violating parameter $\epsilon$ has been well established for thirty
years, any successful model should be able to account for its measured
value. In fact, many models have been built to fit this single parameter.
So that we should first consider the contributions to this parameter in our
model.  The standard definition of $\epsilon$ is
\begin{equation}
\epsilon = \frac{1}{\sqrt{2}}(\frac{Im M_{12}}{2 Re M_{12}} + \xi_{0})
e^{i\pi /4}
\end{equation}
where $\xi_{0}=Im A_{0}/ Re A_{0}$ with $|A_{0}| = (3.314\pm 0.004) \times
10^{-7}$ GeV the isospin-zero amplitude of $K\rightarrow \pi \pi $ decay.
Usually, the $\xi_{0}$ term is relatively small as it is proportional to
the small direct CP-violating parameter $\epsilon'$ (see below).

   In our model $Im M_{12}$ contains several parts and receives
contributions from various induced CP-violating mechanisms. As pointed out
in the previous section, in order to discuss their contributions one should
first specify the phase convention. For convenience in comparison with the
standard KM-model, we redefine the quark fields through
the phase transformation $f_{i} \rightarrow e^{i\theta_{f_{i}}} f_{i}$, so
that $V \rightarrow V_{KM}$ which has the phase convention in the 'standard'
parameterization\cite{SKM} of CKM matrix. In fact, as we discussed in the
previous section, the above  phase convention is very closed to the one in the
'standard' parameterization of CKM matrix. With such a phase redefination of
the fermions, the couplings of the flavor-changing scalar interactions are
correspondingly changed via  $S_{ij}^{f} \rightarrow
\tilde{S}_{ij}^{f}=e^{i(\theta_{f_{j}} -\theta_{f_{i}})} S_{ij}^{f}$.
Note that the diagonal Yukawa couplings
$\xi_{i}$ are unchanged. With this in mind,   we now come to discuss various
contributions to $\epsilon$. The first part comes from the box diagrams through
W-boson and charged-scalar exchanges
\begin{eqnarray}
Im M_{12}^{Box} & = & Im M_{12}^{WW} + Im M_{12}^{HH} + Im M_{12}^{HW}
\nonumber \\
 & = & \frac{G^{2}}{12\pi^{2}} f_{K}^{2} B_{K}m_{K} m_{i} m_{j}
\{ \sum_{i,j}^{c,t} Im (\lambda_{i} \lambda_{j}) Re B_{ij}(m_{i}, m_{j};
\xi_{i}, \xi_{j})  \nonumber \\
& & +  Re (\lambda_{i} \lambda_{j}) Im B_{ij}(m_{i}, m_{j}; \xi_{i}, \xi_{j})
\}
\end{eqnarray}
where $B_{ij}(m_{i}, m_{j}; \xi_{i}, \xi_{j})$ depend on the integral
functions of the box diagrams and their general form is given
in the Appendix. The imaginary part $Im B_{ij}(m_{i}, m_{j}; \xi_{i},
\xi_{j})$ arises from the complex couplings $\xi_{i}$ in our model.

 The second part is due to the FCNSI at tree level
\begin{equation}
Im M_{12}^{H^{0}} = \frac{G^{2}}{12\pi^{2}} f_{K}^{2} \tilde{B}_{K}m_{K}
(\sqrt{\frac{m_{d}}{m_{s}}})^{2} m_{c}^{2}\sum_{k}
(\frac{2\sqrt{3}\pi v m_{K}} {m_{H_{k}^{0}}m_{c}}
\frac{\zeta_{D}}{s_{\beta}})^{2} Im (Y_{k,12}^{d})^{2}
\end{equation}
This provides a contribution to $\epsilon$ in almost
any models which possess CP-violating FCNSI.

  In particular, in our model which likes in the Weinberg 3HDM,  it can
receive large contributions from the long-distance dispersive effects
through the $\pi$, $\eta$ and $\eta'$ poles \cite{DPD}.
For a quantitative estimate of these effects, we follow the analyses in refs.
\cite{DPD,DH,HYC1,HYC2}
\begin{eqnarray}
(Im M'_{12})_{LD} & = & \frac{1}{4m_{K}} \sum_{i}^{\pi,\eta, \eta'} \frac{Im
(<K^{0}|L_{eff}|i><i|L_{eff}|\bar{K}^{0}>)}{m_{K}^{2} - m_{\pi}^{2}}
\nonumber  \\
 & = & \frac{1}{4m_{K}} \frac{2\kappa}{m_{K}^{2} - m_{\pi}^{2}}
<K^{0}|L_{-}|\pi^{0}><\pi^{0}|L_{+}|\bar{K}^{0}>) \\
 & = & \frac{G^{2}}{12\pi^{2}} f_{K}^{2} B'_{K}m_{K} (\frac{m_{K}}
{m_{s}})^{2} \sin\theta m_{s}^{2} (\sqrt{\frac{\pi \alpha_{s}}{2}}
\frac{3\kappa A_{K\pi}}{4m_{s}(m_{K}^{2} - m_{\pi}^{2})}) \nonumber \\
& & \cdot \sum_{i} [Im \lambda_{i} Re P_{i}(m_{i}, \xi_{i}) + Re \lambda_{i}
Im P_{i}(m_{i}, \xi_{i})]
\end{eqnarray}
where $\kappa$ is found to be $\kappa \simeq 0.15$ when considering the
$SU(3)-$ breaking  effects in the $K-\eta_{8}$ transition, and nonet-symmtry-
breaking in $K-\eta_{o}$ as well as $\eta-\eta'$ mixing. We shall not repeat
these analyses, and the reader who is interested in it is refered to the
paper \cite{HYC2} and references therein. $L_{-}$ and $L_{+}$ are CP-odd and
CP-even lagrangians respectively (with convention $L_{eff}= L_{-} + i L_{+}$).
The $L_{-}$ is induced from the gluon-penguin diagram with charged-scalar
\begin{equation}
L_{-} = f_{s} \bar{d}\sigma_{\mu\nu}(1+\gamma_{5})\lambda^{a} s G_{\mu\nu}^{a}
 - f_{d} \bar{d}\sigma_{\mu\nu}(1-\gamma_{5})\lambda^{a} s G_{\mu\nu}^{a}
\end{equation}
with
\begin{equation}
f_{q} = \frac{G}{\sqrt{2}} \frac{g_{s}}{32\pi^{2}} m_{q} \sum_{i}
Im(\xi_{q}\xi_{i} \lambda_{i}) y_{i} P_{T}^{H}(y_{i})
\end{equation}
where $P_{T}^{H}(y_{i})$ is the integral function and presented in the
Appendix. From $f_{s}$ and $f_{d}$ it is not difficult to read off the
$Re P_{i}(m_{i}, \xi_{i})$ and $Im P_{i}(m_{i}, \xi_{i})$ (see Appendix).
In obtaining the last expression of the above equation, we have used the
result $<K^{0}|L_{-}|\pi^{0}> = (f_{s} - f_{d}) A_{K\pi}$ where $A_{K\pi}$
has been computed in the MIT bag model and was found \cite{DHH} to be $A_{K\pi}
=
0.4 GeV^{3}$ for $\alpha_{s}=1$, and the convention
$<\pi^{0}|L_{+}|\bar{K}^{0}>=\frac{1}{2} G f_{K}^{2} B'_{K}m_{K}^{2}
(2m_{K}/m_{s})^{2} \sin\theta $, where $B'_{K}$ is introduced to
fit the experimental value $<\pi^{0}|L_{+}|\bar{K}^{0}>= 2.58 \times 10^{-7}
GeV^{2}$ and is found to be $B'_{K}=1.08$. We then obtain
$\sqrt{\pi \alpha_{s}} 3\kappa A_{K\pi}/[4\sqrt{2}m_{s}(m_{K}^{2} -
m_{\pi}^{2})] \simeq 1.4$.

   It is easily seen that in the parts $Im B_{ij}(m_{i}, m_{j}; \xi_{i},
\xi_{j})$ and $Im M'_{12}$ the dominant contributions come from the loop-charm-
quark because $\lambda_{c} \gg \lambda_{t}$ and $m_{u}\ll m_{c}$.
Neglecting the $t-$ and $u-$quark contributions and also the terms
proportional to $m_{d}$ in comparison with the terms proportional to $m_{s}$,
the total contributions to the CP-violating parameter $\epsilon$ can be
simply calaculated from the following formula
\begin{eqnarray}
|\epsilon| & = & 3.2\times 10^{-3} B_{K} (\frac{|V_{cb}|}{0.04})^{2}
\frac{2|V_{ub}|}{|V_{cb}||V_{us}|}  \sin\delta_{KM} \{ -\frac{1}{4}
[\eta_{cc} B^{WW}(x_{c}) \nonumber \\
& & + \frac{1}{4} \eta_{cc}^{HH} y_{c}
|\xi_{c}|^{4}  B_{V}^{HH}(y_{c}) + 2 \eta_{cc}^{HW} y_{c} |\xi_{c}|^{2}
B_{V}^{HW}(y_{c}, y_{w})] \nonumber \\
& & + (\frac{|V_{cb}| m_{t}}{2m_{c}})^{2}(1-\frac{|V_{ub}|}{|V_{cb}||V_{us}|}
\cos\delta_{KM}) [\eta_{tt} B^{WW}(x_{t}) \nonumber \\
& & + \frac{1}{4}
\eta_{tt}^{HH} y_{t} |\xi_{t}|^{4}  B_{V}^{HH}(y_{t}) + 2 \eta_{tt}^{HW}
y_{t} |\xi_{t}|^{2}  B_{V}^{HW}(y_{t}, y_{w})] \nonumber \\
& & + \frac{m_{t}}{4m_{c}}[\eta_{ct} B^{WW}(x_{c}, x_{t}) + \frac{1}{2}
\eta_{ct}^{HH} \sqrt{y_{c}y_{t}}
|\xi_{c}|^{2} |\xi_{t}|^{2} B_{V}^{HH}(y_{c}, y_{t}) \nonumber \\
& & + 4 \eta_{ct}^{HW} \sqrt{y_{c}y_{t}}Re(\xi_{c}\xi_{t})
B_{V}^{HW}(y_{c}, y_{t}, y_{w}) ]\}    \\
& &  + 2.27\times 10^{-3} \frac{Im (\tilde{Y}_{k,12}^{d})^{2}}{6.4\times
10^{-3}}\tilde{B}_{K}\sum_{k}(\frac{10^{3} GeV} {m_{H_{k}^{0}}}
\frac{\zeta_{D}}{s_{\beta}})^{2}  \nonumber \\
& & - 2.27\times 10^{-3} Im(\xi_{c}\xi_{s})^{2} \frac{6.8 GeV^{2}}
{m_{H^{+}}^{2}}\tilde{B}_{K} (ln\frac{m_{H^{+}}^{2}}{m_{c}^{2}} -2) \nonumber
\\
& & + 2.27 \times 10^{-3} Im(\xi_{c}\xi_{s}) \frac{37 GeV^{2}}{m_{H^{+}}^{2}}
B'_{K} (ln\frac{m_{H^{+}}^{2}}{m_{c}^{2}} -\frac{3}{2})
+ \frac{\xi_{o}}{\sqrt{2}} \nonumber \\
& \equiv & (\epsilon_{W-Box}^{III} + \epsilon_{H^{+}-Box}^{III}) +
\epsilon_{H^{0}-Tree}^{II+IV} + \epsilon_{H^{+}-Box}^{I} +
\epsilon_{H^{+}-LD}^{I}   \nonumber
\end{eqnarray}
where we have used the experimental constraint on $2 Re M_{12}=
\Delta m_{K}^{exp.}$ given in the eq.(45). In the second equality,
each term shortly denotes the corresponding term appearing in the previous
equality.

 From the above equation,  it is manifest that all the induced four types of
CP-violating mechanism can contribute to the parameter $\epsilon$, as it is
explicitly denoted in the second equality, where the
superscripts I, II, III, and IV denote the type of CP-violating mechanism
classified in the previous section and the subscriptions indicate the graphs
and processes from which $\epsilon$ receives contribution.  To see how the
various mechanisms play the role on $\epsilon$, let us consider the
following limit cases:

(I) when $\zeta_{F}/s_{\beta}\ll 1$ with  $(S_{1}^{F})_{ij}\sim O(1)$,
 $m_{H^{+}} < v =246$ GeV and  $|\xi_{i}| \gg 1$ ($i\neq t$), it is
easily seen that only the new type CP-violating mechanism (type-I) plays the
important role,  namely the $\epsilon$ is fitted by the last two terms
\begin{equation}
\epsilon \simeq \epsilon_{H^{+}-Box}^{I} + \epsilon_{H^{+}-LD}^{I}
\end{equation}
 where the former term comes from the short-distance box
diagrams and the latter one from the long-distance dispertive effects.
These two contributions can be of the same order of magnitude.  Even the
$\epsilon_{H^{+}-Box}^{I}$ may become more important for $|\xi_{i}| > 1$,
since this term is proportional to the fourth power of $|\xi_{i}|$.
Furthermore, the contribution from the short distance box diagrams
may concern less uncertainties on the hadronic matrix element, which appears
to make the new type of CP-violating mechanism more attractive. Moreover, the
smallness of the CP-violating effects from the short distance box diagrams
also becomes natural and is attributed to the smallness of the mass ratio
between the strange quark and charm quark, i.e. $m_{s}^{2}/m_{c}^{2}$.
As we have seen from the previous section,  for $|\xi_{c}| \gg 1$, the mass
difference $\Delta m_{K}$ can also be accounted for
by purely short-distance contributions from box graphs through W-boson and
charged-scalar exchanges. The requirement for $|\xi_{c}| > 1$ from these
two phenomena are consistent.

(II)  when $ 10^{-4}m_{H^{0}_{k}}/GeV \alt \zeta_{F}/s_{\beta} < 0.1 $ and
$(S_{1}^{F})_{ij} \sim O(1)$, $|\xi_{i}|\sim O(1)$, both
the new type  of CP-violating mechanism and induced KM-type mechanism become
less important and the parameter $\epsilon$ is then accounted for by the
FCSW-type  mechanism (type-II) incorporating with the SPM mechanism
(type-IV), i.e. $\epsilon \sim \epsilon_{H^{0}-Tree}^{II+IV}$.

It should be pointed out that in this case one may need to fine-tune the
parameters  to fit the observed small value of $\epsilon$. In particular,
when the FCNSI is also demanded to accommodate the mass difference
$\Delta m_{K}$,  one then has to fine-tune the parameters $\delta_{d}$,
$\delta_{s}$, $arg(O_{1k}^{H} + i O_{3k}^{H})$, $\theta_{s}$, $\theta_{d}$ and
$(S_{1}^{D})_{21}$ so that $Im (\tilde{Y}_{k,12}^{d})^{2}/
Re(\tilde{Y}_{k,12}^{d})^{2} \sim 6.4\times 10^{-3}$,
namely the effective CP-violating phase must be very small although  the phases
$\delta$, $\delta_{d}$, $\delta_{s}$ and $arg(O_{1k}^{H} + i O_{3k}^{H})$
 are not necessary to be small. Alternatively, assuming that the
FCNSI is only used to accommodate the CP-violating parameter $\epsilon$,
one then can choose
\begin{equation}
\zeta_{D}/s_{\beta} \sim 0.8\times 10^{-4} m_{H_{k}^{0}}/GeV\ , \qquad
Im (\tilde{Y}_{k,12}^{d})^{2}/Re(\tilde{Y}_{k,12}^{d})^{2}  \sim O(1)
\end{equation}
 In this case, the CP-violating phases are indeed generically
of order unity.

(III)  when $\zeta_{D}/s_{\beta} \agt 0.2 $ and/or $\zeta_{U}/s_{\beta}
\agt 0.6 $ for $m_{t} \sim 150$ GeV (note that the latter condition may
become weak if $|(S_{1}^{U})_{ij}| > 1 $ for $i< j = 3$, unlike the
$(S_{1}^{D})_{ij}$ which are subject to the restriction from the neutral
meson mixings, the $|(S_{1}^{U})_{ij}| $ are not so strictly restricted ) ,
$|\xi_{i}|\sim O(1)$ and  $m_{H^{0}_{k}} \gg v =246$ GeV, namely the neutral
scalars are very heavy, the CP-violating mechanism is
then governed by the induced KM-mechanism, i.e. the first term of the above
equation becomes dominant. Note that it can remain different from the
standard KM-model if the charged scalar is not  heavy and $|\xi_{t}| \sim 1$,
this can be easily seen from the above equation, i.e.  $\epsilon =
(\epsilon_{W-Box}^{III} + \epsilon_{H^{+}-Box}^{III})$, the  contribution
to $\epsilon$ from top-quark box diagrams with
charged scalar can be comparable and even larger than the one from
W-boson diagram . The condition for its happening depends on $\xi_{t}$,
$m_{H^{+}}$ and $m_{t}$, and is the same as the one obtained from the
 $B^{0}-\bar{B}^{0}$ mixing. A similar case
has been discussed in \cite{AGG} for a 2HDM with NFC and
standard KM-phase. The difference between that model and ours
is that the model discussed in \cite{AGG} is equivalent to $\xi_{u} =
\xi_{c} = \xi_{t} \equiv v_{1}/v_{2}$,  this is why the constraint from
$\epsilon$ ($|\xi_{t}| \sim 1$ ) is much stronger than the one
from $\Delta m_{K}$ ($|\xi_{c}| \gg 1$ ) in that model.
In our model, in general $\xi_{u} \neq \xi_{c} \neq \xi_{t}$,
therefore we have more freedom to fit $\epsilon$ and
$\Delta m_{K}$ as well as $\Delta m_{B}$.  With this consideration, we see
that only if the charged scalar also becomes very heavy and/or
$|\xi_{t}| \ll 1$, the induced KM-type mechanism then conincides with the
standard KM-model.

    In general, four types of CP-violating mechanism may simultaneously play
an important role on the CP-violating parameter $\epsilon$, which depends on
the important parameters $\zeta_{F}$ and $\tan\beta$ and the masses of
the scalars. Anyway, we can conclude that in this model the observed
indirect CP-violating parameter $\epsilon$ in kaon decay can be easily
accounted for.

\subsection{Direct CP Violation in kaon Decay ($\epsilon'/\epsilon$)}.

  Direct CP violation in kaon decay ($\epsilon'/\epsilon$) has been studied
extensively by both theoretists and experimentlists.  The original
motivation was to clarify whether the observed
CP violation is due to the generic superweak interaction or a complex phase
in the quark mixing CKM matrix. In our model, this modivation will be
challenged by the new type of CP-violating  mechanism (i.e. type-I
mechanism classified in the previous section). This is because the predicted
value by this new mechanism can be of order $\epsilon'/\epsilon \sim 1.0 \times
10^{-3}$ which is comparable with the prediction \cite{YLWU1} from the
standard KM-model. To further convince ourselves, let us make a more
detailed analysis.

   Evidence for direct CP violation would be established if one could show
that

\[ phase A(K^{0}\rightarrow \pi \pi, I=2) \neq phase A(K^{0}\rightarrow
\pi \pi, I=0) \]
 which is one of the major objectives in the measurements of
$\eta_{+-}$ and $\eta_{00}$. The relative phase is measured by the parameter
\begin{eqnarray}
\epsilon'/\epsilon & = & \frac{1}{|\epsilon|}\frac{1}{\sqrt{2}}
\frac{Im(A_{2}A_{0}^{\ast})}{|A_{0}|^{2}} e^{i(\delta_{2} - \delta_{0} +
\frac{\pi}{4})} \nonumber \\
& \simeq & \frac{1}{|\epsilon|}\frac{1}{\sqrt{2}}
(\frac{ImA_{2}}{Re A_{0}} - \omega \frac{ImA_{0}}{Re A_{0}})
\end{eqnarray}
with $\omega = |A_{2}/A_{0}|=1/22$. Where the fact that $(\delta_{2} -
\delta_{0}) \simeq \pi /4$ has been used.

  In the KM-model, it was noted by Gilman and Wise\cite{GWISE}, Guberina and
Peccei\cite{GP} that large effect on $\epsilon'$ arises from the QCD
corrections due to the existence of gluon "penguin-graphs" with W-boson
exchange . The calculations have recently been improved by considering the
electroweak penguin graphs due to the heavy top quark effects\cite{FR,BBH,PW}
and the next-to-leading order corrections to the hadronic matrix
elements\cite{HPSW,YLWU4} along the line of \cite{BAG2} as well as the
two-loop perturbative QCD corrections\cite{BJLW}. In the standard notation,
the ratio is expressed
\begin{eqnarray}
\frac{(\epsilon')_{W-Pen.}^{III}}{\epsilon^{exp.}} & = & (\frac{1}{|\epsilon|}
\frac{1}{|A_{0}|} \frac{G}{2}) \omega Im\lambda_{t}  \sum_{i\neq 4}^{8}
C_{i} [<Q_{i}>_{0}-\frac{1}{\omega} <Q_{i}>_{2}] \nonumber \\
&  = & 1.25\times 10^{-3} (\frac{Im\lambda_{t}}{10^{-4}})(\frac{C_{6}
<Q_{6}>_{0}}{0.05 GeV^{3}})(\frac{1-\Omega}{0.7})
\end{eqnarray}
Here the superscript and subscript on the $\epsilon'$ have the
similar meanings as those on $\epsilon$ considered in the previous subsection.
Where $|Im\lambda_{t}| = |V_{ub}||V_{cb}|\sin\delta_{KM}$ is a parameter of
CKM matrix. $C_{6}<Q_{6}>_{0}$ arises from the gluon penguin and
was found to be $|C_{6}<Q_{6}>_{0}| \simeq (0.18 - 0.05) GeV^{3}$
for $m_{s} = 125-200$ MeV and $\Lambda_{QCD}= (200-300)$ MeV. $\Omega$
characterizes the relative contributions of electroweak penguins and $\eta
-\eta'$ mixing as well as other operators, and was found $0.3 < \Omega
 < 0.7$ for $100 GeV \alt m_{t} \alt 250$ GeV \cite{YLWU1}.

   When applying this analaysis to the induced KM-type mechanism in our model,
the only point that needs to be noted is that the induced KM-phase depends
on the magnitude of the FCNSI.

   In our model, the first additional contribution arises from the
 penguin graphs  with charged-scalar exchange instead of W-boson. The
analysis is analogous to the previous one. Keeping to the lowest
order of QCD corrections, we have
\begin{equation}
\frac{(\epsilon')_{H^{+}-Pen.}^{III}}{\epsilon^{exp.}}\simeq  1.8\times 10^{-4}
(\frac{Im\lambda_{t}}{10^{-4}})(\frac{\tilde{C}_{6}<Q_{6}>_{0}}{0.005 GeV^{3}})
\end{equation}
with
\begin{equation}
\tilde{C}_{6} = \frac{\alpha_{s}}{12\pi} P_{V}^{H^{+}}(y_{t})
\end{equation}
where $P_{V}^{H^{+}}(y_{t})$ is an integral function (see the
Appendix). Such an effect was also discussed in \cite{AGG} for a 2HDM with NFC.
The dominant contribution comes from the top quark exchange.
For $\alpha_{s}=1$, $|\xi_{t}| =1$ and $m_{t} =150$ GeV, we find that
$\tilde{C}_{6} = 0.01 (0.005)$ for $m_{H^{+}}=50 (100)$ GeV, which is smaller
than $C_{6}$ by an order of magnitude.  Therefore, in this approximation,
the short-distance effect provides at most about $10\%$ contribution to the
ratio $\epsilon'/\epsilon$ in the KM-type mechanism.

  The second additional contribution comes from the FCNSI. Here the effective
$\Delta S=1$ Hamiltonian results from a tree level graph through
the neutral scalar exchange, which is analogous to the $\Delta S=2$
Hamiltonian in calculating the $\Delta m_{K}^{H^{0}}$ and
$(\epsilon)_{H^{0}-Tree}^{II+IV}$. It is not difficult to find
\begin{eqnarray}
H_{eff}^{H^{0}}(\Delta S=1) & = & \frac{G}{\sqrt{2}}
\frac{\zeta_{D}}{s_{\beta}}
\sqrt{\frac{m_{d}}{m_{s}}}\sum_{k=1}^{3}\frac{m_{s}}{m_{H_{k}^{0}}^{2}}
[S_{k,12}^{d} \bar{d}(1+\gamma_{5})s + S_{k,21}^{d \ast} \bar{d}
(1-\gamma_{5})s] \nonumber \\
& & \cdot \{ \frac{1}{2}(m_{d}\eta^{(k)}_{d} + m_{u}\eta^{(k)}_{u})[\bar{d}
(1+\gamma_{5})d + \bar{u}(1+\gamma_{5})u]  \\
& & + \frac{1}{2}(m_{d}\eta^{(k)}_{d}
- m_{u}\eta^{(k)}_{u})[\bar{d}(1+\gamma_{5})d - \bar{u}(1-\gamma_{5})u] +
H.C. \} \nonumber
\end{eqnarray}
with this effective Hamiltonian, we find that
\begin{eqnarray}
\frac{(\epsilon')_{H^{0}-Tree}^{I+II+IV}}{\epsilon^{exp.}} & \simeq &
\frac{1}{|\epsilon|}\frac{1}{\sqrt{2}} (\frac{ImA_{2}}{Re A_{0}}) \nonumber \\
& \simeq & - 2.3\times 10^{-5} Im \eta_{k} (\frac{50GeV}{m_{H_{k}^{0}}})\sqrt{
\frac{\Delta m_{K}^{H^{0}}}{\Delta m_{K}^{exp.}}}\frac{Re X_{k,12}^{d}}{
\sqrt{Re(Y_{k,12}^{d})^{2}}} \\
& \simeq & - 1.22\times 10^{-6} Im \eta_{k}\  (\frac{50GeV}{m_{H_{k}^{0}}})
\frac{Re X_{k,12}^{d}}{\sqrt{Im(Y_{k,12}^{d})^{2}}}
\sqrt{\frac{(\epsilon)_{H^{0}-Tree}^{II+IV}}{10^{-3}}} \nonumber
\end{eqnarray}
where $Im A_{2}$ is expected to be of the same order as $ImA_{0}$, we then
neglect the suppressed term by $\omega =1/22$. In the above result, we have
replaced the parameter $\zeta_{D}/s_{\beta}$ by the values constrained from the
mass difference $\Delta m_{K}$ and/or parameter $\epsilon$. To estimate the
numerical values, the vacuum insertion and factorization approximation has
be used to evaluate the hadronic matrix element
\begin{eqnarray}
Im A_{2} & = & Im (\sqrt{\frac{3}{2}} <\pi^{0} \pi^{+} | H_{eff}^{H^{0}}(\Delta
S=1) | K^{+} > ) \nonumber \\
& = & \frac{G}{\sqrt{2}} \frac{\zeta_{D}}{s_{\beta}}
\sqrt{\frac{m_{d}}{m_{s}}}\frac{m_{s}m_{d}}{m_{H_{k}^{0}}^{2}}
\sqrt{3} f_{\pi}(m_{K}^{2}- m_{\pi}^{2}) (\frac{m_{K}}{m_{s}})^{2}
Re X_{k,12}^{d}Im\ \eta_{k}
\end{eqnarray}
with
\begin{equation}
Im\ \eta_{k} = Im (\eta_{d}^{(k)} - \frac{m_{u}}{m_{d}}\eta_{u}^{(k)})\ ;
\qquad  Re X_{k,12}^{d}  =  Re(S_{k,12}^{d} + S_{k,21}^{d \ast})/2
\end{equation}
where $Re X_{k,12}^{d}$ is expected to be of order unity. Note that the first
term in the curl bracket of the eq.(69) and its hermitian conjugate do not
contribute to $A_{2}$ as they tramsform like $(8_{L,R}, 1_{L,R})$ of
$SU(3)_{L}\times SU(3)_{R}$, it only gives a contribution to the $A_{0}$ with
a suppression factor $O(m_{K}^{2})/\Lambda_{\chi}^{2}$, here $\Lambda_{\chi}
\simeq 1$ GeV is the chiral symmetry breaking scale.

    It is easily seen that even in the case that $\Delta m_{K}$ is dominated
by the $\Delta m_{K}^{H^{0}}$ and $\epsilon$ is fitted by fine-tuning
various parameters, the contribution to $\epsilon'$ from the FCNSI through
the type-(II+IV) CP-violating mechanism is likely up to
$10^{-4}$ for a possible large values $Im\eta^{(k)}\sim 5-10$ and the small
mass of the neutral scalar $m_{H_{k}^{0}} \sim 50$ GeV.
Considering all the possible reasonable ranges
of the parameters, we may expect that from the FCSW-type mechanism
$(\epsilon')_{H^{0}-Tree}^{II+IV}/\epsilon^{exp.} \sim 10^{-7}-10^{-4}$.

    We now discuss the important contributions to $\epsilon'$ from purely new
type of CP-violating mechanism. It can arise from both the short-distance
contribution of a tree level graph through charged-scalar exchange and
the long-distance contribution of the one-loop  penguin graph with
also charged-scalar exchange.
Let us first consider the tree graph contribution,  which is similar to the
one analyzed above for the FCNSI. Here, the effective Hamiltonian is more
simple
\begin{equation}
H_{eff}^{H^{+}}(\Delta S=1) =  \frac{G}{\sqrt{2}}\lambda_{u} \frac{m_{d}m_{s}}
{m_{H^{+}}^{2}} \xi_{s} \bar{u}(1+\gamma_{5})s [\xi_{d}^{\ast}
\bar{d}(1-\gamma_{5})u - \frac{m_{u}}{m_{d}}\xi_{u}\bar{d}
(1+\gamma_{5})u ]
\end{equation}
with a similar consideration, we obtain
\begin{eqnarray}
\frac{(\epsilon')_{H^{+}-Tree}^{I}}{\epsilon^{exp.}} & \simeq &
\frac{1}{|\epsilon|}\frac{1}{\sqrt{2}} (\frac{Im A_{2}}{Re A_{0}}) \nonumber \\
& \simeq & 1.5 \times 10^{-3} Im [\xi_{s}(\xi_{d}^{\ast} + \frac{m_{u}}{m_{d}}
\xi_{u})] \frac{1}{5}  (\frac{50GeV}{m_{H^{+}}})^{2}
\end{eqnarray}
This result  predicts  an observable value for the ratio.
For  reasonable values of $|\xi_{i}| > 1$ ($i=s,d,u$), the ratio is
likely to be of order  $10^{-3}$. We expect that the uncertainties from the
hadronic matrix element will not change significantly this estimation.

  The long-distance contribution to $\epsilon'$ from the penguin graph with
charged-scalar arises from the amplitude $<\pi \pi | L_{-} | K^{0}>$.  Here
$L_{-}$ is the CP-odd lagrangian which has been used in the previous
subsection to calculate the parameter $\epsilon$.
It was first pointed out by Donoghue and
Holstein\cite{DH} that this amplitude involves an additional pole contribution
arising from the strong-interaction scattering $K\pi \rightarrow K\pi$
followed by a $ |K > \rightarrow | 0 >$ (vacuum) weak transition. A general
discussion has been made in \cite{HYC1,HYC2}. Following
those analyses, we have
\begin{eqnarray}
<\pi^{+} \pi^{+} | L_{-} | K^{0}> & = & <\pi^{+} \pi^{+} | L_{-} |
K^{0}>_{direct} + S_{K\pi K\pi} \frac{1}{m_{K}^{2}} < 0| L_{-} | K^{0}>
\nonumber \\
 & \equiv &  <\pi^{+} \pi^{+} | L_{-} | K^{0}>_{direct} D = -i
\frac{1}{\sqrt{2}
f_{\pi}} < \pi^{0} | L_{-} | K^{0}> D
\end{eqnarray}
where $S_{K\pi K\pi}$ is a $K\pi$ strong vertex and $D$ is introduced as
a measure of the amplitude.  Consequently, in the limit of chiral $SU(3)$
symmetry or equivalently to the lowest-order of chiral lagrangian, the two
terms
exactly cancel each other. Namely, the $L_{-}-$induced $K\rightarrow \pi \pi$
vanishes to the lowest-order in chiral symmetry (i.e. $D=0$).

    The high-order contributions to the above two terms are not yet
calculable from our present knowledge. Nevertheless, based on the
observation that the high-order terms involve the four derivative
contributions which are suppressed by factors
of $p^{2}/\Lambda_{\chi}^{2}$,  it is then expected that $D$ is of
order $O(m_{K}^{2}, m_{\pi}^{2})/\Lambda_{\chi}^{2}$  with $\Lambda_{\chi} \sim
1$ GeV being a chiral-symmetry breaking scale.  With these considerations,
we obtain
\begin{eqnarray}
\frac{(\epsilon')_{H^{+}-LD}^{I}}{\epsilon^{exp.}} & = &
\frac{1}{|\epsilon|}\frac{1}{\sqrt{2}} \omega (\frac{Im A_{0}}{Re A_{0}})
=[\frac{(\epsilon')_{H^{+}-LD}^{I}}{(\epsilon)_{H^{+}-LD}^{I}} ]
[\frac{(\epsilon)_{H^{+}-LD}^{I}}{\epsilon^{exp.}}] \nonumber \\
& \simeq & 0.017D  \frac{(\epsilon)_{H^{+}-LD}^{I}}{\epsilon^{exp.}}
\sim (0.4-6.0) \times 10^{-3} [\frac{(\epsilon)_{H^{+}-LD}^{I}}
{\epsilon^{exp.}}]
\end{eqnarray}
Where we have used $(\epsilon)_{H^{+}-LD}^{I} \gg Im A_{0}/Re A_{0} $.
Note that the ratio

\[ (\epsilon')_{H^{+}-LD}^{I}/(\epsilon)_{H^{+}-LD}^{I}
= 0.017D \sim (0.4-6.0)\times 10^{-3} \]
is independent of the CP-odd matrix
element $< \pi^{0} | L_{-} | K^{0}>$.  We then conclude that as long as the
long-distance contribution $(\epsilon)_{H^{+}-LD}^{I}$ to the
$\epsilon^{exp.}$  is significant, which has been seen in the
previous subsection that it is true for appropriate values of $\xi_{s}$ and
$\xi_{c}$,  then the ratio $\epsilon'/\epsilon$
is also likely to be at the observable level, i.e.,
 $\epsilon'/\epsilon \sim 10^{-3}$.

  With these results, we come to the following observations: first,
for the three limit cases discussed in the previous subsection for $\epsilon$,
there are three corresponding predictions on the $\epsilon'$, which can be
easily seen from the above analyses. It is clear that the contribution to
$\epsilon'$ from the FCNSI is likely the smallest one.  Second,  once the
FCNSI is only used to accommodate the CP-violating parameter $\epsilon$ but
not mass difference $\Delta m_{K}$, thus the parameter $\zeta_{D}/s_{\beta}$
will be in general small. If $\zeta_{U} \sim \zeta_{D}$, we see that the
induced KM-phase $\delta_{KM}$ also becomes small and its contribution to
$\epsilon'/\epsilon$ is deminished and will be unobservable small.
In this case, the contribution to  $\epsilon'/\epsilon$ from the new type of
CP-violating mechanism is dominant and can be in the observable level.
Third, in general, without accidental cancellation among various terms,
for $|\xi_{f_{i}}| > 1 $  it is unnatural if the ratio $\epsilon'/\epsilon$
becomes unobservable small.

\subsection{Direct CP Violation in B-System}

   In B system, unambiguous evidence of CP violation comes from oscillatory
behavior in the decays of $B^{0}$ and/or $\bar{B}^{0}$ into a CP eigenstate
\cite{BS,CPRV}. Some processes such as

\[ B_{d}\rightarrow J/\psi\ K_{s},
\pi\ \pi, \rho\ \pi, \cdots ; \qquad
B_{s}\rightarrow \rho\ K_{s}, D_{s}^{\pm}\ K^{\mp}, D^{0}\ \phi \cdots \]
have been suggested to measure three angles (usually denoted by
$\alpha$, $\beta$ and $\gamma$) of the CKM unitary triangles in the SM
\cite{ID}. This is
the case that only if the source of CP violation comes from a single KM-phase.
Any models with additional CP-violating sources and/or additional contributions
to the $B^{0}-\bar{B}^{0}$ and $B_{s}^{0}-\bar{B}_{s}^{0}$ mixings may result
in a significant deviation.  To see this point clearly, let us introduce a
total phase $\phi$ due to direct and/or indirect CP violation,
\begin{equation}
e^{-2i\phi_{f}} \equiv (\frac{p}{q})_{B}\  |\frac{q}{p}|_{B}\  \frac{A_{f}}
{\bar{A}_{f}}
\end{equation}
where $(q/p)_{B} \simeq \sqrt{(M_{12}^{\ast}/M_{12})_{B}}$ for
$\Gamma_{12}/2 \ll M_{12}$. $A_{f}$ and $\bar{A}_{f}$ are the decay amplitudes
with final state $f$.
In a good approximation for $B-$ system, $|q/p|_{B}\simeq 1$, i.e.
the indirect CP violation could be neglected, the CP asymmetry in oscillation
experiments is thus given by
\begin{equation}
A_{CP}^{t} \equiv \frac{\Gamma (B^{0}_{q}(t)\rightarrow f) - \Gamma
(\bar{B}^{0}_{q}(t)\rightarrow f)}{\Gamma (B^{0}_{q}(t)\rightarrow f) + \Gamma
(\bar{B}^{0}_{q}(t)\rightarrow f)} \simeq \mp \sin (2\phi) \sin(\Delta m_{B} t)
\end{equation}
and the CP asymmetry in time-integrated measurement is found to be
\begin{equation}
A_{CP} \equiv \frac{\Gamma (B^{0}_{q}\rightarrow f) - \Gamma
(\bar{B}^{0}_{q}\rightarrow f)}{\Gamma (B^{0}_{q}\rightarrow f) + \Gamma
(\bar{B}^{0}_{q}\rightarrow f)} \simeq \mp \frac{x}{1 + x^{2}}
\frac{\sin (2\phi)}{1 \pm y \cos(2\phi)}
\end{equation}
where $\mp$ corresponds to the CP eigenstate $CP|f> = \pm |f>$. $x = \Delta
m_{B}/\Gamma_{B}$ and $y=\Delta \Gamma_{B}/\Gamma_{B}$.

 It is clear that the situation in our model can be quite different because
of the existance of rich sources of CP violation and the new interactions in
our model. Let us consider the three limit cases discussed in the previous
subsections.

First, when the new type of CP-violating mechanism becomes dominant, which is
the case when $\zeta_{F} \ll 1$,  the above discussed direct CP violation in
B decay will become small and unobservable even in the B-factory although the
$B^{0}-\bar{B}^{0}$ mixing may receive large contribution from the box diagrams
with charged-scalar. This is because the induced KM-phase and the FCNSI are
negligibly small. Nevetheless, its effect in kaon decay can be significant.
Therefore, if the direct CP violation in kaon decay is confirmed to be of
order $10^{-3}$, and direct CP violation in B-meson decay is unobservable
small, it then indicates a signature of the importance of the new type of
CP-violating  mechanism.

Second, when the FCNSI play an important role on the $B^{0}_{d}-
\bar{B}^{0}_{d}$ and $B^{0}_{s}-\bar{B}^{0}_{s}$ mixings, any values
of CP asymmetries are possible in B decay. This is because the phase
$\phi_{M}$ in our model is in general a free parameter. In the limit case
that the FCNSI is dominant and the induced KM-phase is small as well as
the new type of CP-violating mechanism become unimportant, the measured
angles will be approximately equal and are actually the phase of the mixing
matrix $M_{12}$ if it is large enough. If this case occurs, the direct CP
violation in kaon decay should be small. It then implies that the FCNSI
is substantial.

Third, when the induced KM-mechanism becomes dominant, the situation is then
similar to the standard KM-model which has been extensively investigated.

   In general, our model indicates that the relevant measurements of B-meson
decays are likely not to provide the three angles of the CKM unitary
triangle if all the CP-violating mechanisms and the scalar interactions
play an important role. In fact, this can be easily tested by simply
checking the sum of the three angles, i.e. $\Theta = \alpha +
\beta + \gamma$, its deviation from $\pi$ (i.e. $\Theta \neq \pi$) implies
a signal of new physics. Nevertheless, one can remain getting some features
of the CKM matrix from the differences between the angles which contain the
common phase $\phi_{M}$ of the mixing matrix $M_{12}$.

    From these analyses, we may conclude that measurements of direct
CP violation in B-meson decay together with the one in kaon decay are extremely
important in clarifying the mechanisms of CP violation. A more detailed
and quantitative calculation should be of interest.

\subsection{CP Violation in Hyperon Decays}

   Hyperon decays may also provide tests of direct CP violation. Non-leptonic
hyperon decays are usually analyzed by decomposing the amplitudes into
two parts  S and P which correspond to the S-wave (parity-violating) and
P-wave (parity-conserving) final states. The amplitudes S and P are contributed
from different final isospin states and are conveniently written as
\begin{equation}
S = \sum_{I} S_{2I} e^{i\delta_{2I}^{0}}, \qquad
 P = \sum_{I} P_{2I} e^{i \delta_{2I}^{1}}
\end{equation}
with $I$ the isospin and $\delta_{2I}^{i}$ ($i=0,1$) the strong final-state
interaction phases for S and P waves.  The observables are the rates
$\Gamma$, the asymmetry $\alpha$ of the outgoing particles relative to the
initial hyperon spin, and the $\beta$
parameter measuring the polarization of final hyperon transverse to the
plane of the initial hyperon spin and the final momentum. They are usually
defined by
\begin{equation}
\alpha = \frac{2 Re S^{\ast}P}{(|S|^{2} + |P|^{2})}, \qquad
\beta =  \frac{2 Im S^{\ast}P}{(|S|^{2} + |P|^{2})}
\end{equation}
The physical observables of CP violation are defined by comparing these
paramerters for hyperon and its antiparticle
\begin{eqnarray}
\Delta & = & \frac{\Gamma - \bar{\Gamma}}{\Gamma + \bar{\Gamma}} , \nonumber \\
A & = & \frac{\alpha +\bar{\alpha}}{\alpha -\bar{\alpha}}
 = \frac{\sum_{I,I'} Im(S_{2I}^{\ast}P_{2I'}\sin(\delta_{2I'}-\delta_{2I})}
{\sum_{I,I'} |S_{2I}| |P_{2I'}| \cos(\delta_{2I'}-\delta_{2I})} \\
B & = & \frac{\beta +\bar{\beta}}{\beta -\bar{\beta}}
 = \frac{\sum_{I,I'} Im(S_{2I}^{\ast}P_{2I'}\cos(\delta_{2I'}-\delta_{2I})}
{\sum_{I,I'} |S_{2I}| |P_{2I'}| \sin(\delta_{2I'}-\delta_{2I})} \nonumber
\end{eqnarray}
These asymmetry ratios have been studied  in detail by Donoghue, He and
Pakvasa\cite{DHP} in the KM-model, Weinberg 3HDM and left-right symmetric
model.
Their analyses can be directly applied to our model since this model possesses
properties of both the KM-model and Weinberg 3HDM (corresponding to the new
type of CP-violating mechanism , i.e. type-I in our model). Based on their
calculations, we can immediately conclude that in the $\Lambda^{0} \rightarrow
p \pi^{-}$ decay, the CP-violating observables with either KM-type dominant
or new type mechanism dominant  are of the same order of magnitude:
$A\sim 10^{-5}$ and $B\sim 10^{-3}$.  In the $\Xi^{-} \rightarrow
\Lambda^{0} \pi^{-}$, $\Sigma^{\pm} \rightarrow n \pi^{\pm}$ and
$\Sigma^{+} \rightarrow p \pi^{0}$ decays, the CP-violating observables with
the new type of CP-violating mechanism dominant  are larger than those with
the KM-type dominant  and enhanced respectively by a factor of 5,  one order
of magnitude and two orders of magnitude for these three processes . The
values of the CP-violating observables with the new type of CP-violating
mechanism dominant will be significant. In general, we have
\begin{eqnarray}
A(\Sigma^{-} \rightarrow  n \pi^{-}) & \simeq & 10^{-3}, \qquad
B(\Sigma^{-} \rightarrow n \pi^{-}) \simeq 0.1 \ . \nonumber \\
 A(\Sigma^{+} \rightarrow  n \pi^{+}) & \simeq & 10^{-3} \ , \qquad
 B(\Sigma^{+} \rightarrow p \pi^{0}) \simeq 4\times 10^{-2} \ . \\
A(\Xi^{-} \rightarrow  \Lambda ^{0} \pi^{-}) & \simeq & 10^{-4}, \qquad
B(\Xi^{-} \rightarrow  \Lambda^{0} \pi^{-}) \simeq 10^{-3} \ . \nonumber
\end{eqnarray}
it is seen that the CP-violating ratio B is in general large, but the
parameter $\beta$ is difficult to measure. It may be of interest in
measuring the parameter A  at the $10^{-3}$ level in the $\Sigma^{\pm}
\rightarrow n \pi^{\pm}$ decay, which also provides a possible channel to
clarify the different mechanisms.

\section{Neutron and Lepton EDMs $d_{n}$ and $d_{l}$ in the Model}

   The neutron and lepton EDMs are known to be sensitive to the new sources of
CP violation beyond the standard model. This is because the neutron EDM $d_{n}$
and the lepton EDM $d_{l}$ were predicted to be unobservable small in the
standard KM-model.  In fact, in the standard model there exists no CP
violation in the lepton sector.
Therefore, a non-zero EDM of the leptons ($d_{l}$) and the neutron ($d_{n}$)
at the present observable level should clearly provide a signal of new
sources of CP violation. It was recently pointed out by Weinberg \cite{SW2},
Barr and Zee \cite{BZ} that some new class of graphs in the Higgs sector can
provide important contributions to the neutron and lepton EDMs. In particular,
the Barr-Zee mechanism dramatically enhances the electron EDM ($d_{e}$). These
observations renewed interest in this subject. We shall apply those
analyses to our model and show how various contributions to $d_{l}$ and $d_{n}$
arise from our model and distinguish from other models.

\subsection{ $d_{n}$ From One-loop Contributions}

  Let us first consider the contributions to the neutron EDM from the usual
one-loop photon penguin graph with scalar exchange.

 i) {\bf With Charged-scalar Exchange}

     The quark EDM in this case is known \cite{BD} and has the result in our
model
\begin{equation}
d_{d_{i}} = -\frac{G}{\sqrt{2}}\frac{1}{6\pi^{2}} m_{d_{i}} \sum_{j}
Im (\xi_{d_{i}}\xi_{u_{j}}) |V_{ji}|^{2} \frac{y_{j}}{(1-y_{j})^{2}}
(\frac{3}{4} - \frac{5}{4}y_{j} + \frac{1-1.5y_{j}}{1-y_{j}} ln y_{j})
\end{equation}
and
\begin{equation}
d_{u_{i}} = -\frac{G}{\sqrt{2}}\frac{1}{6\pi^{2}} m_{u_{i}} \sum_{j}
Im (\xi_{u_{i}}\xi_{d_{j}}) |V_{ij}|^{2} \frac{y_{j}}{(1-y_{j})^{2}}
(y_{j} + \frac{1-3y_{j}}{2(1-y_{j})} ln y_{j})
\end{equation}

  For an order-of-magnitude estimation, using the nonrelativistic-quark-model
relation
\begin{equation}
d_{n} = \frac{4}{3} d_{d} - \frac{1}{3} d_{u}
\end{equation}
and noticing that $d_{d}\gg d_{u}$ and $|V_{td}| \ll |V_{cd}|$, we obtain
in a good approximation of charm-quark dominance that
\begin{equation}
d_{n}^{q\gamma} \simeq \frac{4}{3} d_{d} = 0.5\times 10^{-26} Im(\xi_{d}
\xi_{c})(\frac{50 GeV}{m_{H^{+}}})^{2} \frac{1}{3} (ln\frac{m_{H^{+}}}{m_{c}}
- \frac{3}{8})\  \mbox{e cm}
\end{equation}

ii) {\bf With Neutral-scalar Exchange}

    From the nonrelativistic-quark-model relation, the neutron EDM from
neutral-
scalar exchange is negligible since it is suppressed by the light quark mass
about order of $m_{d}^{2}/m_{c}^{2}$ comparing to the case with charged-scalar.
A reasonable estimation through considering the nonperturbative effects at low
momenta was suggested by Anselm {\it et al} \cite{ABGU} and has recently been
reconsidered by T.P. Cheng and L.F. Li\cite{CL} in the light of experimental
new information on the scalar and pseudoscalar couplings of the Higgs boson.
Most recently, this has also been examined by H.Y. Cheng \cite{HYC2} in
the Weinberg 3HDM. Following their analyses, we have in our model
\begin{equation}
d_{n}^{n\gamma} \simeq -2.4 \times 10^{-26} \sum_{k=1}^{3} Im(\eta_{d}^{(k)} -
\eta_{u}^{(k)})^{2}(\frac{50 GeV}{m_{H^{0}_{k}}})^{2}\  \mbox{e cm}
\end{equation}
we see that this is not far below to the present experimental limit for
appropriate values of the parameters  $\eta_{u,d}^{(k)}$.

\subsection{$d_{n}$ From Weinberg Gluonic Operator and Quark CEDM}

i) {\bf $d_{n}$ from Weinberg Gluonic Operator}

  It was  pointed out by Weinberg that the dimension-6 $P$- and $T$-
violating three-gluon operator may provide significant contribution to
the neutron EDM, the resulted lagrangian can be written
\begin{equation}
L_{3g} = \frac{G}{\sqrt{2}} C_{3g} O_{3g} \equiv  \frac{G}{\sqrt{2}} C_{3g}
(-\frac{1}{6} f^{abc} G_{\mu \rho}^{a} G_{\nu}^{b\rho} \tilde{G}^{c \mu \nu})
\end{equation}
with $C_{3g} = C_{3g}^{N} + C_{3g}^{C}$, where $C_{3g}^{N}$ comes from
the neutral-scalar exchange and is given in our model
\begin{equation}
C_{3g}^{N} = \frac{1}{64\pi^{4}}\sum_{k=1}^{3} Im(\eta_{t}^{(k)})^{2}
h_{N}(z_{t}^{(k)}) \prod_{n=3}^{5}
(\frac{g_{s}(m_{q_{n}})}{g_{s}(m_{q_{n+1}})})
^{\gamma_{n}^{N}/\beta_{n}}
\end{equation}
and $C_{3g}^{C}$ from the charged-scalar exchange\cite{DICUS}
 \begin{equation}
C_{3g}^{C} = \frac{1}{64\pi^{4}} Im(\xi_{t}\xi_{b})
h_{C}(y_{b}, y_{t}) \prod_{n=3}^{5}
(\frac{g_{s}(m_{q_{n}})}{g_{s}(m_{q_{n+1}})})
^{\gamma_{n}^{C}/\beta_{n}}
\end{equation}
with\cite{GAMA} $\gamma_{3}^{N} =\gamma_{4}^{N}=\gamma_{5}^{N} \equiv
\gamma_{g} = -18$,  $\gamma_{3}^{C} =\gamma_{4}^{C}= \gamma_{g}$,
$\gamma_{5}^{C} = \gamma_{b}= -14/3$ and $\beta_{n} = (33-2n)/6$.
$m_{q_{3}} = \mu$, $m_{q_{4}} = m_{c}$,
$m_{q_{5}} = m_{b}$ and $m_{q_{6}} = m_{t}$. Numerically,
the ratio of the QCD corrections between the charged-scalar-exchange and
the neutral-scalar-exchange is about 3. Where $h_{N}(z_{t}^{(k)})$ and
$h_{C}(y_{b}, y_{t})$ are integral functions (see also Appendix) with
$h_{N}(1)\sim 0.05$ and $h_{C}(y_{b}\ll 1, y_{t}\sim 1)= 1/12$.
 For a numerical estimation, using the naive dimensional analysis and
taking $g_{s}(\mu)/4\pi \simeq 1/\sqrt{6}$ \cite{SW2}, it reads
\begin{eqnarray}
d_{n}^{3g} & = & d_{n}^{3gN} + d_{n}^{3gC} \nonumber \\
d_{n}^{3gN} & = & 1.2 \times 10^{-26} \sum_{k=1}^{3} Im(\eta_{t}^{(k)})^{2}
[20\  h_{N}(z_{t}^{(k)})]\  \mbox{e cm} \\
d_{n}^{3gC} & = & 3.3 \times 10^{-25} Im(\xi_{t}\xi_{b})
[12\  h_{C}(y_{b}, y_{t})]\  \mbox{e cm}
\end{eqnarray}
note that the above results can only be regarded as an order-of-magnitude
estimation within the theoretical uncertainties for the hadronic matrix
element.
A recent reanalysis by
Bigi and Uraltsev\cite{BU} shown that the hadronic matrix element may be
reduced by a factor $30$.  Even for this case, when the CP-violating parameter
$\epsilon$ is fitted to the experimental data in the Weinberg 3HDM, the
resulting neutron EDM $d_{n}$ from the charged-scalar exchange is already
in the ball of the present experimental bound. In our model,
this is avoided by choosing relative small values of $ Im(\xi_{t}\xi_{b})$
in fitting the  neutron EDM $d_{n}$, whereas $\epsilon$ is fitted
independently by $Im(\xi_{c}\xi_{s})$ and $ Im(\xi_{c}\xi_{s})^{2}$
(see previous section).

ii) {\bf $d_{n}$ from Quark Gluonic Chromo-EDM}

   It was first observed by Barr and Zee that a new class of two-loop graphs
due to neutral-scalar exchange can produce a large EDM for light quarks and
leptons. Motivated from this observation,  Chang, Keung and Yuan\cite{CKY},
Gunion and Wyler\cite{GW} investigated the gluonic chromo-electric dipole
moment (CEDM) with photons replaced by gluons in the Barr-Zee mechanism and
found that the contributions to $d_{n}$ from the CEDM of light quarks are
likely to dominate over those arising from the quark EDMs via the Barr-Zee
mechanism. The quark CEDM $d_{q}^{g}$ is defined as
\begin{equation}
L_{qg} = i\frac{1}{2} d_{q}^{g}\ \bar{q}\sigma_{\mu\nu}\frac{1}{2}\lambda^{a}
\gamma_{5} q G^{a\mu\nu}
\end{equation}
In our model, the results for quark CEDM's have the following form
\begin{eqnarray}
 d_{q}^{G} & = & \frac{G}{\sqrt{2}} \frac{m_{q}}{\pi} (\frac{g_{s}}{4\pi})^{3}
\prod_{n=3}^{5}[\frac{g_{s}(\mu)}{g_{s}(m_{q_{n+1}})}]^{74/6\beta_{n}}
\nonumber \\
 & & \cdot \sum_{k=1}^{3} \{Im(\eta_{Q}^{(k)}\eta_{q}^{(k)})[f(z_{Q}^{(k)}) +
g(z_{Q}^{(k)})] - Im(\eta_{Q}^{(k)}\eta_{q}^{(k)\ast})[f(z_{Q}^{(k)}) -
g(z_{Q}^{(k)})] \}
\end{eqnarray}
where $f(z_{Q}^{(k)})$ and $g(z_{Q}^{(k)})$ are integral functions with
$Q$ the loop-quark and for $z_{Q}^{(k)}\sim 1$ $f(1)\sim 0.8$ and $g(1)\sim 1$.
(see also Appendix).
Note that the above form is valid for all the quarks in our parametrization and
considerations of the model. This differs from the usual 2HDM with NFC or
supersymmetric 2HDM in which the results for the up-type and down-type  quarks
are different due to discrete symmetries. In the valence quark model, the
contribution to the $d_{n}$ is given by
\begin{eqnarray}
 d_{n}^{(qg)} & = & \frac{1}{3} e (\frac{4}{3} d_{d}^{G} +
\frac{2}{3} d_{u}^{G})/g_{s} = \frac{G}{\sqrt{2}} \frac{4m_{d}}{9\pi}
(\frac{e}{4\pi})(\frac{g_{s}}{4\pi})^{3} \nonumber \\
& & \cdot \prod_{n=3}^{5} [\frac{g_{s}(\mu)}{g_{s}(m_{q_{n+1}})}]^{74/6
\beta_{n}} 2\sum_{k=1}^{3} \{Re\eta_{Q}^{(k)}Im \eta'_{k}\ f(z_{Q}^{(k))} +
Im\eta_{Q}^{(k)}Re \eta'_{k}\  g(z_{Q}^{(k))} \}  \nonumber \\
&  \simeq & 6 \times 10^{-26} \sum_{k=1}^{3}\{Re\eta_{Q}^{(k)}
Im \eta'_{k}\ f(z_{Q}^{(k)}) +  Im\eta_{Q}^{(k)} Re \eta'_{k}\ g(z_{Q}^{(k)})
 \}\  \mbox{e cm}
\end{eqnarray}
with $\eta'_{k} = \eta_{d}^{(k)} + \eta_{u}^{(k)}m_{u}/2m_{d}$.
where $g_{s}(\mu)/4\pi \simeq 1/\sqrt{6}$ has been used to obtain the numerical
value. Note that the loop-bottom-quark contribution may also give a sizeable
contribution if $|Im\eta_{b}^{(k)}| \gg |Im\eta_{t}^{(k)}|$
 or $|Re\eta_{b}^{(k)}| \gg |Re\eta_{t}^{(k)}|$.

  For comparison, we also present the results via Barr-Zee mechanism
\begin{eqnarray}
 d_{n}^{(BZ)} & = &  \frac{4}{3} d_{d}^{q} -
\frac{1}{3} d_{u}^{q} \nonumber \\
 & \simeq & 1.5\times 10^{-27} \sum_{k=1}^{3}\{Re\eta_{Q}^{(k)}Im \eta'_{k}
\ f(z_{Q}^{(k)}) +  Im\eta_{Q}^{(k)} Re \eta'_{k}\ g(z_{Q}^{(k)}) \}\
\mbox{e cm}
\nonumber \\
& & +  4.5\times 10^{-27} \sum_{k=1}^{3} O_{2k}^{H}
Im (\eta_{d}^{(k)} -\frac{1}{4}\frac{m_{u}}{m_{d}}\eta_{u}^{(k)})
[\frac{3}{5} f(z_{W}^{(k)}) + g(z_{W}^{(k)} ) ]\ \mbox{e cm}
\end{eqnarray}
where the first term is the contribution arising from the Q-quark-loop and
the second term is the one from the W-boson-loop. One sees that it is smaller
than the contribution from the quark CEDM by one order of magnitude.

   Furthermore, it was pointed out by He, Mckeller and Pakvasa\cite{HMP} that
the strange quark contribution to the neutron EDM is significant because the
strange quark CEDM is enhanced by a factor $m_{s}/m_{q}$ ($q=d, u$).
 From the  CP-odd  $K\Sigma n$ vertex generated from the $L_{sg}$ (i.e.
$<K\Sigma | -L_{sg} | n>$)  and the CP-even $K\Sigma n$ vertex, they
obtained an effective lagrangian which induces a strange quark contribution
to the netron EDM. From their numerical relation, we have
\begin{eqnarray}
 d_{n}^{(sg)} & = & 0.027 d_{s}^{G}/g_{s}(\mu) \nonumber \\
 &  \simeq & 3.2 \times 10^{-26} \sum_{k=1}^{3}\{Re\eta_{Q}^{(k)}
Im \eta_{s}^{(k)} f(z_{Q}^{(k)}) +  Im\eta_{Q}^{(k)} Re \eta_{s}^{(k)}
g(z_{Q}^{(k)}) \}\  \mbox{e cm}
\end{eqnarray}

\subsection{$d_{n}$ from FCNSI}

    The new contributions to $d_{n}$ can arise from the FCNSI in our model.
It is easily seen that the dominant contribution comes from the  up-quark
through  one-loop photon penguin with neutral-scalar exchange and virtual
top quark.  This is because such a vertex in the  FCNSI is
proportional to $\sqrt{m_{u} m_{t}}$. With a simple one-loop calculation and
using the relation between the neutron EDM and the quark EDM in the valence
quark model, we obtain
\begin{eqnarray}
d_{n}^{(FCNSI)} & = & \frac{2}{3} d_{u}^{q} = \frac{G}{\sqrt{2}}
\frac{m_{\mu}}{18\pi^{2}}
(\frac{\zeta_{U}}{s_{\beta}})^{2} \sum_{k=1}^{3}Im(S_{k,13}^{u} S_{k,31}^{u})
F(z_{t}^{(k)}) \nonumber \\
& = & 4\times 10^{-26} (\frac{\zeta_{U}}{0.1 s_{\beta}})^{2} \sum_{k=1}^{3}
Im(S_{k,13}^{u} S_{k,31}^{u}) F(z_{t}^{(k)})\  \mbox{e cm}
\end{eqnarray}
with
\begin{equation}
F(z_{t}^{(k)})  =  \frac{z_{t}^{(k)}}{z_{t}^{(k)}-1}
(1 + \frac{1}{1-z_{t}^{(k)}}ln z_{t}^{(k)})
\end{equation}
 and $F(1) = 1/2$ and $F(\infty) = 1$. We see that the above value of
$d_{n}$  can also be closed to the present experimental limit. From this result
together with the one of $D^{0}-\bar{D}^{0}$ mixing discussed in the above
subsection and the corresponding experimental bounds on $d_{n}$ and
$\Delta m_{D}$, it is expected that $S_{k,13}^{u} \sim S_{k,12}^{u} \sim O(1)$
and $\zeta_{U}/s_{\beta} \sim O(10^{-1})$ if CP-violating phases are indeed
generically of order unity (here $\sim$ will be understood an
approximate equality within a factor of about 2).

    With all these results, the total value of the neutron EDM is then the sum
of all the contributions
\begin{equation}
d_{n} = d_{n}^{(q\gamma)} + d_{n}^{(n\gamma)} + d_{n}^{(3g)} + d_{n}^{(qg)}
       + d_{n}^{(sg)} + d_{n}^{(BZ)} + d_{n}^{(FCNSI)}
 \end{equation}

    It should be noted that  various contributions to $d_{n}$ discussed above
may become comparable each other in our  model, which depends on the
paramerters $\xi_{f_{i}}$, $\eta_{f_{i}}^{(k)}$, $\zeta_{U}$ and masses of
the scalars,  this is unlike either the Weinberg 3HDM or the usual 2HDM
with NFC or supersymmetric 2HDM. Note also that it is not
excluded to have cancellations among several terms, nevertheless, without
accidental total cancellation among various terms, the neutron EDM $d_{n}$
should not be far below to the present experimental bound  for the allowed
reasonable range of the parameters.

 Before proceeding,  we would like to comment that the above analyses should
be simply extended to the other baryons. For example,
the $\Lambda$- baryon EDM $d_{\Lambda}$.  It is in general enhanced by a
factor of $m_{s}/m_{d}$ or even more $(\sin\theta_{c})^{-2} m_{s}/m_{d}
\sim (m_{s}/m_{d})^{2}$ comparing to $d_{n}$. Nevertheless, it is still
much far below to the present experimental bound\cite{PDG}
$d_{\Lambda} < 1.5 \times  10^{-16} $ e cm.

\subsection{One-Loop Contribution to $d_{l}$}

  As the mass of the neutrinos is considered to be zero, the one-loop
contribution to $d_{l}$ thus only arises from the neutral-scalar exchange.
For the flavor-conserving part, we obtain
\begin{equation}
d_{l} = \frac{G}{\sqrt{2}}\frac{m_{l}}{8\pi^{2}} \sum_{k=1}^{3}
Im (\eta_{l}^{(k)})^{2} z_{l}^{(k)} [1 + \frac{1}{2z_{l}^{(k)}} ln z_{l}^{(k)}
+ \frac{1 - 2 z_{l}^{(k)}}{2z_{l}^{(k)}\sqrt{1-4z_{l}^{(k)}}}
ln \frac{1 + \sqrt{1-4z_{l}^{(k)}}}{1 - \sqrt{1-4z_{l}^{(k)}}} ]
\end{equation}
with $z_{l}^{(k)} = m_{l}^{2}/m_{H_{k}^{0}}^{2}$. Numerically, we have
 \begin{eqnarray}
d_{e} & \simeq &  4.9\times 10^{-33} \sum_{k=1}^{3}
Im (\eta_{e}^{(k)})^{2} (\frac{50 GeV}{m_{H_{k}^{0}}})^{2} (ln
\frac{m_{H_{k}^{0}}}{50GeV} + 11.6)\  \mbox{e cm} \\
d_{\mu} & \simeq &  2.4\times 10^{-26} \sum_{k=1}^{3}
Im (\eta_{\mu}^{(k)})^{2} (\frac{50 GeV}{m_{H_{k}^{0}}})^{2} (ln
\frac{m_{H_{k}^{0}}}{50GeV} + 6.2)\  \mbox{e cm} \\
d_{\tau} & \simeq &  6.3\times 10^{-23} \sum_{k=1}^{3}
Im (\eta_{\tau}^{(k)})^{2} (\frac{50 GeV}{m_{H_{k}^{0}}})^{2} (ln
\frac{m_{H_{k}^{0}}}{50GeV} + 3.4)\  \mbox{e cm}
\end{eqnarray}
where $d_{e}$ and $d_{\mu}$ are far below to the present experimental
sensitivity \cite{PDG}
\begin{equation}
d_{e} = (-3\pm 8)\times 10^{-27}\  \mbox{e cm}\ , \qquad  d_{\mu} =
(3.7\pm 3.4)\times 10^{-19}\  \mbox{e cm}\ .
\end{equation}

\subsection{$d_{l}$ From Two-loop Barr-Zee Mechanism}

  The smallness of the one-loop contributions to $d_{l}$ is because
three helicity flips are involved in the simple one-loop
graphs.  Recently, Barr and Zee pointed out that the suppression factor of
$m_{l}^{2}/m_{H_{k}^{0}}^{2}$ due to the helicity flip can be overcome in a
new class of two-loop graphs. In these graphs photons are setted on a top-quark
loop or W-boson loop, only one helicity flip occurs through an explicit
neutral-scalar coupling. Consequently, the electron EDM is enhanced
dramatically and can be closed to the present experimental bound. Following
the analysis in \cite{BZ}, we have the result in our model
\begin{eqnarray}
 d_{l}^{Q-loop} & = & \frac{G}{\sqrt{2}} \frac{32}{3}\frac{\alpha}{(4\pi)^{3}}
m_{l} \sum_{k=1}^{3}\{Im(\eta_{Q}^{(k)}\eta_{l}^{(k)}) [f(z_{Q}^{(k)}) +
g(z_{Q}^{(k)})] \nonumber \\
& & - Im(\eta_{Q}^{(k)}\eta_{l}^{(k)\ast}) [f(z_{Q}^{(k)}) -
g(z_{Q}^{(k)})] \} \\
d_{l}^{W-loop} & = & \frac{G}{\sqrt{2}}\frac{40\alpha}{(4\pi)^{3}}
m_{l} \sum_{k=1}^{3} O_{2k}^{H} Im \eta_{l}^{(k)} [ \frac{3}{5}
f(z_{W}^{(k)}) + g(z_{W}^{(k)} ) ]
\end{eqnarray}
Numerically,  we find
\begin{eqnarray}
 d_{l}^{Q-loop} & = & 0.7 \times 10^{-26} \frac{m_{l}}{m_{e}}
\sum_{k=1}^{3} \{ Re\eta_{Q}^{(k)}Im \eta_{l}^{(k)} f(z_{Q}^{(k)}) +
Im\eta_{Q}^{(k)} Re \eta_{l}^{(k)} g(z_{Q}^{(k)}) \} \mbox{e cm} \\
d_{l}^{W-loop} & = & 2\times 10^{-26}
\frac{m_{l}}{m_{e}} \sum_{k=1}^{3} O_{2k}^{H} Im \eta_{l}^{(k)} [ \frac{3}{5}
f(z_{W}^{(k)}) + g(z_{W}^{(k)} ) ]\ \mbox{e cm}
\end{eqnarray}
which should be not far below to the present experimental bound.

\subsection{$d_{l}$ From FCNSI}

    We now discuss a new contribution to $d_{l}$  in our model.
It arises from the FCNSI.   The dominant contribution to $d_{l}$ comes
from the one-loop photon penguin with neutral-scalar exchange and virtual
$\tau$- lepton.  The considerations are analogous to the one discussed
above for the quark EDM from the FCNSI. Following those analyses, it is
easily seen that
\begin{eqnarray}
d_{l_{i}}^{(FCNSI)} & = & \frac{G}{\sqrt{2}}\frac{m_{l_{i}}}{4\pi^{2}}
(\frac{\zeta_{E}}{s_{\beta}})^{2} \sum_{k=1}^{3}Im(S_{k,i3}^{e} S_{k,3i}^{e})
F(z_{\tau}^{(k)})  \\
d_{e}^{FCNSI} & \simeq & 0.43\times 10^{-27} (\frac{\zeta_{E}}{0.1 s_{\beta}})
^{2} \nonumber \\
& & \cdot \sum_{k=1}^{3} Im(S_{k,13}^{e} S_{k,31}^{e})
(\frac{50 GeV}{m_{H_{k}^{0}}})
^{2}( 1 + 0.42 ln \frac{m_{H_{k}^{0}}}{50 GeV}) \mbox{e cm}  \\
d_{\mu}^{FCNSI} & \simeq & 0.86\times 10^{-25} (\frac{\zeta_{E}}
{0.1 s_{\beta}})^{2} \nonumber \\
& & \cdot \sum_{k=1}^{3} Im(S_{k,23}^{e} S_{k,32}^{e})
(\frac{50 GeV}{m_{H_{k}^{0}}})^{2} ( 1 + 0.42 ln \frac{m_{H_{k}^{0}}}{50 GeV})
e cm
\end{eqnarray}
Obviously, their exact values depend explicitly on the parameters
$\zeta_{E}$ and $ S_{k,ij}^{e}$. It is of interest in noticing that
when naively taking $\zeta_{E} \sim \zeta_{U}\sim O(10^{-1})$ and $S_{k,ij}^{e}
\sim S_{k,ij}^{u} \sim O(1) $ for an order-of-magnitude estimation, the
predicted values for $d_{e}$ and $d_{n}$ from the FCNSI, both of them,
are closed to their present experimental sensitivities.

    We can now present the total results of the lepton EDM  $d_{e}$ which is
the sum of all the contributions
\begin{equation}
d_{l} = d_{l}^{W-loop} + d_{l}^{Q-loop} + d_{l}^{FCNSI}
 \end{equation}

With these considerations, we come to the following remarks:  Firstly,
comparing to the experimental data, the present experimental
sensitivity of the electron EDM is already at the level of providing a clean
test of CP violation in the neutral scalar sector. This is unlike the neutron
EDM $d_{n}$ which involves the uncertanties from nonperturbative QCD effects.
Secondly, it should be of interest in improving the
measurement of the muon EDM $d_{\mu}$  to the same experimental sensitivity
as the one for the electron EDM $d_{e}$. This may help us to distingush
our 2HDM model from the other 2HDM with NFC and soft CP violation
or multi-Higgs doublet model with NFC or supersymmetric 2HDM with soft CP
violation.  This is because in the latter
models the CP-violating phases for the electron and muon EDM's are the same,
which leads to $d_{\mu}/d_{e} = m_{\mu}/m_{e}$. Whereas in our model,
as stated in the previous section that Higgs mechanism produces CP-violating
phases for all the fermions, which are in general dishtinguished for different
fermions. Therefore  without accidental coincidence, it should have
$d_{\mu}/d_{e} \neq m_{\mu}/m_{e}$ in our model.

\section{CP Violation in Other Processes}

\subsection{Muon Polarization in $K_{L} \rightarrow \mu^{+} \mu^{-}$ Decays}

  The longitudinal $\mu$ polarization in $K_{L}\rightarrow \mu^{+} \mu^{-}$
decay is defined by
\begin{equation}
P_{L} = \frac{N_{R} - N_{L}}{N_{R} + N_{L}} = \frac{2r Im(ba^{\ast})}{|a|^{2}
+ r^{2}|b|^{2}}
\end{equation}
where $N_{R}$ and $N_{L}$ are the numbers of left-handed and right-handed
outgoing muons respectively, $r^{2} = (1 - 4m_{\mu}^{2}/m_{K}^{2})$. $a$ and
$b$
are defined in an effective $K_{L}\rightarrow \mu^{+} \mu^{-}$ decay
hamiltonian as
\begin{equation}
H_{eff} = \frac{G}{\sqrt{2}} \bar{s}i\gamma_{5} d ( b \bar{\mu}\mu - a
\bar{\mu}i\gamma_{5} \mu ) + h.c.
\end{equation}
where $a$ is in general complex due to the absorptive part from the $2\gamma$
intermediate state which is known dominating the $K_{L}\rightarrow
\mu^{+} \mu^{-}$ decay. $b$ represents the CP-violating part. Using the
unitarity estimate and the experimental values for the decay rates of
$K_{L}\rightarrow \mu^{+} \mu^{-}$ and $K_{L}\rightarrow \gamma \gamma$,
one finds \cite{GN,HM,PH}
\begin{equation}
P_{L} \simeq 2.36 \times 10^{6} Re b
\end{equation}
its value has recently been estimated to be $P_{L} \leq 10^{-3}$ for the
standard model\cite{GN}. The estimations have also been carried out for
the supersymmetric models and left-right models

We now present an estimation for our model. The dominant contribution to
$Re b$ comes from the FCNSI at tree level, it is not difficult to find
in our model
\begin{equation}
Re b = \sum_{k=1}^{3}\frac{\sqrt{m_{s}m_{d}}m_{\mu}}{m_{H_{k}^{0}}^{2}}
\frac{\zeta_{D}}{s_{\beta}} Re[\frac{i}{2}(\tilde{S}_{k,12}^{d} -
\tilde{S}_{k,21}^{d \ast})] 2Re\eta_{\mu}^{(k)}
\end{equation}
 where the CP-violating factor $i(\tilde{S}_{k,12}^{d} -
\tilde{S}_{k,21}^{d \ast})$ may be related to the one appearing in
$\epsilon^{II}_{H^{0}-Tree}$. To see this, we define
\begin{equation}
-\frac{i}{2}(\tilde{S}_{k,12}^{d} - \tilde{S}_{k,21}^{d \ast}) \equiv
|\tilde{Z}_{k,12}^{d}| e^{i\tilde{\delta}_{k,12}^{d}}
\end{equation}
In the vacuum insertion approximation for evaluating the $\Delta S=2$ matrix
element, we have $\tilde{Y}_{k,12}^{d} \simeq \tilde{Z}_{k,12}^{d}$.
In this approximation, we obtain
\begin{equation}
Re b \simeq -0.9 \times 10^{-8} \sum_{k=1}^{3} Re\eta_{\mu}^{(k)}
\frac{50 GeV}{m_{H_{k}^{0}}} \sqrt{\frac{\cos
\tilde{\delta}_{k,12}^{d}}{\sin \tilde{\delta}_{k,12}^{d}}}
\sqrt{\frac{\epsilon^{II}_{H^{0}-Tree}}{\epsilon^{exp.}}}
\end{equation}
we now consider two alternative choices. First, when the FCNSI is demanded
to accommodate both the $\Delta m_{K}$ and $\epsilon$, one must fine-tune
the parameter so that $\sin \tilde{\delta}_{k,12}^{d}/
\cos \tilde{\delta}_{k,12}^{d} \simeq 6.4 \times 10^{-3}$. In this case,
$P_{L}$ is given by
\begin{equation}
|P_{L}|  \simeq 0.35 \sum_{k=1}^{3} |Re\eta_{\mu}^{(k)}|
\frac{50 GeV}{m_{H_{k}^{0}}}
\end{equation}
Second, CP-violating phases are indeed generically of order unity,
thus, the FCNSI is only used to fit the CP-violating parameter
$\epsilon$. For this case, if taking  $\sin 2\tilde{\delta}_{k,12}^{d}
\sim \cos 2\tilde{\delta}_{k,12}^{d}$, we then obtain
\begin{equation}
|P_{L}| \simeq 10^{-2} \sum_{k=1}^{3} |Re\eta_{\mu}^{(k)}|
\frac{50 GeV}{m_{H_{k}^{0}}}
\end{equation}
where the coupling constant $Re\eta_{\mu}^{(k)}$ is in principle
free parameter and  can be much larger than one in our model. Note that
$Re\eta_{\mu}^{(k)}$ is directly related to the muon EDM (see previous
section), nevertheless, the present experimental data do not yet
provide more stringent restriction on it.
Therefore for both cases the value
of the $P_{L}$ could be closed to the experimental limit $|P_{L}|
\leq 0.5$ which is extracted \cite{HVGN} from the measured braching ratio
$B(K_{L}\rightarrow \mu^{+} \mu^{-})$ and the unitarity bound
$Br(K_{L}\rightarrow \mu^{+} \mu^{-})_{2\gamma}$.
In general, it is expected that  the value of
the $P_{L}$ in this model is larger than the one predicted from the standard
model  by at least one order of magnitude.

 \subsection{T-odd and CP-odd Triple Momentum Correlations  in
$B\rightarrow D^{\ast}(\rightarrow D\pi) \tau \nu_{\tau}$ Decay}

 T-odd and CP-odd correlations  arise from interferences among different
currents which have non-zero relative CP-violating phases. In the standard
model, the lowest-order contribution to the b- semileptonic decays
is via W-boson exchange. Therefore the interference necessary for the T-odd
and CP-odd correlations in the standard model can only come from
high-order radiative corrections.
The resulted effects are expected to be small. Thus, the observation of
such correlations can  also provide  an evidence for new physics. Unlike the
standard model, such correlations can occur in our  model
at the lowest-order  because of the existence of the CP-violating scalar
interactions. As the scalar CP-violating interactions are energy dependence,
it will be of importance to probe CP violation in the heavy fermion system.
Obviously, the system of hadrons containing a b-quark should be of
interest. Let us examine two interesting processes in this paper. One is in the
exclusive semi-leptonic decay $B\rightarrow D^{\ast}(\rightarrow D\pi)
\tau \nu_{\tau}$, and another is in the inclusive semi-leptonic decay
$b \rightarrow c \tau \nu_{\tau}$.

 The first process has been detailed studied in the paper\cite{KSW} with a very
general CP-violating effective Hamiltonian.  As one of the cases, the T-odd and
CP-odd triple momentum correlation occurs through the
interference between the transverse vector current due to the W-exchange and
the pseudoscalar interaction due to the charged-scalar exchange.  It
becomes clear from the following effective Hamiltonian induced in our
model
\begin{eqnarray}
H_{eff} & = & \frac{G}{\sqrt{2}}V_{cb} \{ \bar{c}\gamma_{\mu}(1-\gamma_{5})b
\bar{\tau}\gamma^{\mu}(1-\gamma_{5})\nu_{\tau} + \frac{m_{b}m_{\tau}}
{m_{H^{+}}^{2}}\tilde{\xi}_{b} \xi_{\tau}^{\ast} \bar{c}(1+\gamma_{5})b
\bar{\tau}(1-\gamma_{5})\nu_{\tau} \nonumber \\
& & - \frac{m_{c}m_{\tau}}
{m_{H^{+}}^{2}}\tilde{\xi}_{c}^{\ast} \xi_{\tau}^{\ast} \bar{c}(1-\gamma_{5})b
\bar{\tau}(1-\gamma_{5})\nu_{\tau} + H.C. \}
\end{eqnarray}
with
\begin{equation}
\tilde{\xi}_{c}^{\ast} = \xi_{c} + \frac{\zeta_{U}}{s_{\beta}}
\sqrt{\frac{m_{t}}{m_{c}}}\frac{1}{|V_{cb}|} (\frac{V_{cb}}{|V_{cb}|}
\tilde{S}_{32}^{u})^{\ast}; \qquad  \tilde{\xi}_{b} = \xi_{b} +
\frac{\zeta_{D}}{s_{\beta}}\sqrt{\frac{m_{s}}{m_{b}}}\frac{1}{|V_{cb}|}
(\frac{V_{cb}^{\ast}}{|V_{cb}|}
\tilde{S}_{23}^{d})
\end{equation}
where the $\zeta_{F}$ terms arise from the flavor-changing scalar interactions.
Their effects can become substantial when $\zeta_{U}/s_{\beta}
\sim O(10^{-1})$ (see below).

   Before discussing the CP-asymmetry measure of the correlations among the
three momenta ${\bf p_{\tau}}$, ${\bf p}_{D}$ and  ${\bf p_{\pi}}$ of the
$\tau$ lepton,  $D$- and $\pi$- mesons respectively, it is useful to
first define a conventional
frame. Such a frame is found to be that  $B$ is at the rest, $D^{\ast}$ is
along the $z$- direction, ($\tau-\bar{\nu}_{\tau}$) and ($D-\pi$) are in
their CM systems. Let $\theta$ and $\chi$ are the polar and azimuthal angles
of the $\tau$- lepton in the ($\tau-\bar{\nu}_{\tau}$) CM system,
$\theta^{\ast}$ is the angle between the $D$ and the $D^{\ast}$ system
in the ($D-\pi$) CM system.

   The experimental observable of CP-asymmetry ratio of the triple momentum
correlation is then defined by
\begin{equation}
A_{V_{T}P}^{exc.} = \frac{\Gamma (B\rightarrow D^{\ast}(D\pi)
\tau \nu_{\tau})|_{0\leq \chi < \pi}^{0 \leq \theta^{\ast} < \pi/2}
- \Gamma (B\rightarrow D^{\ast}(D\pi)
\tau \nu_{\tau})|_{\pi \leq \chi < 2\pi}^{\pi/2 \leq \theta^{\ast} < \pi}}
{\Gamma (B\rightarrow D^{\ast}(D\pi)
\tau \nu_{\tau})|_{0\leq \chi < \pi}^{0 \leq \theta^{\ast} < \pi/2}
+ \Gamma (B\rightarrow D^{\ast}(D\pi)
\tau \nu_{\tau})|_{\pi \leq \chi < 2\pi}^{\pi/2 \leq \theta^{\ast} < \pi}}
\end{equation}
Numerically, we find that
\begin{eqnarray}
A_{V_{T}P}^{exc.} & \simeq &  10^{-2}[\frac{m_{b}m_{\tau}}{m_{H^{+}}^{2}}
Im(\tilde{\xi}_{b} \xi_{\tau}^{\ast})- \frac{m_{c}m_{\tau}}{m_{H^{+}}^{2}}
Im(\tilde{\xi}_{c}^{\ast} \xi_{\tau}^{\ast}) ] \simeq 2.6\times 10^{-4} \{
\frac{0.04}{|V_{cb}|}\frac{\zeta_{U}}{0.1s_{\beta}} \nonumber \\
& & \cdot \sqrt{\frac{m_{t}}{150GeV}}
Im(\frac{V_{cb}}{|V_{cb}|}\tilde{S}_{32}^{u}\xi_{\tau})
+ 0.13 [Im(\tilde{\xi}_{b} \xi_{\tau}^{\ast})
+ 0.3 Im(\xi_{c} \xi_{\tau}) ] \} (\frac{50 Gev}{m_{H^{+}}})^{2}
\end{eqnarray}
It is seen that  this asymmetry ratio can be of order $10^{-4} - 10^{-3}$
for the allowed range of the parameters, for instance,  when $\tan\beta =
v_{2}/v_{1} \sim m_{t}/m_{b} \sim 30 $, $|\xi_{\tau}| \sim \tan\beta$ and
the CP-violating phase is of order unity.
We see that the flavor-changing scalar inteaction may give dominant
contributions to the asymmetry measure.  Experimental study for such an
asymmetry should be of interest since  the
branch ratio of the $B\rightarrow D^{\ast}(D\pi) \tau \nu_{\tau}$ decay
is expected to be about $2\%$ \cite{MWW}.

 \subsection{Triple Spin-Momentum Correlations
in the Inclusive Decay $b \rightarrow c \tau \nu_{\tau}$}

   We consider here the case of two momenta and a spin correlation in the
inclusive decay $b \rightarrow c + \tau + \bar{\nu}_{\tau}$. The experimental
observable is defined  as \cite{GV}
\begin{equation}
A^{inc.}_{VP} = \frac{N_{events}({\bf p\cdot p_{\tau}\times s_{\tau}} > 0) -
N_{events}({\bf p\cdot p_{\tau}\times s_{\tau}} < 0)}{N_{events}({\bf p\cdot
p_{\tau}\times s_{\tau}} > 0) + N_{events}({\bf p\cdot p_{\tau}\times
s_{\tau}} < 0)}
 \end{equation}
where the $\tau$ lepton has momentum ${\bf p}_{\tau}$ and is supposed to
have a definite spin $s_{\tau}$. {\bf p} is the momentum of the final quark
jet. Averaging over the initial quark spin, and sum over the final quark
jet and $\bar{\nu}_{\tau}$ spins, we have
\begin{eqnarray}
A^{inc.}_{VP} & = & 0.1 \frac{m_{b}m_{\tau}}{m_{H^{+}}^{2}}
\frac{Im(\tilde{\xi}_{b} \xi_{\tau}^{\ast}) - 0.3 Im(\tilde{\xi}_{c}^{\ast}
\xi_{\tau}^{\ast})}{1+ 1.15 \frac{m_{\tau}^{2}}{m_{H^{+}}^{2}}
Re(\xi_{b}\xi_{\tau}^{\ast}) + 0.23 (\frac{m_{c}^{2}}{m_{H^{+}}^{2}}
|\xi_{b}\xi_{\tau}^{\ast}|)^{2}} \simeq 2.6\times 10^{-3} \{
\frac{0.04}{|V_{cb}|}\frac{\zeta_{U}}{0.1s_{\beta}} \nonumber \\
& & \cdot \sqrt{\frac{m_{t}}{150GeV}}
Im(\frac{V_{cb}}{|V_{cb}|}\tilde{S}_{32}^{u}\xi_{\tau})
+ 0.13 [Im(\tilde{\xi}_{b} \xi_{\tau}^{\ast})
+ 0.3 Im(\xi_{c} \xi_{\tau}) ] \} (\frac{50 Gev}{m_{H^{+}}})^{2}
\end{eqnarray}

  Comparing to the CP-asymmetry ratio $A_{V_{T}P}^{exc.}$ in the triple
momentum correlation of the exclusive semileptonic decay of the B-meson,
 it is easily seen that the CP-asymmetry ratio $A^{inc.}_{VP}$
in the triple spin-momentum correlation of the inclusive semileptonic decay
is larger by an order-of-magnitude. Within the allowed range of the parameters
considered above, we see that the CP-asymmetry $A^{inc.}_{VP}$ can be as large
as few per cent.  Nevertheless, triple momentum correlations have advantages
for experimental measurement.

   To be sure that an observed triple-product correlations arise from the truly
CP-violating effects, not from the unitarity effects, one can investigate the
sums and differences of the above asymmetries between the particle and
antiparticle decays. This is because the unitarity phases are the same for
particle and antiparticle decays, whereas the CP-violating phases have
opposite sign between the particle and antiparticle decay modes.

\section{Are Higgs Bosons Heavy or Light ?}

  It is well-known that in a general multi-Higgs doublet model
there exist FCNSI, without imposing any conditions, the most stringent
constriant on the mass of the Higgs bosons then arises from the
$K^{0}-\bar{K}^{0}$ and $B^{0}_{d}-\bar{B}^{0}_{d}$ mixings.  In general,
the masses of the Higgs bosons were found to be of order few TeV or
even hundreds TeV.  The situation in our model is quite different, this is
because  the FCNSI are characterized by the parameters $\zeta_{F}$ which
can be naturally small in the assumption of the AGUFS.
So that the mass of the Higgs bosons is less constrained from
the $K^{0}-\bar{K}^{0}$ and  $B^{0}_{d}-\bar{B}^{0}_{d}$ mixings in our model.
We now further show that the other direct and/or indirect experimental
measurements do not yet exclude the existence of the exotic light Higgs
bosons for our model. The mass of the Higgs bosons in our model still opens
to the direct experimental searches at both $e^{+}e^{-}$ and hadron colliders.

\subsection{Implications From Experiments at LEP}

The strongest direct search limits on the Higgs bosons are provided by the CERN
$e^{+}e^{-}$ collider LEP. The mass of the charged Higgs $H^{\pm}$
is restricted to be \cite{LEP,PDG} $m_{H^{\pm}} > 41$ GeV  in all 2HDM,
this limit is independent of the Higgs  branching ratio. The bounds on the
mass of the exotic neutral Higgs bosons remain model-dependent.
In the standard model, there is only one
single neutral physical Higgs and its interacting coupling constants are
fixed by the known parameters and the fermion masses.  Whereas in general
multi-Higgs doublet models, the bounds usually depend on the additional unkown
Yukawa couplings and the mixing angles among the Higgs bosons. These
parameters could have different properties in various models depending on
physical considerations. Even in the simplest 2HDM, there are several schemes
considered in the literature: a very general 2HDM, 2HDM with NFC in which
there exist two scenarios (i.e. so-called model I and model II),
supersymmetric 2HDM or  Minimal Supersymmetric Standard Model (MSSM),
2HDM with VCPV and AGUFS considered in this paper {\it et. al.}.

    Supersymmetric 2HDM has the most stringent constraint on the parameters and
so does on the mass of the Higgs bosons, at tree level one has $m_{H^{\pm}}
\geq m_{W}$, $m_{H^{0}} \geq m_{Z}$, $m_{A^{0}} \geq m_{h^{0}}$ and
$m_{h^{0}} \leq m_{Z}$ \cite{KANE}.  The parameters in the 2HDM with
NFC have also been strongly constrained by the discrete symmetry.  These two
models have been extensively studied by both the theoretists and
experimentlists \cite{LEP,PDG,KANE}. Some  limits on the mass of the  Higgs
bosons in these two models have also been provided by the CERN
$e^{+}e^{-}$ collider LEP\cite{LEP}.  A more stringent constraint on the
mass of the Higgs bosons for these two models has
recently been found \cite{HBBP} from the inclusive $b\rightarrow s \gamma$
decay (see below).

  Nevertheless, our 2HDM with VCPV and AGUFS could be
very different. This is because the AGUFS only act on the fermions and
suppress the FCNSI. Therefore both the diagonal Yukawa couplings and
the mixing angles among the Higgs bosons are in principle free parameters
in this model. Some constraints on the Yuakawa couplings may be extracted
from  other physical phenomena, but they usually remain having a large range
due to uncertainties  from experimental data and /or theoretical estimations.
Obviously,  to fit the current existed experimental data at LEP via our model,
the resulted limits on the mass of the Higgs bosons are expected to
be different from those obtained from fitting the supersymmetric 2HDM
and 2HDM with NFC. In order to see this, let us look at the processes from
which experiments at LEP have been searching for the Higgs bosons

(i) $e^{+} e^{-} \rightarrow Z \rightarrow H_{k}^{0} + Z^{\ast}  \rightarrow
H_{k}^{0} + f \bar{f} $

(ii) $e^{+} e^{-} \rightarrow Z \rightarrow H_{k}^{0} + H_{k'}^{0}  \rightarrow
 f \bar{f} + f' \bar{f'} $

(iii) $e^{+} e^{-} \rightarrow Z \rightarrow H^{+} H^{-} \rightarrow
   \tau^{+} \nu \tau^{-} \nu$, $\tau \nu cs$, $c\bar{s}\bar{c} s$

It is easily seen from the scalar interactions of the gauge bosons
(eqs. (26) and (27) ) that
the branching ratio in the Bjorken process (i) is proportional to the mixing
matrix elements $(O_{2k}^{H})^{2}$ with $\sum_{k=1}^{3} (O_{2k}^{H})^{2} = 1 $,
and in the process (ii) is proportional to the factor
$(O_{1k}^{H}O_{3k'}^{H} - O_{1k'}^{H}O_{3k}^{H})$.
The three mixing angles in the $3\times 3$ orthogonal
matrix  $O_{ij}^{H}$ are in general independent and free parameters since the
Higgs potential in our model is the most general one with only subject to the
gauge invariance and VCPV. Whereas in the
2HDM with NFC and supersymmetric 2HDM, it has $O^{H}_{3i} = 0$, i.e. there
is no mixings among the scalar and pseudoscalar Higgs bosons due to the
discrete symmetry and supersymmetry, so that only one mixing angle
(usually denoted by $\sin(\beta - \alpha)$) exists in those two models.

    On the other hand,  additional new effects
in this  model also arise from the fermionic decays of the Higgs bosons,
this is because the additional Yukawa couplings of the exotic scalars to the
fermions are in principle all free parameters in this model. This is
different from those two models which  only depend on additional
parameters $\tan\beta$ and $\sin \alpha$. Therefore the limits which rely on
the branching ratio of the Higgs bosons could be changed.

  These new features suggest that a more general analysis, from our point of
view, is needed  in order to obtain more general limits on the mass of
the Higgs bosons  from the existed data of the LEP experiments at CERN.
Such a reanalysis within the 2HDM with VCPV
and AGUFS is worthwhile. This is because, on the one hand, it is the simplest
extension of the standard model, and on the other hand, as we have shown that
it may provide a simple scheme to understand the origin and mechanisms of CP
violation and has rich physical phenomena in the presently accessible
energy range.

\subsection{Implications From $b\rightarrow s\gamma$ Decay}

   Recently, two papers\cite{HBBP} by Hewett and Barger {\it et al}  have shown
that the present limit from the CLEO collaboration on the inclusive decay
$b\rightarrow s \gamma$ provides strong constraints on the parameters of
the charged-Higgs boson sector in 2HDM with NFC. They found that in a
supersymmetric type 2HDM the charged Higgs mass is restricted to be
$m_{H^{\pm}}
> 110$ GeV at large $\tan\beta$ with $m_{t} = 150$ GeV,  and even stronger
bounds on $m_{H^{\pm}}$ result for small values of $\tan\beta$ and larger
top quark mass $m_{t}$. In the case of MSSM, when $m_{t} = 150$ GeV the
excluded
region for the mass spectrum of Higgs bosons is comparable to what can be
explored by LEP I and LEP II.  If the CLEO bound is further improved to be
smaller by about $30\%$ and $m_{t} = 150$ GeV, this then largely excludes a
region that would be inaccessible to MSSM Higgs boson searches at both
$e^{+} e^{-}$ and hadron colliders.

    Nevertheless, unlike a supersymmetric 2HDM, our 2HDM with VCPV and
AGUFS remains allowing a relative light Higgs bosons which is accessible
to searches at  $e^{+} e^{-}$ and hadron colliders.
To see this, let us make an analysis for our model.
Following the analyses in refs. \cite{HBBP,GSW}, the ratio of the decay rates
between $b\rightarrow s \gamma$ and $b\rightarrow c e\nu$ is given in a good
approximation by
\begin{equation}
\frac{\Gamma (b\rightarrow s \gamma)}{\Gamma (b\rightarrow c e\nu)}
\simeq 0.031 |c_{7}(m_{b})|^{2}
\end{equation}
with
\begin{eqnarray}
c_{7}(m_{b}) & = & (\frac{\alpha_{s}(m_{W})}{\alpha_{s}(m_{b})})^{16/23}\{
c_{7}(m_{W}) - \frac{8}{7} c_{8}(m_{W}) [ 1- (\frac{\alpha_{s}(m_{b})}{
\alpha_{s}(m_{W})})^{2/23}] \nonumber \\
& & + \frac{232}{513} [1 - (\frac{\alpha_{s}(m_{b})}{
\alpha_{s}(m_{W})})^{19/23}] \}
\end{eqnarray}
where for our model $c_{i}(m_{W})$ have the following results
\begin{eqnarray}
c_{7}(m_{W}) & = & -\frac{1}{2} A(x_{t}) + \tilde{\xi}_{t}\xi_{b}
B(y_{t}) - \tilde{\xi}_{t}\xi_{t}^{\ast} \frac{1}{6} A(y_{t}) \\
c_{8}(m_{W}) & = & -\frac{1}{2} D(x_{t}) + \tilde{\xi}_{t}\xi_{b}
E(y_{t}) - \tilde{\xi}_{t}\xi_{t}^{\ast} \frac{1}{6} D(y_{t})
\end{eqnarray}
with
\begin{equation}
\tilde{\xi}_{t} = \xi_{t} + \frac{\zeta_{U}}{s_{\beta}}
\sqrt{\frac{m_{c}}{m_{t}}}\frac{1}{|V_{ts}|} (\frac{V_{ts}}{|V_{ts}|}
\tilde{S}_{23}^{u})
\end{equation}
where $x_{t} = m_{t}^{2}/m_{W}^{2}$  and $y_{t} = m_{t}^{2}/m_{H^{\pm}}^{2}$.
The functions A, B, D and E are defined in Ref.\cite{GSW}. It is quite distinct
from the 2HDM with NFC for which it is equivalent to the cases
$\xi_{b} = \xi_{t} = \cot\beta$ and $\zeta_{F} = 0$ (i.e. so-called model I)
or $\xi_{b} =\tan\beta$ and $\xi_{t} = -\cot\beta$  (i.e. so called model
II or in supersymmetric 2HDM). In our model,  $\xi_{b}$ and
$\xi_{t}$ are in principle free parameters.  A stringent constraint on
$\xi_{t}$ comes from the $B^{0}-\bar{B}^{0}$ mixing, which we have discussed
in the subsection 3.3. It is likely that $|\xi_{t}| < 1$ for
$m_{t} = 150$ GeV and $m_{H^{\pm}} \sim 100$ GeV if the $B^{0}-\bar{B}^{0}$
mixing is mainly accounted for by the box diagram with W-boson exchange.
Nevertheless, its precise bounds from fitting the $B^{0}-\bar{B}^{0}$ mixing
remain depending on many parameters, such as the values of CKM matrix element
$V_{td}$, hadronic matrix element $\eta^{QCD}B_{B}f_{B}^{2}$ and
top quark mass $m_{t}$. The second term in the $\tilde{\xi}_{t}$ is in general
smaller than one for $\zeta_{U} \sim O(10^{-1})$, $\tilde{S}_{23}^{u}
\sim O(1)$, $|V_{ts}| \sim 0.04$ and $m_{t} = 150$ GeV.

  In general, it is seen that when $|\xi_{t}| < 1 $ and/or  $|\xi_{b}| < 1$ the
constraint on the charged-Higgs mass from $b\rightarrow s \gamma $ could
become weaker than the one from the direct experiment at LEP, i.e.
$m_{H^{\pm}} >41$ GeV. We also note that if $Re(\tilde{\xi}_{t}\xi_{b}) > 0$
and $Re(\tilde{\xi}_{t}\xi_{b}) \gg Im(\tilde{\xi}_{t}\xi_{b})$, a destructive
interferences between the $H^{\pm}$ and $W$ contributions may occur for
some values of the $\xi_{b}$ and $\xi_{t}$. In particular, we see that when
$\tilde{\xi}_{t}\xi_{b}\simeq \tilde{\xi}_{t}\xi_{t}^{\ast}$ and
$|\tilde{\xi}_{t}\xi_{t}^{\ast}| < 4$  and
$m_{t} = 150$ GeV, the bound on $m_{H^{\pm}}$ from $b\rightarrow s \gamma $
becomes weaker than the bound $m_{H^{\pm}} > 41$ GeV.

    It is clear that even if the inclusive decay
$b\rightarrow s \gamma $ is further improved to be small and closed to
the standard model prediction, it can still be fitted by choosing
relatively small Yukawa coupling $\tilde{\xi}_{t}$ and/or $\xi_{b}$.

 Based on these considerations, we then conclude that in our 2HDM with VCPV and
AGUFS, the mass of the Higgs bosons  can be relatively light and still
leave to  the direct experimental seraches at both $e^{+} e^{-}$ and hadron
colliders.  Nevertheless, the inclusive
decay $b\rightarrow s \gamma $ really provides indirectly a stringent
constraint either for the Yukawa coupling constants or for the mass of the
charged scalar.

\subsection{Where are the Higgs Bosons ?}

  In this subsection, we shall eximine some interesting processes from which
the Higgs bosons are expected to be searched for at $e^{+}e^{-}$ and hadron
colliders. In addition,  by giving their explicit decay rates, one can see
how their results in our model could be distinguished from those
in other models.

(i) Bjorken process: $e^{+} e^{-} \rightarrow Z \rightarrow H_{k}^{0}
( \rightarrow  f' \bar{f'} ) + Z^{\ast}( \rightarrow  f \bar{f} ) $

\begin{equation}
\frac{BR(Z \rightarrow H_{k}^{0} + f \bar{f})}{BR(Z \rightarrow f \bar{f})}
=  \frac{g^{2}}{192 \pi^{2} \cos^{2}\theta_{W}} (O_{2k}^{H})^{2} F(y_{k})
\end{equation}
with $F(y_{k})$ the function\cite{KANE} of $y_{k}=m_{H_{k}^{0}}^{2}/m_{Z}^{2}$.
 This is actually searched for at LEP.  Unlike the
2HDM with NFC for which the limits on standard model Higgs boson production
can be simply converted into a limit on $\sin^{2}(\beta - \alpha)$. In our
model, the situation is more complicated since one more independent mixing
element (note that $\sum_{k=1}^{3} (O_{2k}^{H})^{2} = 1$) and three
neutral Higgs bosons are involved in this process due to CP-violating
interactions. Nevertheless, from this single measurement, the following
conclusion remains valid, that is, the Higgs boson which associates with the
small mixing element $O_{2k}^{H}$ is allowed to have  a small mass.

(ii) Two scalar process: $e^{+} e^{-} \rightarrow Z \rightarrow H_{k}^{0} +
H_{k'}^{0}  \rightarrow  f \bar{f} + f' \bar{f'} $

\begin{equation}
\frac{ \Gamma(Z \rightarrow H_{k}^{0} + H_{k'}^{0})}{\Gamma(Z \rightarrow
\nu \bar{\nu})}  =  (\frac{2|p|}{m_{Z}})^{3}\frac{1}{4}
(O_{1k}^{H}O_{3k'}^{H} - O_{1k'}^{H}O_{3k}^{H})^2
\end{equation}
This is being actively searched for at LEP. Where $|p|$ is the
magnitude of the three-momentum of one of the final Higgs bosons. It is also
unlike the 2HDM with NFC and supersymmetric 2HDM, all the three neutral
scalars can be produced in our model depending on the mixings, instead of
single ($h$, $A$) plane in the 2HDM with NFC, one needs to consider three
possible ($H_{k}^{0}$, $H_{k'}^{0}$) planes in our model.

  Where the decay rate of $H_{k}^{0}$ to the fermions is
\begin{equation}
\Gamma (H_{k}^{0}\rightarrow f \bar{f})  =  \frac{3g^{2} m_{f}^{2}
m_{H_{k}^{0}}}{32\pi m_{W}^{2}} (1 -
\frac{4m_{f}^{2}}{m_{H_{k}^{0}}^{2}})^{1/2}
 [ (Re\eta_{f}^{(k)})^{2} (1 - \frac{4m_{f}^{2}}{m_{H_{k}^{0}}^{2}})^{2} +
(Im \eta_{f}^{(k)})^{2} ]
\end{equation}
which depends on parameters $\eta_{f}^{(k)}$.

(iii) Scalar-induced top decay:  $t\rightarrow H^{+} b$ and
$t\rightarrow H^{0}_{k} c$.

This may provide  the most interest processes searching for the light
Higgs boson  when the top quark is discovered at FNAL in the near future.
We could then conclude that the minimal standard model was not nature's choice.
In particular, if $t\rightarrow H^{+} b$ is detected, then the
MSSM might also not nature's choice because, as mentioned in the previous
subsection,  the limit of the inclusive decay $b\rightarrow s\gamma$
reported recently by CLEO nearly closes such a decay channel in the MSSM.
It seems that observation of the $t\rightarrow H^{+} b$ decay channel
might be in favor of our model.

   If top quark mass is less than W-boson mass, these decay modes will
become the dominant ones \cite{HOU,HW}.  For $m_{t} > m_{W} + m_{b}$,
 the ratio of the decay rates of these two channels to the one of the
decay $t\rightarrow W^{+} b$ is given  by
\begin{eqnarray}
\frac{\Gamma (t\rightarrow H^{+} b)}{\Gamma (t\rightarrow W^{+} b)}
& = & \frac{p_{H^{+}}}{p_{W^{+}}} \frac{(m_{t}^{2}+ m_{b}^{2} - m_{H^{+}}^{2})
(m_{b}^{2}|\xi_{b}|^{2} + m_{t}^{2}|\xi_{t}|^{2}) - 4m_{b}^{2}m_{t}^{2}
Re(\xi_{b}\xi_{t}^{\ast})}{(m_{t}^{2}+ m_{b}^{2} - 2m_{W^{+}}^{2})m_{W}^{2}
+ (m_{t}^{2} - m_{b}^{2})^{2}} \\
\frac{\Gamma (t\rightarrow H^{0}_{k} c)}{\Gamma (t\rightarrow W^{+} b)}
& = & \frac{p_{H^{0}_{k}}}{p_{W^{+}}} m_{c}m_{t}\zeta_{U}^{2} \nonumber \\
& & \frac{(m_{t}^{2}+ m_{c}^{2} - m_{H^{0}_{k}}^{2})[(S^{u}_{k,32})^{2} +
(S^{u \ast}_{k,23})^{2}] - 4m_{c}m_{t}
Re (S^{u}_{k,32}S^{u \ast}_{k,23})}{(m_{t}^{2}+ m_{b}^{2} -
2m_{W^{+}}^{2})m_{W}^{2} + (m_{t}^{2} - m_{b}^{2})^{2}}
\end{eqnarray}
The flavor-changing decay modes $t\rightarrow H^{0}_{k} c$ and
$H^{+}\rightarrow c\bar{b}$ could become significant only when the
flavor-changing Yukawa coupling $(S_{1}^{U})_{32}$ becomes unexpected
large in this model.

  In our model,  $H^{+}$ mainly decays to the fermionic states $c\bar{s}$
and $\tau \nu$
\begin{eqnarray}
\Gamma (H^{+}\rightarrow u_{i} \bar{d}_{i}) & = & \frac{3g^{2}}
{32\pi m_{W}^{2} m_{H^{+}}^{3}} \lambda^{1/2}
 \{ (m_{H^{+}}^{2} - m_{d_{i}}^{2} - m_{u_{i}}^{2}) \nonumber \\
& & \cdot (m_{d_{i}}^{2}|\xi_{d_{i}}|^{2} + m_{u_{i}}^{2}|\xi_{u_{i}}|^{2}) +
4m_{d_{i}}^{2}m_{u_{i}}^{2}Re(\xi_{d_{i}}\xi_{u_{i}}^{\ast}) \}  \\
 \Gamma (H^{\pm}\rightarrow \tau^{\pm} \nu) & = &
\frac{3 g^{2}m_{\tau}^{2}}{32\pi m_{W}^{2}} m_{H^{\pm}} (1 -
\frac{m_{\tau}^{2}}{m_{H^{\pm}}^{2}})^{2} |\xi_{\tau}|^{2}
\end{eqnarray}
with $\lambda^{1/2} = [(m_{1}^{2} + m_{2}^{2} - m_{3}^{2})^{2} - 4m_{1}^{2}
m_{2}^{2}]^{1/2}$.

(iv) Higgs boson Decays.

     The fermionic decays of Higgs bosons may be the most important
measurements to distinguish our model
from the usual 2HDM with NFC and supersymmetric 2HDM. As one knows that in
the latter two models, for $m_{H^{0}_{k}} < 2 m_{t}$, the dominant decay
channel of $H_{k}^{0}$ is $b\bar{b}$. Whereas, in our model the dominant
decays of $H_{k}^{0}$ could be no longer the $b\bar{b}$. It is quite possible
that $c\bar{c}$ and $\tau\bar{\tau}$ become the important channels of
$H^{0}_{k}$ decay, this is because $|\xi_{c}|m_{c}$ and
$|\xi_{\tau}| m_{\tau}$ could be larger than $|\xi_{b}|m_{b}$ in this model.
It is also of importance to measure the ratio $BR(H_{k}^{0}\rightarrow c
\bar{c})/BR(H_{k}^{0}\rightarrow b \bar{b})$ and  $BR(H_{k}^{0}\rightarrow \tau
\bar{\tau})/BR(H_{k}^{0}\rightarrow b \bar{b})$ as well as
$BR(H^{-}\rightarrow s \bar{c})/BR(H^{-}\rightarrow \tau \bar{\nu})$.
In the 2HDM with NFC (i.e. model II) and supersymmetric 2HDM, as
$\tan\beta > 1$ the ratio $BR(H_{k}^{0}\rightarrow \tau \bar{\tau})
/BR(H_{k}^{0}\rightarrow b \bar{b})$ should be much larger than the ratio
$BR(H_{k}^{0}\rightarrow c \bar{c})/BR(H_{k}^{0}\rightarrow b \bar{b})$, and
the ratio $BR(H^{-}\rightarrow s \bar{c})
/BR(H^{-}\rightarrow \tau \bar{\nu})$ will be smaller than one. In our model,
all of these ratios could be dramatically changed. In fact, as we have seen
in the sections 4.1 and 5.1,  from fitting the $\Delta m_{K}$ and $\epsilon$,
it favors to have large $|\xi_{c}|$. While to accommodate the
inclusive decay $b \rightarrow s \gamma$, it is likely to have a relatively
small $|\xi_{b}|$.

(v) Probably the following channels are also worthwhile for searching for
the light Higgs bosons

 \[ \mbox{(a)} \qquad Z \rightarrow H^{0}_{k} H^{0}_{k} +
Z^{\ast}( \rightarrow l\bar{l}) \]

 \[ \mbox{(b)} \qquad W^{\pm} \rightarrow H^{0}_{k} H^{0}_{k} + W^{\pm \ast}(
\rightarrow l\nu) \]

with $k=1,2,3$. Note that in our model the  $Z Z H^{0}_{k} H^{0}_{k}$ and
$ W W H^{0}_{k} H^{0}_{k}$ vertexes with two identical Higgs bosons have a
precise prediction from the standard model, i.e.  they are independent of
any additional parameters because of the unitarity condition.

 (vi)  It may be of interest in directly searching for the
charged Higgs boson from purely bosonic interactions since it is less model
dependent. For instance,

\[ Z \rightarrow (H^{+}H^{-})^{\ast} + \gamma \rightarrow 2\gamma + \gamma \]

\[ Z \rightarrow (H^{+}H^{-})^{\ast} + Z^{\ast} \rightarrow 2\gamma +
l\bar{l} \]

\[ W^{\pm} \rightarrow (H^{+}H^{-})^{\ast} + W^{\pm \ast} \rightarrow 2\gamma +
l^{\pm} \nu \]

\[ W^{+} + W^{-} \rightarrow H^{+} + H^{-},  \qquad  \gamma + \gamma
\rightarrow H^{+} + H^{-} \]

\[ Z + Z  \rightarrow H^{+} + H^{-}, \qquad  Z + \gamma  \rightarrow
H^{+} + H^{-} \]

the vertex of  the  above processes in our model is also well known  from
the precise prediction in the standard model.

   There also exists a four-particle vertex which contains the charged
and neutral scalars as well as gauge bosons. This interaction term
may also be interesting in seaching for the light Higgs bosons

\[ Z \rightarrow W^{- \ast} (\rightarrow l\nu ) + H^{+} + H_{k}^{0}
 \qquad  or \qquad W^{\pm} \rightarrow Z^{\ast}( \rightarrow l\bar{l}) +
H^{\pm} + H_{k}^{0} \]

In general, any signal of the existance of the charged scalars indicates a new
physics beyond the standard model.

\section{Conclusions and Remarks}

  With the above detailed analyses and discussions for a 2HDM with VCPV and
AGUFS, we are now in the position to make conclusions and remarks.

   In general, we have observed that
\begin{enumerate}
\item The mechanism of spontaneous symmetry breaking provides not only a
mechanism for giving mass to the bosons and the fermions, but also
a mechanism for generating CP-phase of the bosons and the fermions.
It indicates that if one Higgs doublet is necessary for the genesis of mass,
two Higgs doublets are then needed for origin of CP violation.

\item Origin and mechanisms of CP violation could be understood in a simple
two Higgs-doublet model within the framework of the standard $SU(2)\times
U(1)$ electroweak gauge theory.

\item To prevent the domain-wall problem from arising explicitly at the
weak scale, it is simple to relax the Spontaneous CP Violation (SCPV)
to a Vacuum CP Violation (VCPV). A general criterion for the VCPV has been
discussed.

\item It has shown that from a single CP-phase in the vacuum, CP violation can
occur everywhere it can after spontaneous symmetry breaking. All the weak CP
violations can be classified into four types of CP-violating mechanism
according to their origins and/or interactions.

\item Any successfull theory at the electroweak scale can only possess
Approximate Global $U(1)$ Family Symmetries (AGUFS) based on the fact of the
small Cabibbo-Kobayash-Maskawa quark mixings.
Without imposing any additional conditions, the AGUFS naturally
lead to a   Partial Conservation of Neutral Flavor (PCNF). The Yukawa
coupling matrices have been well-parameterized in a general but very useful
form  which makes the AGUFS and PCNF more manifest. This parameterization
has been found to be very powerful in analyzing various physical phenomena
arising from this model.

\item It has become clear that the smallness of the quark mixing CKM angles
and the established and reported CP-violating effects as well as the observed
suppression of the FCNC can be attributted to the AGUFS and PCNF

\item It has been seen that the masses of the exotic scalars could be weakly
constrained from the established $K^{0}-\bar{K}^{0}$ and $B^{0}-\bar{B}^{0}$
mixings.  This is because the parameters $\zeta_{F}$ characterizing the
Flavor-Changing Neutral Scalar Interactions (FCNSI) can become natural small
by assuming the AGUFS and PCNF.

\item As the AGUFS only act on the fermions and suppress the off
diagonal gauge and scalar interactions of the fermions, thus both
the diagonal Yukawa couplings and the mixing angles among the neutral Higgs
bosons are in principle free parameters. It is because of these freedoms,
we have found that this model can provide a consistent application to
the observed physical phenomena, such as neutral meson mixings and
CP violations as well as the neutron and lepton electric dipole moments
{\it et al}. In addition, the mass of the Higgs bosons in this model still
opens to the direct experimental searches.

In particular, we have shown that

\item  As long as $\zeta_{F} \ll |V_{ub}|$ for $(S_{1}^{F})_{ij} \sim O(1)$,
the effects arising from the FCNSI
at tree level become unobservable small. In the meantime, the induced KM-type
CP violation also becomes negligible. Nevertheless, it is of interest in
observing that the new type of  CP-violating mechanism (type-I) can
consistently accommodate both the indirect- and direct-CP violation in
kaon decay and the neutron and electron EDMs. In particular, the direct
CP violation in kaon decay is expected to be of order $\epsilon'/\epsilon
\sim 10^{-3}$ from  both the long-distance and short-distance contributions
if there is no any accidental cancellations. Furthermore,
The $K^{0}-\bar{K}^{0}$ and $B^{0}-\bar{B}^{0}$ mixings can be fitted
by the short-distance box graphs with additional charged-scalar exchanges.
In addition, the CP violation in hyperon decays can be significant from the
new-type of CP-violating mechanism.

\item  In general, four types of CP-violating mechanism may simultaneously play
an important role on CP violation depending on the parameters and the masses of
the scalars.  It has been seen that the induced KM-type mechanism becomes
important when the scalars are relatively heavy. This is because it favors
large
values of $\zeta_{F} c_{\beta}$, specifically, for example $\zeta_{D}
c_{\beta} \agt 0.1$ and $\zeta_{U} c_{\beta}\agt 0.3$ for $m_{t} \sim 150$ GeV,
which associates to the requirement of a large mass of the scalars from
the established $K^{0}-\bar{K}^{0}$ and $B^{0}-\bar{B}^{0}$ mixings.

The mass differece $\Delta m_{K}$ and the indirect CP-violating parameter
$\epsilon$ in K-system may be accommodated by FCNSI through fine-tuning
 the parameters. Nevertheless, its contribution to the $\epsilon'/\epsilon$
is in general small. In particular, if the CP-violating phases are indeed
generically of order unity, the Flavor-Changing SuperWeak (FCSW)-type
mechanism is then considered to accommodate only the indirect CP violation
($\epsilon$). In this case, the predicted $\epsilon'/\epsilon$ will be
unobservable small (but somehow still larger than the one predicted from
a generic superweak theory).  It has also been noticed that the FCNSI could
also provide sizeable contributions to the neutron and electron EDMs
depending on the parameters $\zeta_{U}$ and $\zeta_{E}$.

It has been seen that

\item  Precisely measuring the direct CP violation ($\epsilon'/\epsilon$) in
kaon decay and the direct CP violation in B-meson system as well as
the electric dipole moments of the neutron ($d_{n}$)
and the electron ($d_{e}$)  are very important for clarifying origin and
mechanisms of CP violation. It is clear that detecting CP-violating effects
at B-factory and colliders is extremely attractive for the purposes of either
extracting the CKM angles or probing new physics which can arise at least
from our model. In particular, the most likely values of $\epsilon'/\epsilon$
and $d_{e}$ as well as $d_{n}$ are already at the level of the present
experimental sensitivity.

\item It should be of interest in measuring CP violation in hyperon decays
and the T-odd and CP-odd triple-product correlations in B-system.

\item It is also worthwhile to measure the
$D^{0}-\bar{D}^{0}$ and $B^{0}_{s}-\bar{B}^{0}_{s}$ mixings
(i.e. $\Delta m_{D}$ and $\Delta m_{B_{s}}$) as well as the muon longitudinal
polarization ($P_{L}$) in the decay $K_{L}\rightarrow \mu
\bar{\mu}$,  this is because $\Delta m_{D}$ and  $P_{L}$ could be much large
than the predictions by the standard model if the FCNSI play an important role.
$\Delta m_{B_{s}}$ may also significantly deviate the prediction from the
standard model due to both the box graphs with charged scalar exchange and
FCNSI.

\item In this model, the mass of the Higgs bosons is likely not to be strictly
constrained from the present indirect experimental measurements, such as
$b \rightarrow s \gamma$ and the neutral meson mixings. Therefore, seaching
for these exotic scalars is worthwhile at both $e^{+}e^{-}$ and hadron
colliders.  The existed experimental data at LEP
should be of interest to be reanalysed in this model. It is of importance
to probe the scalar decays of the gauge bosons and the top quark (after
discovery) for the searches of the possible exotic light Higgs bosons.

\item  It is expected that the mechanisms of CP violation discussed
in this model should also  play an important role in understanding the
baryogenesis at the electroweak scale \cite{TS}.

\end{enumerate}

 Before ending this paper, we would like to make the following remarks:

First, we only examined  some of interesting physical phenomena resulting
from the 2HDM with VCPV and AGUFS. It is believed that more interesting
physical phenomena can arise from this model, for example, CP violation in
purely bosonic \cite{CG} and leptonic interactions. In fact, the leptonic
interactions may provide a useful way to test the FCNSI and the related
CP-violating mechanism, and the scalar-pseudoscalar mixing mechanism may
be directly probed from the purely bosonic interactions.

Second, we are not yet able to provide quantitative predictions for most of the
observables since all the additional parameters appearing in the new
interactions are in general free parameters. Nevertheless, the implications
to the new physical phenomena make the model interest. In fact,
the unkown parameters open a new window for experiments.
Many of the parameters have actually been
restricted to a small range from the present experimental data.

Third, we are also limited to
have a precise prediction concerning the hadronic processes
even if the parameters are fixed. Sizeable uncertainties may arise
from the nonperturbative QCD corrections. Nevertheless, the well-known low
energy effective theories, such as  PCAC and the Chiral Perturbative Theory
(CHPT),  Heavy Quark Effective Theory (HQEFT) and  QCD sume rule, have
provided very useful approaches in dealing with the hadronic matrix elements.
In addition,  the lattice gauge theory is expected to give a more
reliable results.

It should be noted that we have restricted ourselve to
the weak CP violation in this paper and have ignored the strong CP
problem. Solving this problem is also believed to result in
interesting physical phenomena. In fact, Peccei-Quinn mechanism\cite{PQ}
provides a very attractive scheme for solving this problem,
we expect to discuss it in detail elsewhere.

 Last but not least, we should keep in mind that the model discussed above may
remain an effective one in the sense that it contains many parameters. These
parameters are in principle free, which is analogous to the standard model.
Actually, it is the simplest extension of the standard model, therefore
we do not expect to be able to answer the questions which also appear in the
standard model. Nevertheless, the model discussed here is renormalizable,
and may be considered internally consistent.  What we have shown is that
such a simple two-Higgs doublet model with VCPV and AGUFS possesses very
rich phenomenological features, in particular the phenomenology of CP
violation. What we may do is to determine and/or
restrict all these physical parameters from the direct and/or indirect
experimental measurements, just like what we have been doing \cite{ET}
for the standard model and for the effective chiral perturbation theory
as well as for the heavy quark effective theory.

\vspace{5mm}

{\bf Acknowledgments}

 The author has greatly benefited from discussions with R.F. Holman,
L.F. Li, E.A. Paschos and L. Wolfenstein. He also wishes to thank Professor
K.C. Chou for having introduced him to this subject. This work was
supported by DOE grant \# DE-FG02-91ER40682.

\newpage

{\bf Appendix}

\vspace{1cm}

  In this appendix, we present some functions and quantities appearing in the
text.

A. \   The integral functions from the box diagrams with W boson and charged
scalar exchanges

\begin{eqnarray*}
   B^{WW}(x, x') & = & \sqrt{xx'}\{ (\frac{1}{4} + \frac{2}{3}\frac{1}{1-x} -
\frac{3}{4}\frac{1}{(1-x)^{2}}) \ln \frac{x}{x-x'} \\
& & + (x \leftrightarrow x') - \frac{3}{4} \frac{1}{(1-x)(1-x')} \} \\
   B^{WW}(x) & = & \frac{1}{4} + \frac{9}{4}\frac{1}{1-x} -
\frac{3}{2}\frac{1}{(1-x)^{2}} +\frac{3}{2}\frac{x^{2}}{(1-x)^{3}} \ln x \\
   B^{HH}_{V}(y, y') & = & 16\pi^{2} m_{H}^{2} I_{4} \\
 & = & \frac{y^{2}}{(y-y')(1-y)^{2}} \ln y + (y \leftrightarrow y')
+ \frac{1}{(1-y)(1-y')}  \\
 B^{HH}_{V}(y) & = & 16\pi^{2} m_{H}^{2} I_{1} \\
 & = & \frac{1+y}{(1-y)^{2}} +  \frac{2y}{(1-y)^{3}} \ln y  \\
   B^{HW}_{V}(y, y', y_{W}) & = & 16\pi^{2} m_{H}^{2} (\frac{1}{4}I_{6}
+ m_{W}^{2} I_{5}) =  \frac{(y_{W}-1/4)\ln y_{W}}{(1-y)(1-y')(1-y_{W})} \\
& & + \frac{y(y_{W}-y/4)}{(y-y')(1-y)(y-y_{W})} \ln \frac{y_{W}}{y} + (y
\leftrightarrow y')  \\
   B^{HW}_{V}(y, y_{W}) & = & 16\pi^{2} m_{H}^{2} (\frac{1}{4}I_{3}
+ m_{W}^{2} I_{2})  =  \frac{y_{W}-1/4}{(1-y)(y-y_{W})} \\
& & + \frac{(y_{W}-y/4)}{(1-y)^{2}(1-y_{W})} \ln y
+ \frac{3}{4}\frac{y_{W}^{2}}{(y_{W}-y)^{2}(1-y_{W})} \ln \frac{y_{W}}{y}
  \\
  B^{HH}_{S}(y, y') & = & -\frac{y\ln y}{(y-y')(1-y)^{2}} \ln y -
(y \leftrightarrow y') - \frac{1}{(1-y)(1-y')}  \\
 B^{HH}_{S}(y) & = & -\frac{1}{(1-y)^{2}}[\frac{1+y}{1-y}  \ln y + 2 ] \\
   B^{HW}_{S}(y, y', y_{W}) & = & 16\pi^{2} m_{H}^{2} I_{6} =
- \frac{\ln y_{W}}{(1-y)(1-y')(1-y_{W})} \\
& & +  \frac{y^{2}}{(y-y')(1-y)(y-y_{W})} \ln \frac{y_{W}}{y} + (y
\leftrightarrow y')   \\
 B^{HW}_{S}(y, y_{W}) & = & 16\pi^{2} m_{H}^{2} I_{3}  =
-\frac{y}{(1-y)(y-y_{W})} \\
& &  -\frac{\ln y}{(1-y)^{2}(1-y_{W})}
-\frac{y_{W}^{2}}{(y_{W}-y)^{2}(1-y_{W})} \ln \frac{y_{W}}{y}
  \\
   B^{HW}_{T}(y, y', y_{W}) & = & 16\pi^{2} m_{H}^{2} m_{W}^{2} I_{5} =
 y_{W} \{ \frac{\ln y_{W}}{(1-y)(1-y')(1-y_{W})}  \\
& & + \frac{y\ln y/y_{W}}{(y-y')(1-y)(y-y_{W})}  + (y
\leftrightarrow y')  \\
   B^{HW}_{T}(y, y_{W}) & = & 16\pi^{2} m_{H}^{2} m_{W}^{2} I_{2} =
y_{W} \{ \frac{1}{(1-y)(y-y_{W})} \\
& & + \frac{\ln y}{(1-y)^{2}(1-y_{W})}
- \frac{y_{W} \ln y/y_{W}}{(y_{W}-y)^{2}(1-y_{W})}
\end{eqnarray*}
where the functions $I_{i}$ ($i=1, \cdots , 6$) are the euclidean integrals
of \cite{ASW}.

B. \   The quantities used for calculating the CP-violating parameter
$\epsilon$.

\begin{eqnarray*}
B_{ij}(m_{i}, m_{j}) & = & \sqrt{y_{i}y_{j}}\{ \frac{1}{4}
 |\xi_{i}|^{2}|\xi_{j}|^{2} B_{V}^{HH}(y_{i}, y_{j})
+ 2  Re(\xi_{i}\xi_{j}^{\ast}) B_{V}^{HW}(y_{i}, y_{j}, y_{W}) \}
\\
& & + \eta_{ij} B^{WW}(x_{i}, x_{j}) + \frac{m_{s}^{2}}{4m_{i}m_{j}}
\{ \frac{\tilde{B}_{K}}{B_{K}}
(\frac{m_{K}}{(m_{d}+m_{s})})^{2} \sqrt{y_{i}y_{j}} B_{S}^{HH}(y_{i}, y_{j})
\xi_{ij}^{2}  \\
& & + 2 \frac{m_{d}}{m_{s}} \sqrt{y_{i}y_{j}} B_{V}^{HH}(y_{i}, y_{j})
[\xi_{s}\xi_{d}^{\ast}\xi_{i}\xi_{j}^{\ast} + \frac{1}{2} \sqrt{\frac{m_{d}
m_{s}}{m_{i}m_{j}}} (\xi_{s}\xi_{d}^{\ast})^{2}] \}
\end{eqnarray*}
with
\[ \xi_{ij}^{2} = (\xi_{i}\xi_{s} - \frac{m_{d}}{m_{s}}\xi_{i}^{\ast}
\xi_{d}^{\ast})(\xi_{j}\xi_{s} - \frac{m_{d}}{m_{s}}\xi_{j}^{\ast}
\xi_{d}^{\ast}) - 2 \tilde{r}_{K} \frac{m_{d}}{m_{s}} \xi_{i}\xi_{s}
(\xi_{j}\xi_{d})^{\ast} \]

The involved functions arise from the box graphs. From gluonic penguin graph
with
charged scalar exchange, we have

\[ P^{H}_{T}(y_{i}) = \frac{1}{2(1-y_{i})} + \frac{1}{(1-y_{i})^{2}}
+ \frac{1}{(1-y_{i})^{3}} \ln y_{i} \]

and

\[ P^{H}_{i}(m_{i}, \xi_{i}) = (\xi_{s} \xi_{i} - \frac{m_{d}}{m_{s}} \xi_{d}
\xi_{i}) y_{i} P^{H}_{T}(y_{i}) \]

C.  \   The function used in estimating the CP-violating parameter
$\epsilon'/\epsilon$.

\[ P^{H}_{V}(y_{t}) = \frac{1}{2}|\xi_{t}|^{2} \frac{y_{t}}{1-y_{t}} \{
 \frac{1-3y_{t}}{(1-y_{t})^{3}} \ln y_{t} \frac{5}{6} -
\frac{2y_{t}}{(1-y_{t})^{2}} \}  \]

D.  \   The functions appearing in the calculations of the neutron and
electron EDMs.

\[ h_{NH} = \frac{1}{4} (z_{t}^{(k)})^{2} \int_{0}^{1} dx
\int_{0}^{1} dy \frac{x^{3}y^{3}(1-x)}{[z_{t}^{(k)} x(1-xy) +
(1-x)(1-y)]^{2}} \]

\[ h_{CH} \simeq \frac{1}{4} \frac{y_{t}}{(1-y_{t})^{3}} (-\ln y_{t}
-\frac{3}{2} + 2 y_{t} - \frac{1}{2} y_{t}^{2} ) \]

\[ f(z_{t}^{(k)}) = \frac{1}{2}z_{t}^{(k)} \int_{0}^{1} dx
\frac{1-2x(1-x)}{x(1-x)-z_{t}^{(k)}} \ln \frac{x(1-x)}{z_{t}^{(k)}} \]

\[ g(z_{t}^{(k)}) = \frac{1}{2}z_{t}^{(k)} \int_{0}^{1} dx
\frac{1}{x(1-x)-z_{t}^{(k)}} \ln \frac{x(1-x)}{z_{t}^{(k)}} \]
where $y_{t} = m_{t}^{2}/m_{H^{+}}^{2}$ and $z_{t}^{(k)} = m_{t}^{2}
/m_{H^{0}_{k}}^{2}$. In $h_{CH}$ we have neglected the bottom quark mass.

\end{document}